\documentclass[fleqn,usenatbib]{mnras}

\usepackage{newtxtext,newtxmath}

\usepackage[T1]{fontenc}
\usepackage{ae,aecompl}

\usepackage{graphicx}	
\usepackage{amsmath}	
\usepackage{amssymb}	
\usepackage{comment}    
\usepackage{gensymb}    
\usepackage{lscape}
\usepackage{booktabs}

\title[Kick velocity distribution of BHXBs]{Potential Kick Velocity distribution of black hole X-ray binaries and implications for natal kicks}

\author[P. Atri et al.]
{P. Atri,$^{1}$\thanks{Email: pikky.atri@icrar.org (PA)}
J. C. A.~Miller-Jones,$^{1}$
A. Bahramian,$^{1}$
R. M. Plotkin,$^{1}$
P. G. Jonker,$^{2,3}$
\newauthor 
G. Nelemans,$^{3,4}$
T. J. Maccarone,$^{5}$
G. R. Sivakoff,$^{6}$
A. T. Deller,$^{7}$
S. Chaty,$^{8}$
\newauthor
M. A. P. Torres,$^{9,3}$
S. Horiuchi,$^{10}$ 
J. McCallum,$^{11}$ 
T. Natusch,$^{12}$
C. J. Phillips,$^{13}$ 
\newauthor
J. Stevens,$^{13}$ 
S. Weston$^{12,13}$ 
\\ \\
$^{1}$International Centre for Radio Astronomy Research, Curtin University, GPO Box U1987, Perth, WA 6845, Australia\\
$^{2}$SRON, Netherlands Institute for Space Research, Sorbonnelaan 2, NL-3584 CA Utrecht, the Netherlands\\
$^{3}$Department of Astrophysics/IMAPP, Radboud University Nijmegen, POBox 9010, NL-6500 GL, Nijmegen, the Netherlands\\
$^{4}$Institute of Astronomy, KU Leuven, Celestijnenlaan 200D, B-3001 Leuven, Belgium\\
$^{5}$Department of Physics, Box 41051, Science Building, Texas Tech University, Lubbock, TX 79409-1051, USA\\
$^{6}$Department of Physics, University of Alberta, CCIS 4-183, Edmonton, AB T6G 2E1, Canada\\
$^{7}$Centre for Astrophysics and Supercomputing, Swinburne University of Technology, Mail Number H11, P.O. 218, Hawthorn, VIC 3122 Australia\\
$^{8}$AIM, CEA, CNRS, Universit\`{e} Paris-Saclay, Universit\`{e} Paris Diderot, Sorbonne Paris Cit\`{e}, F-91191 Gif-sur-Yvette, France\\
$^{9}$Departamento de Astrof\'{i}sica, Universidad de La Laguna, E-38206, Santa Cruz de Tenerife, Spain\\
$^{10}$CSIRO Astronomy and Space Science, Canberra Deep Space Communication Complex, Tuggeranong, ACT 2901, Australia\\
$^{11}$University of Tasmania, Private Bag 37, Hobart, Tasmania, 7001, Australia\\
$^{12}$Institute for Radio Astronomy \& Space Research, AUT University, 1010, Auckland, New Zealand\\
$^{13}$CSIRO, P.O. Box 76, Epping, NSW 1710, Australia\\
}

\date{Accepted XXX. Received YYY; in original form ZZZ}

\pubyear{2019}

\hypersetup{draft}

\begin{document}
\label{firstpage}
\pagerange{\pageref{firstpage}--\pageref{lastpage}}
\maketitle

\begin{abstract}
We use Very Long Baseline Interferometry to measure the proper motions of three black hole X-ray binaries (BHXBs). Using these results together with data from the literature and \textit{Gaia}--DR2 to collate the best available constraints on proper motion, parallax, distance and systemic radial velocity of 16 BHXBs, we determined their three dimensional Galactocentric orbits. We extended this analysis to estimate the probability distribution for the potential kick velocity (PKV) a BHXB system could have received on formation. Constraining the kicks imparted to BHXBs provides insight into the birth mechanism of black holes (BHs). Kicks also have a significant effect on BH-BH merger rates, merger sites, and binary evolution, and can be responsible for spin-orbit misalignment in BH binary systems. 75\% of our systems have potential kicks $>70\,\rm{km~s^{-1}}$. This suggests that strong kicks and hence spin-orbit misalignment might be common among BHXBs, in agreement with the observed quasi-periodic X-ray variability in their power density spectra. We used a Bayesian hierarchical methodology to analyse the PKV distribution of the BHXB population, and suggest that a unimodal Gaussian model with a mean of 107$\pm$16$\,\rm{km~s^{-1}}$  is a statistically favourable fit. Such relatively high PKVs would also reduce the number of BHs likely to be retained in globular clusters. We found no significant correlation between the BH mass and PKV, suggesting a lack of correlation between BH mass and the BH birth mechanism. Our Python code allows the estimation of the PKV for any system with sufficient observational constraints.
\end{abstract}




\section{Introduction}\label{Section 1}
Theoretical models suggest the formation of a stellar mass black hole (BH) could take place either by (a) direct collapse, whereby the massive progenitor star collapses directly into a BH with almost no mass ejection \citep[e.g.][]{Reynolds2015,Adams2017a}, or (b) delayed formation in a supernova (SN), wherein matter ejected during the SN falls back onto the proto-neutron star, creating a BH \citep[e.g.][]{Gourgoulhon1991,Woosley1995,Brandt1995}. The theoretical mass distribution derived by \citet{Fryer2001} suggests that beyond a critical progenitor mass of $40M_{\odot}$ (assuming no mass loss through winds), BHs form via direct collapse and a supernova explosion does not occur. However, more recent work suggests that the formation of BHs by direct collapse (implosion) has a wider progenitor mass range, interspersed by narrow pockets of progenitor masses that undergo a supernova explosion to become a BH \citep{Sukhbold2016}. Obtaining observational evidence about how a BH was born for a range of BH masses could help in understanding the relation between the BH birth model and BH mass.\par
Isolated stellar mass BHs are difficult to observe. Thus, black hole X-ray binaries (BHXBs) can be used as probes to understand the birth mechanism of BHs in binary systems. If a BH in a binary system is formed due to a SN explosion, the sudden mass loss from the binary system will change the centre of mass of the remnant binary system. Even if the SN explosion is symmetric in the frame of reference of the progenitor, the change in the centre of mass of the binary system will give the remnant binary system a recoil kick. This kick is referred to as the Blaauw kick \citep{Blaauw1961}. The ejected mass in the SN needs to be less than half the total mass of the system in order for the remnant binary system to stay bound. This corresponds to a maximum recoil kick of a few tens of \,km\,s$^{-1}$, depending on the binary mass, progenitor velocity, orbital period, and ejected mass of the system \citep{Nelemans1999}.\par
BHXBs that have peculiar velocities higher than the Blauuw kick the system could have acquired due to symmetric mass loss, need some other mechanism to explain the high velocities. The space velocity distribution of pulsars was shown to have a mean of 450\,km\,s$^{-1}$ with a high-velocity tail \citep{Lyne1994}. \citet[][V17 hereafter]{Verbunt2017} used accurately measured pulsar proper motions and parallaxes applying a Bayesian method to estimate distances from parallaxes. They found that the velocity distribution was wider than can be fit by a single Maxwellian, and has two prominent peaks, at 120\,$\rm{km~s^{-1}}$ and 540\,$\rm{km~s^{-1}}$. The high velocity kicks \citep{Lyne1994} have been attributed to a variety of mechanisms during the SN explosion. Even if the matter ejected in the SN explosion is symmetric, the anisotropy in the neutrino emission caused by strong magnetic fields could accelerate pulsars to high velocities \citep{Chugai1984,Sagert2008}. Other mechanisms that could be responsible for high kick velocities include supersonic jet generation during a stellar core collapse \citep{Khokhlov1999}, or an intrinsically asymmetric explosion due to hydrodynamic perturbations in the SN core \citep{Janka2017}.\par
Since these kick mechanisms occur prior to the fallback of matter on the proto-neutron star and consequent BH formation \citep{Gourgoulhon1993}, it is plausible that BHs formed by the fallback process should also receive such strong natal kicks. It is expected that the velocity kicks BHs receive will be smaller than NS velocity kicks because BHs are more massive than NSs. On the other hand, fallback of slower moving ejecta from the SN explosion could accelerate the BH \citep{Janka2013}. This would imply that BHs born in a SN explosion could also get large kicks at birth. BHs born by direct collapse are expected to receive kicks lower than those obtained due to birth in SN explosion \citep{Belczynski2002}. Hence, natal kick measurements provide effective probes to differentiate between the SN and direct collapse birth pathways. 
\par
In the absence of direct kick measurements, the height of known BHXBs above the Galactic plane has been used as a proxy \citep[e.g.][]{White1996,Jonker2004} given that the majority of the progenitor systems are closely confined to the plane. Various natal kick distributions were used to simulate BHXB populations and it was found that natal kicks are essential to explain the large displacement of a number of systems from the Galactic plane \citep{Repetto2012,Repetto2017}. However, \citet{Mandel2016} showed that for extreme assumptions (i.e. all systems are currently at maximum z) BH kicks of $>$100\,km\,s$^{-1}$ are not required to explain the observed distribution of heights ($|z|$) above the Galactic plane. Measuring space velocities and positions of BHXBs provides extra information that helps to relate the $|z|$-distribution to the natal kicks received by the BHs and in turn resolve this discrepancy.
\subsection{BH-BH mergers} \label{Section 1.1}
With the discovery of BH mergers from gravitational wave (GW) events \citep{Abbott2016a}, we are achieving new insights into the stellar-mass BH mass distribution \citep{LIGO2018,Abbott2016b} and stellar mass BH formation. According to theoretical models, BH-BH binaries are formed due to dynamical interactions in high density clusters \citep{Fabian1975,Goodman1993,Sigurdsson1993a,Sigurdsson1993b,Benacquista2013} eg., in globular clusters (GCs) \citep{Rodriguez2016} and Galactic nuclear clusters. They can also be formed due to hierarchical three body interactions \citep{Antonini2018}, or in the Galactic field, which hosts the astrophysical stellar binaries that evolve into BH-BH binaries \citep{Belczynski2002,Postnov2014}.\par
Strong natal kicks could lead to the BHs being kicked out of a GC before they could become BH-BH binaries, or result in unbinding the binary system. This would hinder BH-BH binary formation and in turn reduce the rate of BH-BH mergers. Natal kicks are thus a deciding factor in the rate of BH-BH mergers \citep{Wysocki2018}. Natal kicks are also considered to be responsible for spreading the BH binary merger sites further away from the host halo \citep{Kelley2010}. While it is currently poorly constrained, the BH natal kick distribution is a key parameter in N-body simulations of GCs, used to estimate the number of BHs that are retained or ejected upon formation and due to subsequent dynamical interactions \citep[e.g.][]{Strader2012,Heggie2014,Giesler2018}. Thus, the natal kick distribution has important implications for the likelihood of finding BH-BH binaries, and also BHXBs in GCs. 
\subsection{Existing natal kick constraints}\label{Section 1.2}
While we know of $\sim$\,60 strong candidate BH systems \citep{Tetarenko2016a,Corral-Santana2016}, only seven BHXB candidates have been analysed to determine if they were born in a supernova with substantial mass ejection, rather than a direct collapse. System parameters like component masses, orbital period, donor effective temperature and the three dimensional motion of BHXBs need to be well constrained if the complete evolutionary history of these BHXB systems is to be mapped. The Galactocentric orbits of such systems with well known parameters could be integrated back to a reasonable range of their ages, thus the velocity of the system immediately after the BH birth could be determined. Such an analysis has been conducted for three systems, namely, XTE\,J1118+40 \citep{Willems2005}, GRO\,J1655--40 \citep{Fragos2009} and Cyg\,X--1 \citep{Wong2012}. \par
The post SN peculiar velocity of XTE\,J1118+40 along with the assumption that the system was born in the Galactic plane indicated that the system received an asymmetric natal kick \citep{Fragos2009}. The high peculiar velocity could also mean that the system was a GC escapee \citep{Mirabel2001,Gualandris2005}, in which case the assumption that it was born in the Galactic plane would not hold. It was suggested that GRO\,J1655--40 was born in a SN and received a kick when the BH was formed \citep{Brandt1995,Mirabel2002,Willems2005}, though the strength of the kick did not require an asymmetric explosion \citep{Nelemans1999}. Cyg\,X--1 was found to have received a small kick velocity at birth \citep{Mirabel2003,Wong2012}.\par 
Only the peculiar velocities were used to infer whether the kicks BHXB systems received were high enough to suggest a SN origin for the other four systems \citep[e.g.][]{Mirabel2001,Mirabel2003,Dhawan2007,Miller-Jones2009a,Gandhi2019}. This has been done by measuring the proper motion of the BHXBs, and combining them with the line-of-sight velocities and the distances to the systems to measure the velocities of the systems with respect to the local standard of rest (i.e. the peculiar velocity). This gives the full three dimensional motion of the BHXBs, which enables us to trace the Galactocentric orbits through the Galactic potential. A low peculiar velocity was measured for GRS\,1915+105 \citep{Dhawan2007,Reid2014}, which suggested that the system could have been born by direct collapse. The inferred peculiar velocities of V404\,Cyg \citep{Miller-Jones2009a}, VLA\,J2130+12 \citep{Tetarenko2016} and MAXI\,J1836--194 \citep{Russell2015} were more consistent with a symmetric SN explosion. This is a small sample size, with estimations of the minimum natal kick having been made using different methods. Thus in order to draw statistically robust conclusions about the natal kick distribution of BHs, we need to increase this sample size and adopt a more systematic approach. In this paper we increase the sample size of estimated kick velocities of BHXBs to 16, by combining measured proper motions, systemic radial velocities and distances to these systems.\par
\subsection{Peculiar velocity and Potential kick velocity}\label{Section 1.3} 
Using the current peculiar velocity of a system to infer the strength of the kick the system might have received at birth has a few limitations. Peculiar velocity is the current three dimensional velocity of a system relative to the local standard of rest and thus changes based on the epoch of observation. Also, using peculiar velocity relative to that expected from Galactic rotation as an indicator of natal kick velocity could be misleading for sources that are far away from the Galactic plane. Our analysis tries to minimise these shortcomings by instead using the peculiar velocity of the system when it crosses the Galactic plane, which we refer to as the potential kick velocity (PKV). \par
In this paper we have combined the capabilities of VLBI and {\it{Gaia}} to get observationally constrained BH potential kick velocity probability distributions for 16 systems. In Section \ref{Section 2} we identify the BHXB sample we used for our analysis and mention the observational biases this sample suffers from. In Section \ref{Section 3} we discuss the data acquisition and reduction procedures followed to measure proper motions for GX\,339--4, GRS\,1716--249 and Swift\,J1753.5--0127, which is followed by Section \ref{Section 4} where we present the proper motion measurements of the above mentioned three systems. In Section \ref{Section 5} we go through the details of distance estimation from parallax measurements using a Milky Way prior (Section \ref{Section 5.1}) and the systemic radial velocity estimates for the systems that did not have measured systemic radial velocity in literature (Section \ref{Section 5.2}). Section \ref{Section 6} explains the Monte Carlo (MC) code we developed to determine the PKV probability distributions and present the potential kick velocity distributions for all 16 sources . Section \ref{Section 7} discusses the implications of these results.
\section{Data}\label{Section 2}
\subsection{BHXB sample}\label{Section 2.1}
To determine the three dimensional motion of BHXBs, we need to combine the proper motion, systemic radial velocity and distances to these systems. The literature already contains estimates of the current peculiar velocity for seven of our systems, XTE\,J1118+480 \citep{Mirabel2001}, GRO\,1655--40 \citep{Mirabel2002}, Cyg\,X--1 \citep{Mirabel2003}, GRS\,1915+105 \citep{Dhawan2007}, V404\,Cyg \citep{Miller-Jones2009a}, MAXI J1836--194 \citep{Russell2015} and VLA\,J2130+12 \citep{Tetarenko2016}. {\it{Gaia}} in its second data release (DR2) measured the proper motions of 11 BHXBs, three of which were improvements on previous measurements \citep[Cyg X--1, GRO\,J1655--40 and XTE\,J1118+480;][]{Mirabel2001,Mirabel2002,Mirabel2003}. \citet{Gandhi2019} estimated the current peculiar velocities for BHXBs for which \textit{Gaia}-DR2 \citep{Brown2018} measured proper motions and parallaxes. We use these proper motion and parallax measurements along with the Very Long Baseline Interferometry \citep[VLBI;][]{Shapiro1979} measurements present in the literature (based on whichever was more precise) for our analysis.\par
Typical BHXBs have proper motions of a few mas\,yr$^{-1}$, which can be measured by the {\it{Gaia}} satellite \citep{Lindegren2016} or with VLBI. The quiescent optical brightness of BHXBs is usually near to or below the limiting magnitude of the {\it{Gaia}} satellite, and they only get bright enough for high precision astrometry with {\it{Gaia}} when they are in outburst. This makes proper motion measurements challenging for many of these sources. {\it{Gaia}} would also not be able to detect BHXBs towards the Galactic bulge due to high extinction. Thus, triggered VLBI observations when a BHXB goes into an outburst can probe systems not accessible to {\it{Gaia}}. We used VLBI to measure the proper motions of three new sources; GX\,339--4, Swift\,J1753.5--0127 and GRS\,1716--249.
\subsection{Sample biases}\label{Section 2.2}
Our sample set of 16 BHXB suffers from certain observational biases. There are a few systems in our sample that are not dynamically confirmed BHs but are BHXB candidates (see \citet{Tetarenko2016} for a summary of all BHXB candidates). Notably, the nature of VLA\,J2130+12 that is a BHXB detected in quiescence \citep{Kirsten2014} is still under debate \citep{Tetarenko2016a}. We include Cyg\,X--1 in our analysis even though it is a younger, high mass X-ray binary (HMXB) with a potentially more massive donor star than that of a typical low mass X-ray binary. This is because it has well constrained parameters and thus is a good test source. \par
We note that almost all BHXBs in our sample have been in outburst at some point, except VLA\,J2130+12. During outbursts typical BHXBs reach X-ray peak luminosities of $>$10$^{37-39}$\,ergs\,s$^{-1}$. Thus the sample is devoid of any very faint X-ray transients \citep{Wijnands2006}, which have peak luminosities of 10$^{36-37}$\,ergs\,s$^{-1}$ \citep{Heise1999}. We are biased against observing BHs that received kicks strong enough to unbind the binary system, as we can only observe BHs in X-ray binaries. We may also be observationally biased against distant, low kick systems as they will be very close to the Galactic plane and therefore will be highly extincted.
\section{VLBI Observations and Data reduction}\label{Section 3}
\begin{table*}
        \begin{center}
                \caption{Summary of the VLBI observations. Station codes: Ak - ASKAP; At - phased up ATCA; Br - Brewster; Cd - Ceduna; Ef - Effelsberg; Fd - Fort Davis; Gb - Green Bank; Hh - Hartebeesthoek; Hn - Hancock; Ho - Hobart; Jb - Jodrell Bank Mk II; Ke - Katherine; Kp - Kitt Peak; La - Los Alamos; Mc - Medicina; Mk - Mauna Kea; Mp - Mopra; Nl - North Liberty; Nt - Noto; On - Onsala; Ov - Owens Valley ; Pa - Parkes; Pt - Pie Town;  Sc - St. Croix; Td - the 34\,m DSS36 antenna at Tidbinbilla; Ti - the 70\,m DSS43 antenna at Tidbinbilla; Tr - Torun; Wa - Warkworth 30m \citep{Woodburn2015}; Wb - Westerbork; Ww - Warkworth 12m; Yg - Yarragadee; Ys - Yebes \label{tab:table1}}
                \begin{tabular}{l c c c c c c l}
                \hline \hline
                        Target & Array & Code & Date & MJD & Freq. & Time & Stations \\
                        &  &  & &  &  & on &\\ 
                        &  &  & &  &  & Source &\\ 
                         &  &  & & (UTC) & (GHz) & (mins) &\\
                \hline
                        GX 339-4 & LBA & V430A & 2011 Apr 03 & 55654.84 & 8.4 & 368.2 & At, Cd, Ho, Mp, Pa, Ti \\
                        && V486B & 2013 Aug 15 & 56520.20 & 8.4 & 193.9 & Ak, At, Cd, Hh, Ho, Ke, Mp, Pa, Ti, Ww, Yg\\
                        && V447A & 2014 Nov 22 & 56983.17 & 8.4 & 254.4 & At, Cd, Ho, Mp\\
                        GRS 1716-249 & LBA & V447D & 2017 Feb 21 & 57805.41 & 8.4 &  283.0 & At, Cd, Ho, Mp \\
                        & & V447E & 2017 Apr 22 & 57865.73 & 8.4 & 329.3 &  At, Cd, Ho, Mp\\
                        & & V447F & 2017 Aug 13 & 57978.48 & 8.4 & 279.5 & At, Cd, Ho, Ke, Mp, Pa, Td, Ti, Wa, Yg\\ 
                        Swift J1753.5-0127 & VLBA & BM331A & 2009 Dec 16 & 55181.81 & 8.4 & 149.7 & Br, Fd, Hn, Kp, La, Mk, Nl, Ov, Pt, Sc \\
                        & & BM331B & 2009 Dec 19 & 55184.79 & 8.4 & 147.9 & Br, Fd, Hn, Kp, La, Mk, Nl, Ov, Pt, Sc \\
                        & & BM331C & 2009 Dec 22 & 55187.81 & 8.4 & 148.6 & Br, Fd, Hn, Kp, La, Mk, Nl, Ov, Pt, Sc \\
                        & & BM331D & 2009 Dec 24 & 55189.79 & 8.4 & 147.9 & Br, Fd, Hn, Kp, La, Mk, Nl, Ov, Pt, Sc \\
                        & VLBA+GBT & BM326 & 2010 Mar 30 & 55285.47 & 8.4 & 189.1 & Br, Fd, Gb, Hn, Kp, La, Mk, Nl, Ov, Pt, Sc \\
                        & EVN & EM101A & 2012 Nov 13 & 56245.03 & 5.0 & 156.9 & Ef, Jb, Mc, Nt, On, Tr, Ys, Wb, Hh \\
                        & & EM101B &  2013 Mar 20 & 56371.18 & 5.0 & 117.4 & Ef, Jb, Mc, Nt, On, Tr, Ys, Wb, Hh \\
                        & & EM101C &  2013 Jun 18 & 56461.99 & 5.0 & 174.2 & Ef, Jb, Mc, Nt, On, Tr, Ys, Wb, Hh \\
                        & & EM101D &  2013 Sep 17 & 56552.74 & 5.0 & 192.1 & EF, Jb, Mc, Nt, On, Tr, Ys, Wb, Hh \\
                        & & EM101E &  2013 Dec 03 & 56629.75 & 5.0 & 220.8 & Ef, Jb, Mc, Nt, On, Tr, Ys, Wb, Hh \\\hline
                \end{tabular}
        \end{center}
        
\end{table*}
\begin{table*}
        \begin{center}
                \caption{Information about the calibrators used. The R. A. and Dec. positions of the calibrators are those mentioned in the Astrogeo VLBI calibrator search website \url{http://astrogeo.org/calib/search.html} (rfc2019a catalogue). The integrated flux density is the average of the cleaned flux of all epochs of the calibrator. $\theta_{sep}$ is the angular separation of the target from the calibrator. \label{tab:table2}}
                \begin{tabular}{l c c c c c l}
                \hline \hline
                        Target & Array & Calibrators & R. A. (J2000) & Dec. (J2000) & $\theta_{sep}$ & Integrated flux density \\
                        & & & (h m s) & ($^\circ $ $^\prime$ $^{\prime\prime}$) & $^\circ $ & (Jy) \\ \hline
                        GX 339--4 & LBA & J1711--5028 & 17$\rm{^h}$11$\rm{^m}$40$\rm{^s}$.9927 & -50$^\circ$28$^\prime$17$^{\prime\prime}$.409 & 2.21 & 0.08 \\
                        GRS 1716--249 & LBA & J1711--2509 & 17$\rm{^h}$11$\rm{^m}$23$\rm{^s}$.1020 & -25$^\circ$09$^\prime$01$^{\prime\prime}$.564& 1.87 & 0.1 \\
                        Swift J1753.5--0127 & VLBA & J1743--0350 & 17$\rm{^h}$43$\rm{^m}$58$\rm{^s}$.8591 & -03$^\circ$50$^\prime$04$^{\prime\prime}$.617 & 3.36 & 3 \\
                        & EVN & J1743--0350 & 17$\rm{^h}$43$\rm{^m}$58$\rm{^s}$.8591 & -03$^\circ$50$^\prime$04$^{\prime\prime}$.617 & 3.36 & 2.4  \\
                        & & J1752--0147 & 17$\rm{^h}$52$\rm{^m}$18$\rm{^s}$.3640 & -01$^\circ$47$^\prime$16$^{\prime\prime}$.685 & 0.45 & 0.14  \\
                        \hline
                \end{tabular}
        \end{center}
\end{table*}
To measure the proper motions of GX\,339--4, GRS\,1716--249 and Swift\,J1753.5--0127, we used the Very Long Baseline Array (VLBA), the European VLBI Network (EVN) and the Australian Long Baseline Array (LBA). The observational setups are summarised in Table \ref{tab:table1}. The hard X-ray spectral state of BHXBs is associated with the rising and decaying phases of the outbursts \citep{Belloni2016}. The radio jets during the hard spectral state are compact, steady and are causally connected to the BH \citep{Fender2009}. Thus the hard state is an ideal phase during the outburst to conduct astrometry. High precision astrometry can be conducted even if the target image is resolved \citep{Reid2011}. The observations need to be separated in time by at least a couple of months to provide a sufficient time baseline for a proper motion measurement. Thus, depending on how long the source stays in a hard state, we need to observe it over one or multiple outbursts. \par
The data were correlated using the DiFX software correlator \citep{Deller2011} and reduced using the Astronomical Image Processing System \citep[AIPS 31DEC17;][]{Greisen2003}. Observations of bright fringe finder sources were used to correct for non-zero instrumental delays and rates. The observations of the target were bracketed by shorter observations of a phase reference calibrator, located as close to the target as possible \citep[preferably <2$^\circ$;][]{Pradel2006}. The details of the calibrators used for each target are summarised in Table \ref{tab:table2}.\par
For observations made at frequencies higher than 5\,GHz using the VLBA, numerous widely separated calibrators were observed in quick succession (geodetic blocks) for $\sim$30 minutes at the start and end of each observing run. Geodetic blocks help in determining any error in the estimated zenith tropospheric delay and thus improve astrometric accuracy. The multi-band delays were corrected using standard \textit{AIPS} procedures (\textit{AIPS} Memo 110). The overheads associated with observing a geodetic block with the LBA and the EVN were high due to the large dish sizes and consequent slow slew rates of the dishes involved, and hence we did not observe geodetic blocks with the LBA and the EVN. The stations included in the LBA and the EVN are not solely for the purpose of VLBI, and so have infrequent formal VLBI observing sessions. For any transients that have to be observed out of the standard observing sessions, the source is observed with whatever stations are available. This gives varying sensitivity and resolution from epoch to epoch.
\subsection{GX 339--4}\label{Section 3.1}
GX\,339--4 is a low mass black hole X-ray binary (LMXB), which was first detected in 1973 \citep{Markert1973}. This system goes into frequent outbursts, with 20 outbursts since its detection \citep{Tetarenko2016a}. Hence, astrometry over multiple outbursts is possible. The optical emission from the accretion disc of the system even during the low luminosity quiescent phase makes the detection of the donor star difficult \citep{Shahbaz2001}. Recently, NIR spectra of GX\,339--4 were obtained using the VLT/X-Shooter during quiescence, which led to a systemic radial velocity measurement of $\gamma = 26\pm2$\,km\,s$^{-1}$ \citep{Heida2017} and a distance estimate of 9$\pm$4\,kpc for the system.
GX\,339--4 was observed on three epochs (Table \ref{tab:table1}) using the LBA during the hard states of three different outbursts \citep{Homan2005,Buxton2013,Yan2014} spanning over 4 years. We observed at a frequency of 8.4\,GHz in all three epochs for maximum sensitivity.
\subsection{GRS 1716--249}\label{Section 3.2}
GRS\,1716--249 (Nova Oph 1993) was discovered \citep{Ballet1993} as an X-ray transient in Ophiuchus  in 1993. \citet{Della1994} conducted a photometric and spectroscopic analysis of GRS\,1716--249 and concluded that the source is at a distance of 2.4$\pm$0.4\,kpc. The mass of the primary in the system is $>$4.9 M$_{\odot}$ \citep{Masetti1996}. The systemic radial velocity of this system is not yet known. After a prolonged quiescent period \citep{Negoro2016} following some renewed activity in 1995 \citep{Karitskaya1995}, the source went into outburst again in December 2016, but did not reach a soft state. The source was observed at three epochs under the LBA program V447 (Table \ref{tab:table1} and Table \ref{tab:table2}), and the observations were spaced a few months apart. The last two observations were taken right after the source finished brief softening periods \citep{Bassi2019}. Ceduna did not have valid data for three hours of observations for the first epoch. The last epoch was obtained during a scheduled VLBI run and hence had access to all telescopes in the LBA other than Hartebeesthoek.
\subsection{Swift J1753.5--0127}\label{Section 3.3}
Swift\,J1753.5--0127 is a high Galactic latitude system which was first discovered when it went into outburst in 2005 \citep{Palmer2005}. It has one of the shortest known orbital periods (3.2443 $\pm$ 0.001 hr) for a BHXB \citep{Zurita2008}. This system remained in outburst for 11 years before it ultimately faded to quiescence in 2016 November \citep{Russell2015}. The distance to this source is not well known; the best available estimation being 4--8\,kpc \citep{Cadolle2007}. The mass of the BH in this system is $> 7.4\,M_{\odot}$ \citep{Shaw2016}.\par
Swift\,J1753.5--0127 was observed using the EVN, the VLBA and the HSA (High Sensitivity Array) for a total of nine epochs. The four VLBA observations in December 2009 were conducted with the dual 13/4-cm recording mode (2.3 and 8.4\,GHz), but we only use the 8.4\,GHz data to measure the position due to reduced scattering, reduced ionospheric effects that could give rise to systematic errors in astrometry, and higher sensitivity (Table \ref{tab:table1}). In addition to the standard calibration steps, a global model of the phase calibrator was made by stacking the calibrated data sets of the phase calibrator from all four epochs to prevent minor differences in the model (due to varying uv-coverage) from introducing astrometric systematics. The calibrated data sets of the target from 2008 were also stacked, as the array does not have the capability to detect a week's motion of the system. Swift\,J1753.5-0127 was detected in the stacked image with a significance of $\sim$5.2$\sigma$. GBT was a part of the HSA in addition to the standard VLBA dishes for the 8.4\,GHz observations in March 2010, where the target was strongly detected ($\sim$124$\sigma$). \par
The EVN observations of Swift\,J1753.5--0127 were taken as part of a parallax measurement campaign (EM101) in 2012--2013, and consisted of 5 epochs taken over a period of one year. The phase calibrator J1743--0350, which was used in the earlier VLBA observations, was suitable for measuring the proper motion of the target but not the parallax, due to its large angular separation (3.36$^{\circ}$) from the target. We therefore conducted an X-ray binary calibrator survey on the VLBA (program code BS206) to find closer compact calibrators. We used the multi-phase centre capability of the DiFX software correlator \citep{Deller2011} to correlate on the positions of all NVSS sources within 30$^{\prime}$ of the target, observing at 1.6\,GHz to give a field of view such that only four pointings were required to cover the desired sky area.  Within 30$^{\prime}$ of Swift\,J1753.5--0127 we detected two potential calibrators, J1752--0147 (17$^{\rm h}$52$^{\rm m}$18.364$^{\rm s}$, $-01^{\circ}$47$^{\prime}$16.685$^{\prime\prime}$) and J1753--0102 (17$^{\rm h}$53$^{\rm m}$10.4488$^{\rm s}$, $-01^{\circ}$02$^{\prime}$48.854$^{\prime\prime}$). While the latter turned out to be a widely-separated double, the former was a compact, single source of peak flux density 6--9\,mJy, and was therefore adopted as a secondary phase calibrator for our EVN parallax campaign, and was $0.44^{\circ}$ away from our target.\par
To make sure that we were using the same source models to solve for the delays and rates, we concatenated the calibrated data sets of the primary calibrator for all epochs. This model was then used to derive (epoch-wise) the phase, delay and rate solutions that were interpolated to the secondary calibrator and the target. The secondary calibrator was then imaged and the data were stacked to obtain a global model \citep[as done by][]{Miller-Jones2018}. The global model was again used to derive epoch-wise phase solutions that were applied to the target data, which were then imaged. The target positions were then measured by fitting a point source model in the image plane.
\section{Results - Proper motion measurements}\label{Section 4}
\begin{table*}
        \begin{center}
        
        \caption{Summary of the detections of GX\,339--4, GRS\,1716--249 and Swift\,J1753.5--0127. MJD is taken as the middle of the observing run and the error on MJD is calculated as half the length of the observation. The positions, peak intensities and errors on each parameter are calculated by fitting an elliptical Gaussian to the target in the image plane.\label{tab:table3}}
                \begin{tabular}{l c c c c c c l}
                \hline \hline
                        Target & Project Code & MJD & R. A. (J2000) & Dec. (J2000)  & Peak Intensity \\
                         &  & (UTC) & (h m s) &($^\circ$ $^\prime$ $^{\prime\prime}$) & ($\mu$Jy\,bm$^{-1}$) \\ \hline
                         GX 339--4 & \phantom{B}V430A & 55654.84$\pm$0.25& 17$\rm{^h}$02$\rm{^m}$49$\rm{^s}$.38260$\pm$0.00004 & -48$^\circ$47$^\prime$23$^{\prime\prime}$.1413$\pm$0.0003 & \phantom{1}197$\pm$18\phantom{1} \\
                         & \phantom{B}V486B & 56520.21$\pm$0.24 & 17$\rm{^h}$02$\rm{^m}$49$\rm{^s}$.38164$\pm$0.00001 & -48$^\circ$47$^\prime$23$^{\prime\prime}$.1534$\pm$0.0001 & 2280$\pm$163\\
                         & \phantom{B}V447A & 56983.17$\pm$0.16 & 17$\rm{^h}$02$\rm{^m}$49$\rm{^s}$.38114$\pm$0.00001 & -48$^\circ$47$^\prime$23$^{\prime\prime}$.1591$\pm$0.0001 & 2629$\pm$203\\
                         GRS 1716--249 & \phantom{B}V447D & 57805.41$\pm$0.31 & 17$\rm{^h}$19$\rm{^m}$36$\rm{^s}$.92008$\pm$0.00002 & -25$^\circ$01$^\prime$04$^{\prime\prime}$.1215$\pm$0.0005 & 1488$\pm$156 \\
                         & \phantom{B}V447E & 57865.73$\pm$0.23 & 17$\rm{^h}$19$\rm{^m}$36$\rm{^s}$.92003$\pm$0.00001  & -25$^\circ$01$^\prime$04$^{\prime\prime}$.1286$\pm$0.0001 & 1135$\pm$105 \\
                         & \phantom{B}V447F & 57978.48$\pm$0.19 & 17$\rm{^h}$19$\rm{^m}$36$\rm{^s}$.91994$\pm$0.00003 & -25$^\circ$01$^\prime$04$^{\prime\prime}$.1294$\pm$0.0002 & \phantom{1}290$\pm$42\phantom{1} \\
                         Swift J1753.5--0127 & \phantom{B}BM331 & 55186.05$\pm$0.10& 17$\rm{^h}$53$\rm{^m}$28$\rm{^s}$.29060$\pm$0.00001 & -01$^\circ$27$^\prime$06$^{\prime\prime}$.2916$\pm$0.0002 & \phantom{1}290$\pm$56\phantom{1} \\
                         & \phantom{B}BM326 & 55285.47$\pm$0.14 & \phantom{1}17$\rm{^h}$53$\rm{^m}$28$\rm{^s}$.29061$\pm$0.000003 & -01$^\circ$27$^\prime$06$^{\prime\prime}$.2919$\pm$0.0001 & 2614$\pm$21\phantom{1} \\
                         & EM101A & 56245.03$\pm$0.48 & 17$\rm{^h}$53$\rm{^m}$28$\rm{^s}$.29074$\pm$0.00002 & -01$^\circ$27$^\prime$06$^{\prime\prime}$.3011$\pm$0.0003 & \phantom{1}280$\pm$27\phantom{1} \\
                         & EM101B & 56371.18$\pm$0.08 & 17$\rm{^h}$53$\rm{^m}$28$\rm{^s}$.29077$\pm$0.00005 & -01$^\circ$27$^\prime$06$^{\prime\prime}$.3037$\pm$0.0006 & \phantom{1}161$\pm$31\phantom{1} \\
                         & EM101C & 56461.99$\pm$0.09 & 17$\rm{^h}$53$\rm{^m}$28$\rm{^s}$.29078$\pm$0.00003 & -01$^\circ$27$^\prime$06$^{\prime\prime}$.3036$\pm$0.0004 & \phantom{1}141$\pm$20\phantom{1} \\
                         & EM101D & 56552.74$\pm$0.13 & 17$\rm{^h}$53$\rm{^m}$28$\rm{^s}$.29081$\pm$0.00002 & -01$^\circ$27$^\prime$06$^{\prime\prime}$.3053$\pm$0.0004 & \phantom{1}207$\pm$27\phantom{1} \\
                         & EM101E & 56629.75$\pm$0.10 & 17$\rm{^h}$53$\rm{^m}$28$\rm{^s}$.29082$\pm$0.00004 & -01$^\circ$27$^\prime$06$^{\prime\prime}$.3054$\pm$0.0006 & \phantom{1}109$\pm$21\phantom{1} \\
                \hline
                \end{tabular}
        \end{center}
\end{table*}
Our VLBI data were used to find the proper motions of GX\,339--4 and GRS\,1716--249 for the first time. We also measured the proper motion of Swift\,J1753.5--0127, for which \textit{Gaia} also had made a measurement. The position measurements of the three sources at all epochs are summarised in Table \ref{tab:table3}.
\subsection{GX 339--4}
\begin{figure}
\centering
\includegraphics[width=0.4\textwidth]{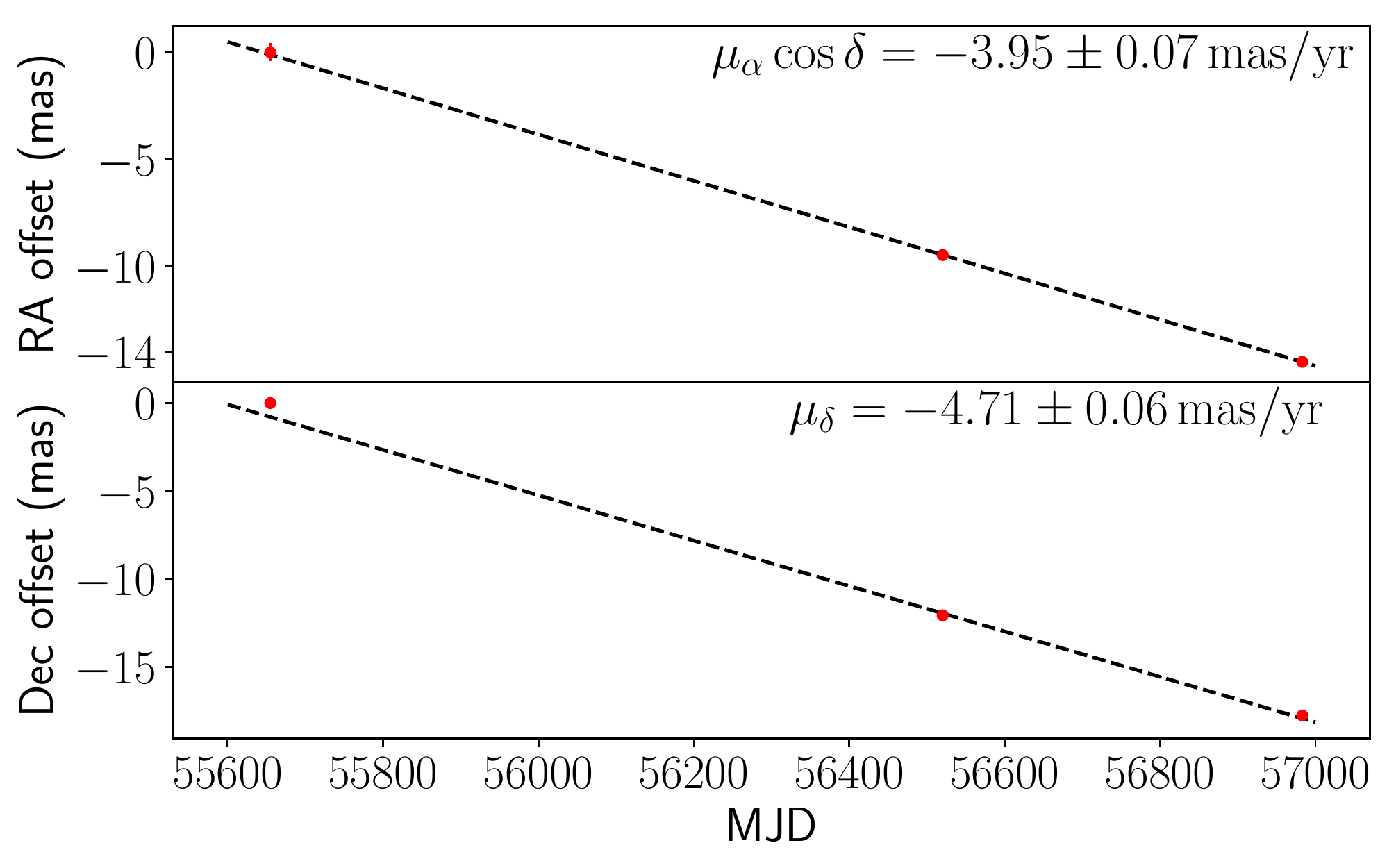}
 \caption{Proper motion fit for GX\,339--4. GX\,339--4 was tracked for three epochs over three different outbursts. Top panel: Offset of measured positions  relative to the first epoch in Right Ascension. Bottom panel: Offset of measured positions in Declination  relative to the first epoch.}
\label{gx339propermotion}
\vspace{0.3cm}
\end{figure}
As shown in Figure \ref{gx339propermotion}, for GX\,339--4 a linear fit for the proper motion of the system gives:
\begin{equation}
\begin{aligned}
\mu_{\alpha}\cos\delta=-3.95\pm0.07\,\mathrm{mas}\,\mathrm{yr^{-1}} \\
\mu_{\delta}= -4.71\pm0.06\,\mathrm{mas}\,\mathrm{yr^{-1}}. 
\end{aligned}
\end{equation}
This gives an overall proper motion for GX\,339--4 of 6.15$\pm$0.06\,mas\,yr$^{-1}$.
\subsection{GRS 1716--249}
\begin{figure}
\centering
\includegraphics[width=0.45\textwidth]{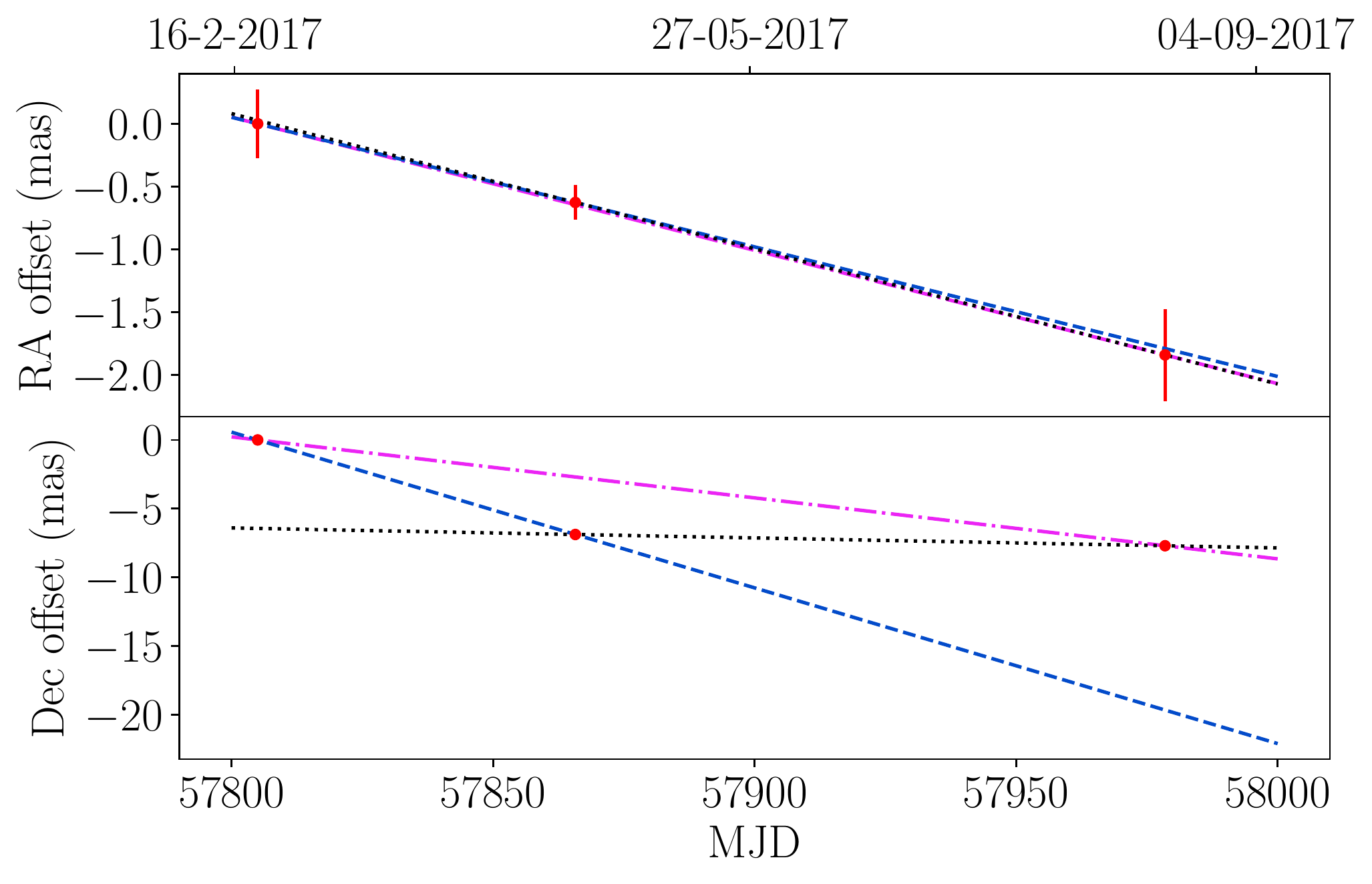}
\includegraphics[width=0.45\textwidth]{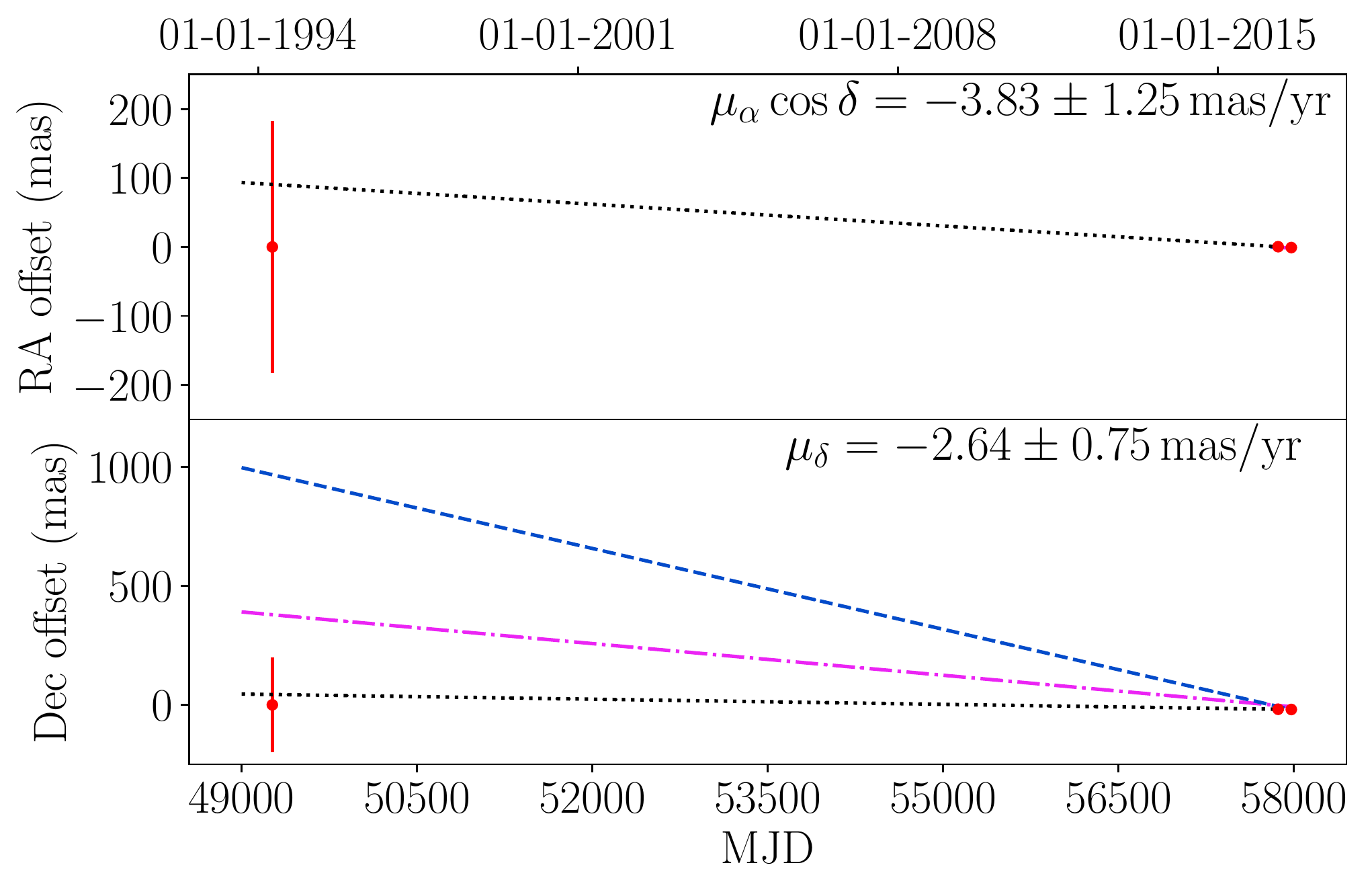}
\caption{Proper motion fits for GRS\,1716--249. GRS 1716--249 was tracked for three epochs over one outburst using the LBA, with positions shown with respect to the first LBA epoch. Top panel: fits by considering position measurements of two epochs at a time in RA and Dec - epoch 1 and 2 (dashed line); epoch 2 and 3 (dotted line); epoch 1 and 3 (dotted dashed line). The three declination measurements do not lie on a straight line. Lower panel: fitting for proper motion by increasing the time baseline using the 1994 K-band image position measurement (dotted line), which breaks the degeneracy and shows that the epoch 1 LBA position is in error (blue and green dashed lines). This was then removed from the fit.}
 \label{grs1716propermotion}
  \vspace{0.3cm}
\end{figure} 
The measured offsets of GRS\,1716--249 in RA and Dec for our three epochs of LBA observations have been plotted in Figure \ref{grs1716propermotion}. Measured declinations in these epochs do not appear to follow a linear trend (Top panel of Fig. \ref{grs1716propermotion}). Choosing different subsets of measurements yield drastically different results. The proper motions of the different fits (considering two epochs at a time) in RA agree with each other within the 1$\sigma$ error limit. On the contrary, the proper motion fits in Dec vary between $-41.4\,\rm{mas\,yr^{-1}}$ and $-2.5\,\rm{mas\,yr^{-1}}$. This suggests that the measured position in Declination for one out of the three epochs is unreliable. \par
To ascertain which of the epochs were giving accurate positions, we obtained an archival position measurement of GRS\,1716--249 from a near infrared observation to increase the time baseline. We use the position of GRS\,1716--249 as measured from the K-band image taken on MJD 49263 during the 1994 outburst \citep{Chaty2002}. We fixed the image astrometry to the \textit{Gaia}--DR2 frame by cross matching it with $\sim$140 sources from the \textit{Gaia}--DR2 catalogue using \textit{PyRAF}. The proper motions of these sources as reported by \textit{Gaia}--DR2 were used to correct the positions of the sources. The position of GRS\,1716--249 was determined with an uncertainty of $\sim$0.2\,arcsec. This shows that the position from the first LBA epoch is the cause of the discrepancy. As mentioned in Section \ref{Section 2.2}, Ceduna was not available for observing during most of that observing run. Thus the uv coverage was sparse and could explain the wrong position measurement. To determine the true proper motion, we therefore fit using the last two LBA epochs (V447E and V447F) and the position measurement from the 1994 K-band image. We obtain the following proper motion
\begin{equation}
\begin{aligned}
\mu_{\alpha}\cos\delta=-3.83\pm1.25\,\mathrm{mas}\,\mathrm{yr^{-1}} \\
\mu_{\delta}= -2.64\pm0.75\,\mathrm{mas}\,\mathrm{yr^{-1}}. \\
\end{aligned}
\end{equation}
This gives an overall proper motion of 4.65$\pm$1.12\,mas\,yr$^{-1}$ for GRS\,1716--249.
\subsection{Swift J1753.5--0127}
\begin{figure}
\centering
\includegraphics[width=0.48\textwidth]{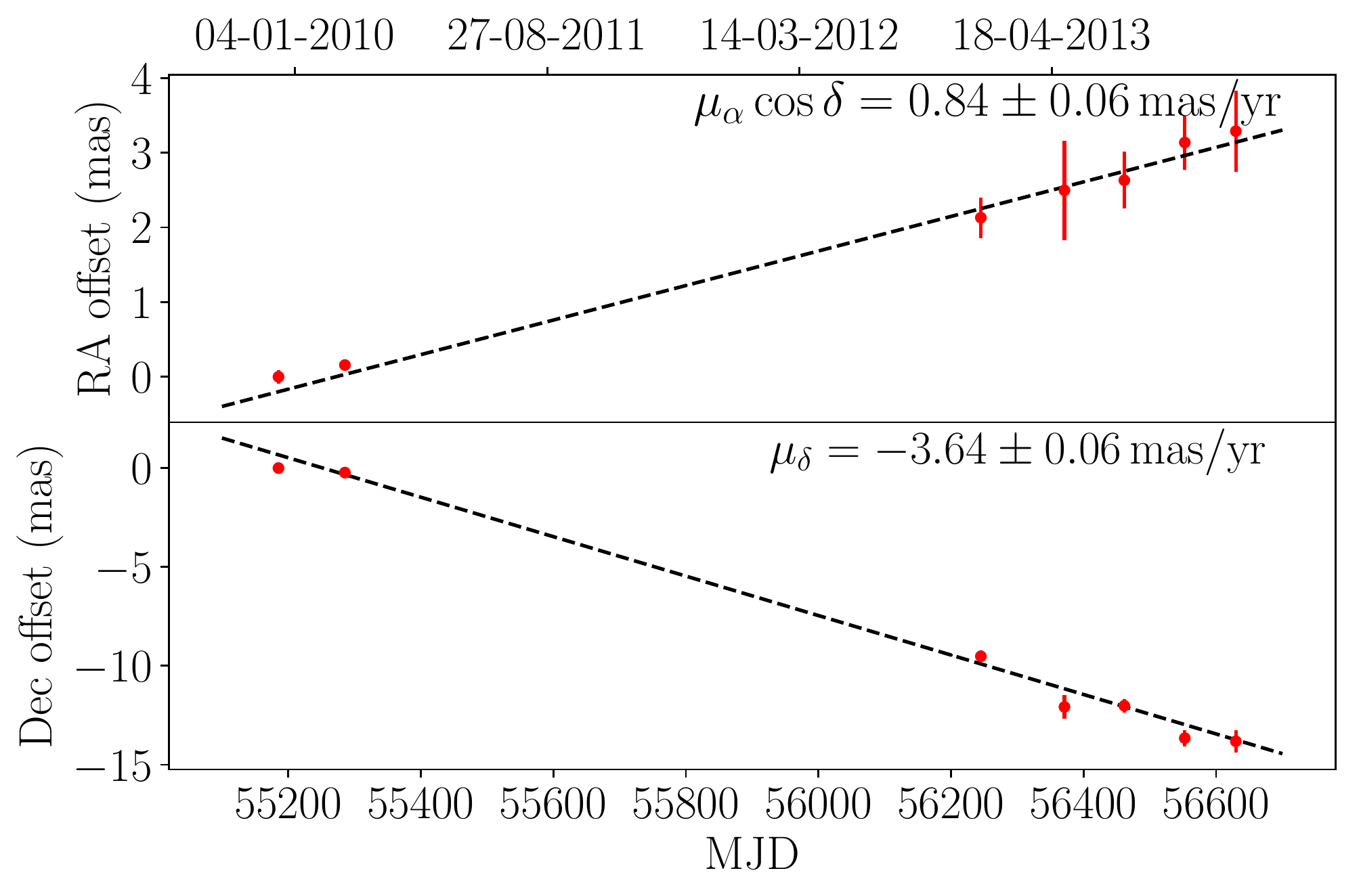}
\caption{Proper motion fit for Swift\,J1753.5--0127. Swift\,J1753.5--0127 was observed for ten epochs spanning four years. The four observations in 2009 December were concatenated into one position measurement (MJD 55181 to MJD 55190). Top panel: offset of measured positions in Right Ascension of Swift\,J1753.5--0127 as a function of MJD relative to the first epoch. Bottom panel: offset of measured positions in Declination of Swift\,J1753.5--0127 as a function of MJD relative to the first epoch. }
\label{swiftj1753propermotion}
\vspace{0.3cm}
\end{figure}
Figure \ref{swiftj1753propermotion} shows a linear fit to the measured positions of Swift\,J1753.5--0127 over a span of 4 years.  Although we used a different calibration scheme for the VLBA and EVN epochs, the nearby secondary calibrator used for the EVN (J1752--0147) was observed using the VLBA primary calibrator J1743--0350 as a phase reference source, such that the two data sets should be referenced to the same absolute reference frame. The absolute astrometric systematics should therefore be of order 0.09 and 0.27\,mas, respectively in R.A.\ and Dec. The fit gives a proper motion of 
\begin{equation}
\begin{aligned}
\mu_{\alpha}\cos\delta=0.84\pm0.06\,\mathrm{mas}\,\mathrm{yr^{-1}} \\
\mu_{\delta}= -3.64\pm0.06\,\mathrm{mas}\,\mathrm{yr^{-1}}. \\
\end{aligned}
\end{equation}
This gives a net proper motion of $3.73\pm0.06\,\rm{mas\,yr^{-1}}$. The proper motion determined above agrees within 2$\sigma$ error limits with that reported by \textit{Gaia}--DR2, which is
\begin{equation}
\begin{aligned}
\mu_{\alpha}\cos\delta=1.13\pm0.16\,\mathrm{mas}\,\mathrm{yr^{-1}} \\
\mu_{\delta}= -3.53\pm0.15\,\mathrm{mas}\,\mathrm{yr^{-1}}. \\
\end{aligned}
\end{equation}
\section{Distance and radial velocity estimates}\label{Section 5}
We will fold the proper motions measured by our VLBI data for the three systems as reported in Section \ref{Section 4}, and those available in archival data (using VLBI and {\textit{Gaia}}) with distance and systemic radial velocities of these systems to get their potential kick velocity distribution. 
\subsection{Distance - Milky way prior}\label{Section 5.1}
The distances to some of the sources in our sample have been inferred from a parallax measurement made by \textit{Gaia}--DR2 \citep{Brown2018} or VLBI. Every \textit{Gaia} measurement of a parallax has an uncertainty associated with it. The fractional errors associated with these measurements are significant, even if they are small in absolute value. Thus, it is essential that these errors are properly considered when converting parallax to distance. In order to make sure that the errors from the measured quantity are appropriately propagated to the inferred quantity, we use a Bayesian inference approach \citep{Astraatmadj2016}. \par
We use the likelihood function as described in \citet{Bailer-Jones2015} and \citet{Gandhi2019}. We follow the work by \citet[][GR02 hereafter]{Grimm2002} to make an analytical model of the LMXB density distribution in our Galaxy to use as the prior. This considers the density of the bulge, disc and spheroid of the Galaxy, as described by the following equations (equations 4,5,6 in GR02).
\begin{equation}
\begin{aligned}
\rm{\rho_{Bulge}} = &~ \rho_{\rm{0,Bulge}}\cdot\left(\frac{\sqrt{r^{2}+\frac{z^{2}}{q^{2}}}}{r_{0}}\right)^{-\gamma}\cdot\exp\left({-\frac{r^{2}+\frac{z^{2}}{q^{2}}}{r_{\rm{t}}^{2}}}\right),\\
\rm{\rho_{Disk}} = &~ \rm{\rho_{0,Disk}}\cdot\exp\left({-\frac{r_{\rm{m}}}{r_{\rm{d}}}-\frac{r}{r_{\rm{d}}}-\frac{|z|}{r_{\rm{z}}}}\right),\\
\rm{\rho_{Sphere}} = &~ \rm{\rho_{0,Sphere}}\cdot\frac{\exp\left({-b\cdot(\frac{R}{R_{\rm{e}}})^{\frac{1}{4}}}\right)}{(\frac{R}{R_{\rm{e}}})^{\frac{7}{8}}}.
\end{aligned}
\end{equation}
Here \textit{r} is the distance from the Galactic centre of the projected position of the source on the Galactic plane, $z$ is the height of the source above the Galactic plane, and $R$ is the distance of the source from the Galactic centre. $q$, $r_{\rm{0}}$, $r_{\rm{t}}$, $r_{\rm{d}}$, $r_{\rm{m}}$, $r_{\rm{z}}$ and $R_{\rm{e}}$ are scale parameters. GR02 derived these scale parameters for LMXBs by constructing the X-ray luminosity function of these systems. We use the values of these constants as summarised in Table 4 of GR02. $\rm{\rho_{0,Bulge}}$, $\rm{\rho_{0,Disk}}$ and $\rm{\rho_{0,Sphere}}$ are normalisation constants. We estimate the density constants $\rm{\rho_{0,Bulge}}$, $\rm{\rho_{0,Disk}}$ and $\rm{\rho_{0,Sphere}}$ for LMXBs using the Disk:Bulge:Sphere mass ratio of 2:1:0.5 as derived by GR02, and assuming the mass of the bulge as $\rm{1.3 \times 10^{10}M_{\odot}}$ (GR02). This gave the values of $\rm{\rho_{0,Bulge}}$, $\rm{\rho_{0,Disk}}$ and $\rm{\rho_{0,Sphere}}$ as 1.1\,$\rm{M_{\odot}\,pc^{-3}}$, 2.6\,$\rm{M_{\odot}\,pc^{-3}}$ and 13.1\,$\rm{M_{\odot}\,pc^{-3}}$ respectively. We note that the spatial distribution model of GR02 is not exclusively for BHXBs as their work considered neutron star X-ray binaries as well, but GR02 is currently the best available model for the spatial distribution of X-ray binaries as existing stellar spatial distributions do not account for kicks.\par
The parallaxes measured by \textit{Gaia}--DR2 have a global zero-point offset of -0.029\,mas \citep{Luri2018}. Hence the parallax values from Gaia were corrected before being used for the distance estimation using the Milky Way prior. These corrected parallax values along with the estimated values of distances of 11 BHXBs for which \textit{Gaia}--DR2 measured parallaxes has been reported in Table \ref{tab:table4} and Table \ref{tab:table5}. There are also reports that the \textit{Gaia}--DR2 parallaxes on average have a systematic offset of -0.075$\pm$0.029\,mas \citep{Xu2019}, but we use the more established -0.029\,mas offset \citep{Luri2018} for the correction of the parallaxes. 11 of the 16 systems in our sample also have distance estimations in the literature (see Table \ref{tab:table4} and \ref{tab:table5} for more details), and {\it{Gaia}}-DR2 measured the parallax of six of those systems (1A\,0620--00, XTE\,J1118+480, GS\,1124--684, GRO\,J1655--40, Swift\,J1753.5--0127 and SAX\,J1819--2525). We compare the distance estimates derived from \textit{Gaia} parallaxes to the ones available in the literature and use the better constrained distance for determining and analysing the PKV distributions.
\subsection{Systemic radial velocity estimates}\label{Section 5.2}
We have four systems with poorly constrained systemic radial velocities (Swift\,J1753.5--0127, MAXI\,J1820+070, VLA\,J2130+12 and GRS\,1716--249). Using the best distance estimates for these systems, we project them onto the Galactic plane. We estimate the expected systemic radial velocity ($\bar{\gamma}$) of the system at this projected distance onto the Galactic plane and assuming that the system is undergoing pure Galactic rotation about the Galactic centre. We note that this estimated value of systemic radial velocity is not a true indicator of the systemic radial velocity of the system as it might have received kicks, and thus is probably not following a pure Galactic rotation about the Galactic centre. Thus for each system, we estimate five probable systemic radial velocity Gaussian distributions with medians of $\bar{\gamma}$, $\bar{\gamma} \pm$ 50\,$\rm{km~s^{-1}}$ and $\bar{\gamma} \pm$ 100\,$\rm{km~s^{-1}}$ (see Table \ref{tab:table5}). We limit our assumed distributions of the systemic radial velocities based on the fact that out of the systems for which systemic radial velocity is measured, the values lie between -142$\pm$1.5\,$\rm{km~s^{-1}}$ for GRO\,J1655--40 \citep{Shahbaz1999} and 107$\pm$2.9\,$\rm{km~s^{-1}}$ for SAX\,J1819.3--2525 \citep{Orosz2001}.
\section{Analysis - Potential kick velocity}\label{Section 6}
As explained in Section \ref{Section 1.3}, peculiar velocity of a system when it crosses the Galactic plane is a better probe to understand the kick a BHXB received when the BH was born. Since all the parameters, namely proper motion, systemic radial velocity and parallax (or distance), have error bars associated with them, it is crucial to propagate these errors appropriately to estimate the Galactocentric orbits of the systems. The age of most BHXBs is not known, which makes integrating the Galactocentric orbits back to the time of birth of the BH uncertain. We thus developed a Monte Carlo (MC) methodology that accounts for the errors on the measured quantities and determines the peculiar velocity every time the system crosses the Galactic plane. \par
The code integrates the Galactocentric orbit of every system in our sample back for 10\,Gyrs and records the velocity of the system at every plane crossing. Instead of a delta function for the measurement, it involves using Gaussian distributions of the measured parameters with reported uncertainties as standard deviation. Random values are picked from these Gaussian distributions as inputs to \textit{galpy} \citep{Bovy2014} to create instances of Galactocentric orbits for $\sim5000$ random draws to make sure we have sampled the input distribution properly. We assume a Galactic potential given by the \textit{galpy} model MWPotential2014 \citep[see ][]{Bovy2014}. The Galactocentric orbits are integrated back in time for 10\,Gyrs, which exceeds the likely ages of LMXB systems. GRO\,J1655--40 \citep{Shahbaz2003}, GS\,1354--64 \citep{Casares2009} and SAX\,J1819--2525 \citep{MacDonald2014} are intermediate-mass systems and may have a shorter lifetime than standard LMXBs. 
Cyg\,X-1 is a HMXB and has not lived long enough to cross the Galactic plane \citep{Wong2012}. So this calculation, while indicative of the kick, does not represent the system's true history. To check if the integration of the systems back for 10\,Gyrs is valid for these systems, we integrated the orbits of these systems back in time for 1\,Gyr with 50,000 random draws and found that our results were not affected and were the same as those using 5000 draws and 10\,Gyr orbits. Thus, for uniformity we have used a time integration of 10\,Gyrs for all systems. \par
The peculiar velocity at each Galactic plane crossing (i.e. z=0, where z is the height above the Galactic plane) is calculated as
\begin{equation}
v_{\rm peculiar} = \left[(U - U_0)^{2}
+ (V - V_0)^{2}
+ (W - W_0)^{2}\right]^{0.5}
\end{equation}
Here $U$, $V$ and $W$ are Galactic space velocities towards the Galactic centre, in the direction of Galactic rotation and towards the North Galactic Pole respectively \citep{Johnson1987}. $U_0, V_0$ and $W_0$ are the $U$, $V$, $W$ components of the Galactocentric space velocities of the local standard of rest at a time when the system crosses the Galactic plane.\par
This approach allows us to estimate potential kick velocity distributions even if all the parameters required to construct the three dimensional motion are not accurately known. This will also help in estimating the potential kick velocities for newly discovered BHXBs that go into outburst, and when uncertain parameters of known systems are updated with new measurements. The code detailing the MC simulation methodology for estimating the potential kick velocity probability distribution and the distance estimation using our Milky way prior is available at the github link \url{https://github.com/pikkyatri/BHnatalkicks}.
\subsection{Potential kick velocity distributions of the new VLBI sources }\label{Section 6.1}
We estimated the potential kick velocity (PKV) probability distributions for the three sources for which we made proper motion measurements; GX\,339--4, Swift\,J1753.5--0127 and GRS\,1716--249. The proper motion input for the MC simulation is a Gaussian distribution based on the measured proper motion and error bars.
\subsubsection{GX 339--4}
\begin{figure}
\centering
\includegraphics[width=0.48\textwidth]{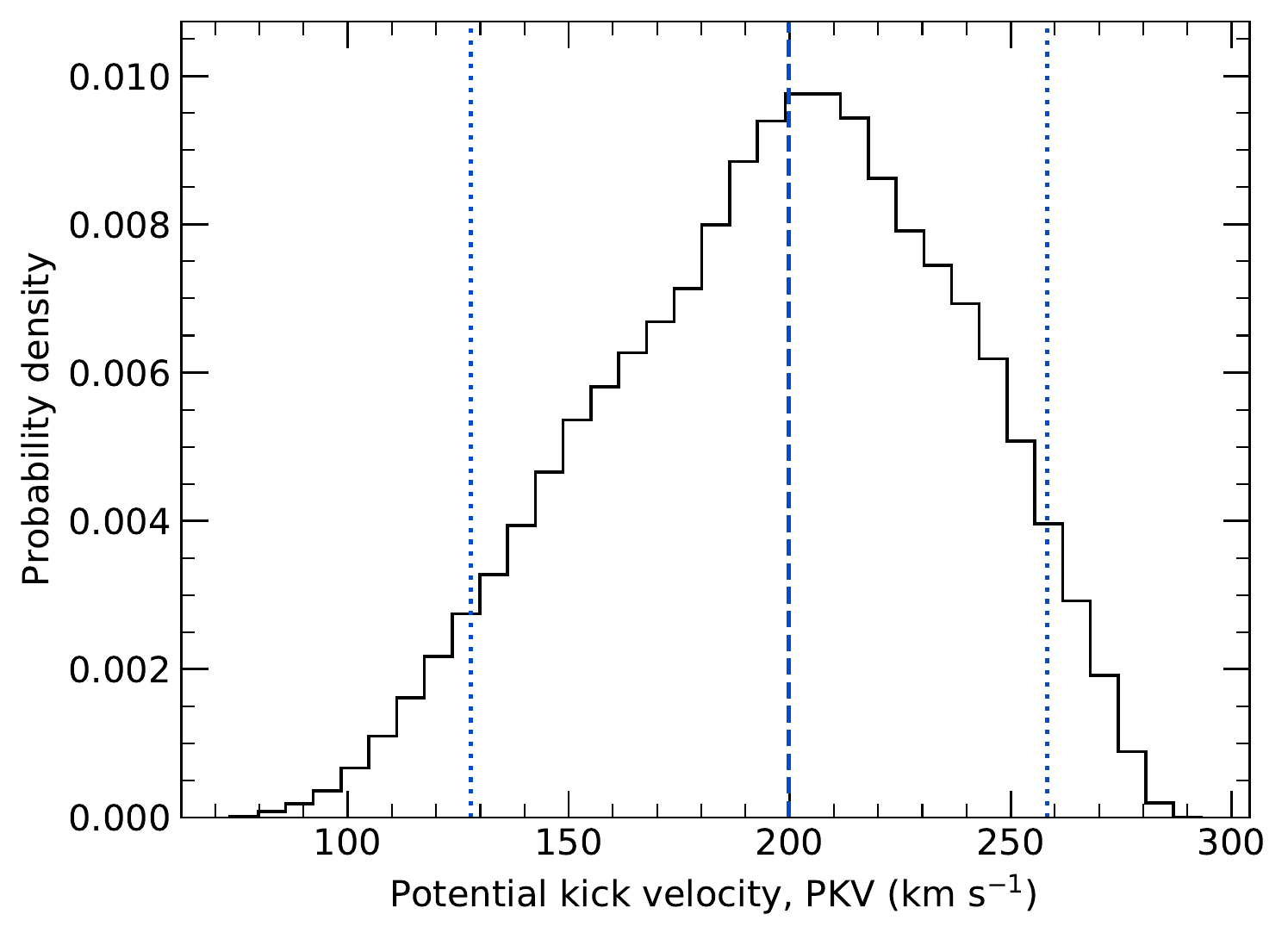}
\caption{Potential kick velocity probability distribution of GX\,339--4 with a median of 200\,$\rm{km~s^{-1}}$ (dashed line) and a 5$^{th}$ and 95$^{th}$ percentile of 122\,$\rm{km~s^{-1}}$ and 258\,$\rm{km~s^{-1}}$ (in dotted lines), respectively.}
\label{gx339natal}
\vspace{0.3cm}
\end{figure}
\textit{Gaia} could not make a parallax measurement for GX\,339--4. For the distance input we used a uniform distribution between 5 and 13\,kpc as estimated by \citet{Heida2017}. The system velocity input for the code was a Gaussian distribution based on the measurement of \citet{Heida2017} (Table \ref{tab:table4}). Running MC simulations for these distributions gives a PKV probability distribution as shown in Figure \ref{gx339natal} with a median of 200\,$\rm{km~s^{-1}}$ snd the 5$^{th}$ percentile at 122\,$\rm{km~s^{-1}}$.
\subsubsection{GRS 1716--249}
\begin{figure*}
\centering
\includegraphics[width=0.48\textwidth]{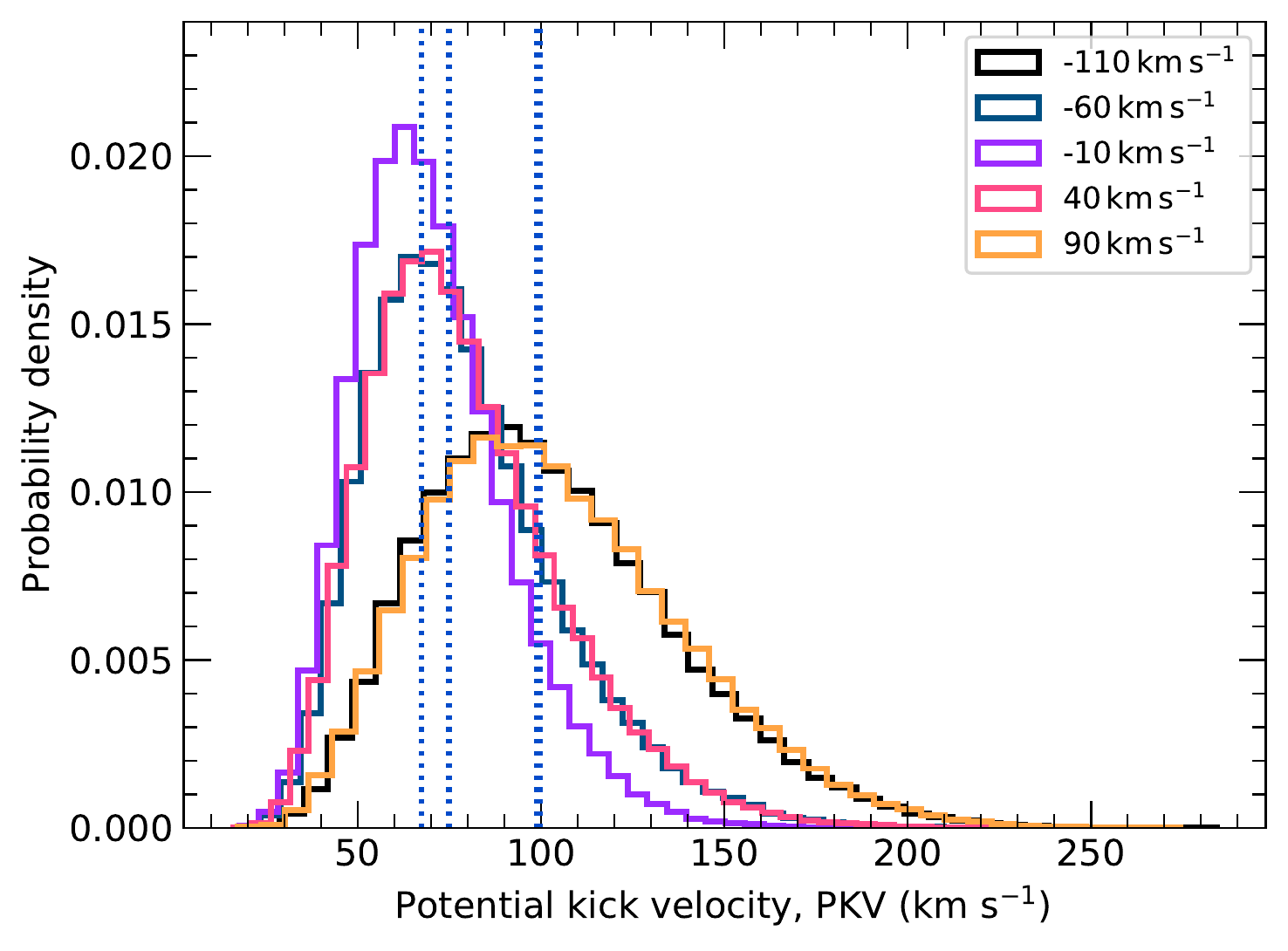}
\includegraphics[width=0.48\textwidth]{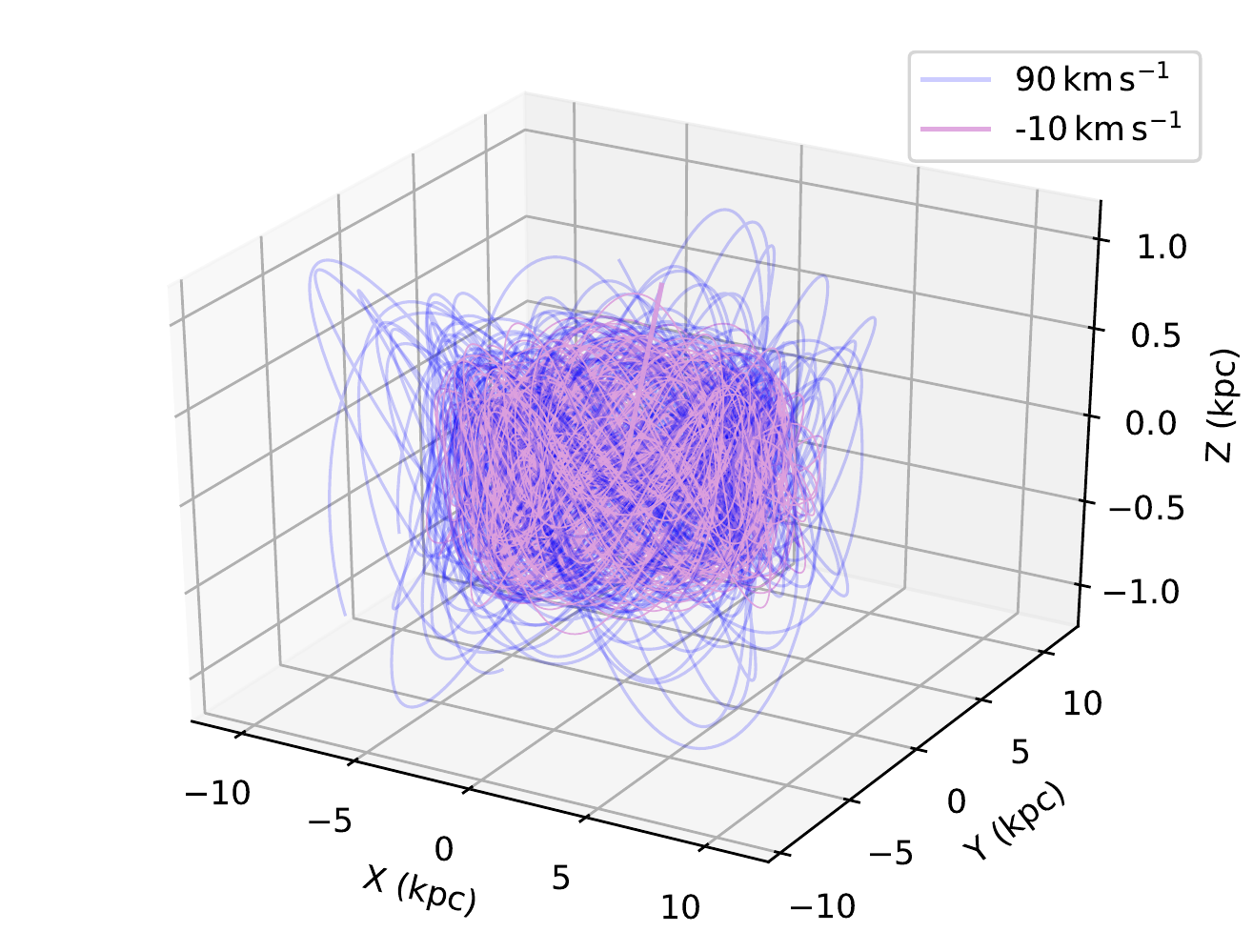}
 \caption{Left panel: PKV probability distribution of GRS\,1716--249 using Gaussian systemic radial velocity ($\bar{\gamma}$) distributions with means of $-110$\,$\rm{km~s^{-1}}$, -60\,$\rm{km~s^{-1}}$, -10\,$\rm{km~s^{-1}}$, 40\,$\rm{km~s^{-1}}$ and 90\,$\rm{km~s^{-1}}$ all with a 1$\sigma$ of 50\,$\rm{km~s^{-1}}$. The medians of all the PKV distributions are the blue dashed vertical lines. The lowest PKV probability distribution median is of $\sim$70\,$\rm{km~s^{-1}}$ and corresponds to the systemic radial velocity of -10$\pm$50\,$\rm{km~s^{-1}}$. Right panel: a 3-D visualisation of the Galactocentric orbit of GRS\,1716--249, integrated for 1\,Gyr for 20 orbit instances each of the lowest ($67^{+41}_{-27}$\,$\rm{km~s^{-1}}$) and highest ($100^{+68}_{-47}$\,$\rm{km~s^{-1}}$) PKV corresponding to systemic radial velocities of -10$\pm$50$\,\rm{km~s^{-1}}$ and 90$\pm$50\,$\rm{km~s^{-1}}$, respectively. All three axes are in kpc. The system does not go beyond 1\,kpc above the Galactic plane in both the cases.}
\label{grs1716natal}
\vspace{0.3cm}
\end{figure*}
\citet{Della1994} estimated the distance to GRS\,1716--249 as $\rm{2.4\pm0.4}$\,kpc, which we used as a Gaussian input prior. The systemic radial velocity of this system has not been measured yet. A system at the Galactocentric radius of GRS\,1716--249 and in the Galactic plane is expected to have a systemic radial velocity $\bar{\gamma}$ of $\sim$-10\,$\rm{km~s^{-1}}$. We thus run the MC simulations assuming five different Gaussian systemic radial velocity distributions with means of -110\,$\rm{km~s^{-1}}$, -60\,$\rm{km~s^{-1}}$, -10\,$\rm{km~s^{-1}}$, 40\,$\rm{km~s^{-1}}$ and 90\,$\rm{km~s^{-1}}$, all with a 1$\sigma$ of 50\,$\rm{km~s^{-1}}$. All five PKV probability distributions have a median above $\sim$ 70\,$\rm{km~s^{-1}}$ (see Figure \ref{grs1716natal}, left panel and Table \ref{tab:table4}). In figure \ref{grs1716natal} (right panel) we plotted twenty orbits for the maximum ($100^{+68}_{-47}$\,$\rm{km~s^{-1}}$) and minimum ($67^{+41}_{-27}$\,$\rm{km~s^{-1}}$) PKVs corresponding to systemic radial velocities of 90\,$\rm{km~s^{-1}}$ and -10$\,\rm{km~s^{-1}}$, respectively. The orbits stay within the vertical thick stellar disc limits of $\sim$1\,kpc \citep{Gilmore1983}. 
\subsubsection{Swift J1753.5--0127}
\begin{figure*}
\centering
\includegraphics[width=0.48\textwidth]{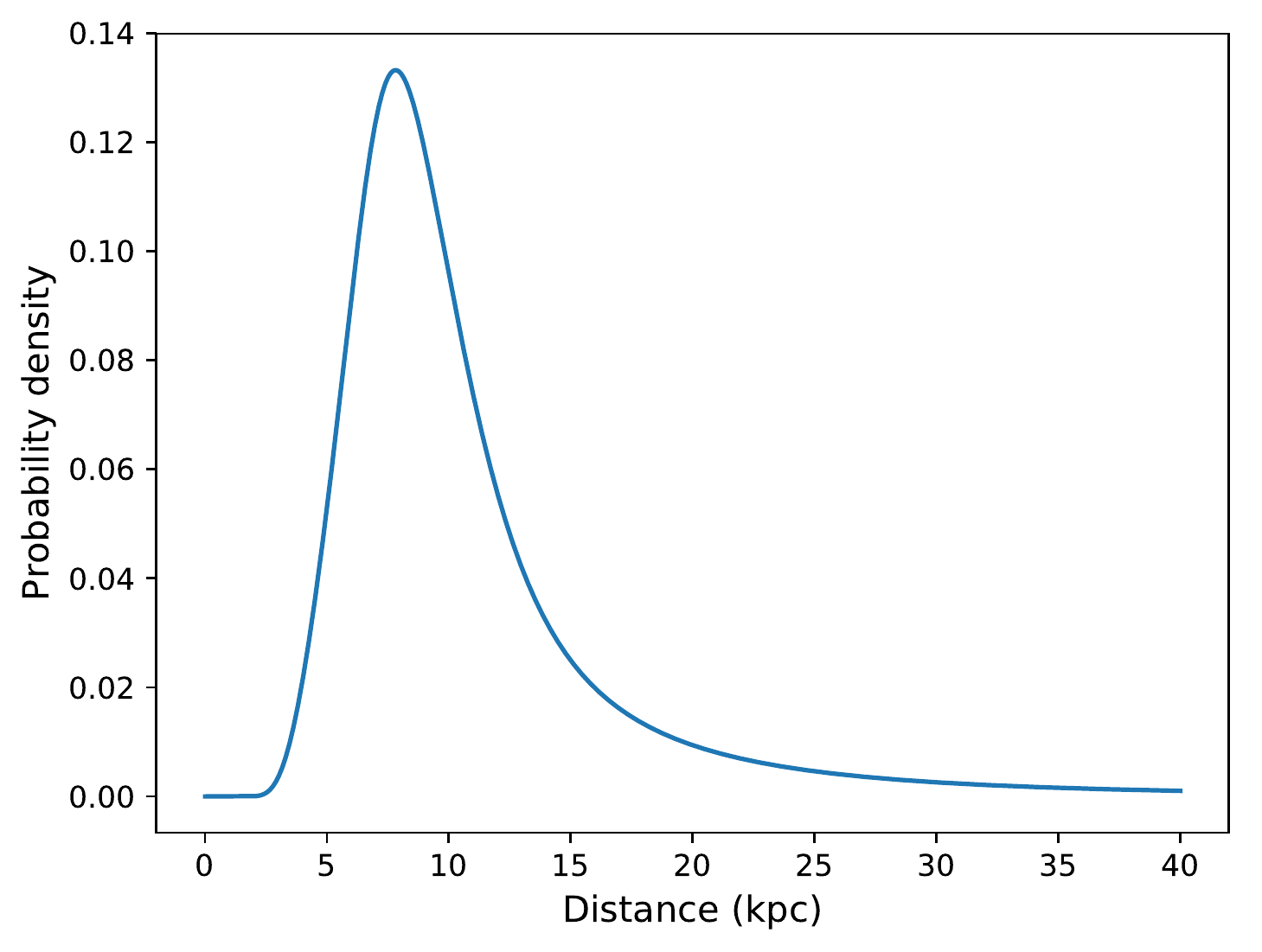}
\includegraphics[width=0.48\textwidth]{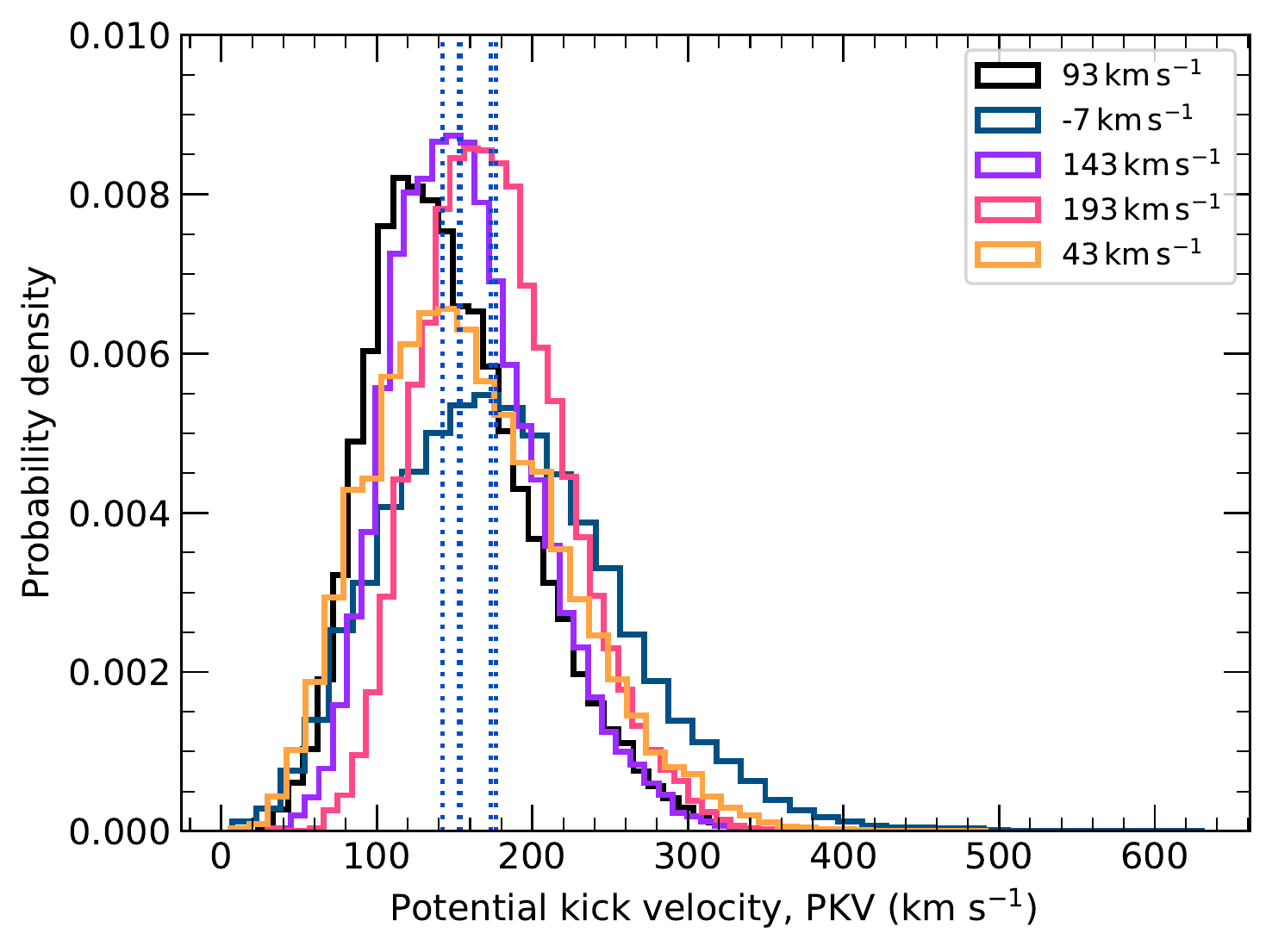}
 \caption{Left panel: Posterior distribution function for the distance to Swift\,J1753.5--0127, constructed from the \textit{Gaia}-DR2 parallax measurement using the Milky Way LMXB distribution prior (see Section \ref{Section 5.1} for details). The mode of the distribution is at 7.7\,kpc. The 5$^{th}$, 50$^{th}$ and 95$^{th}$ percentiles of the distance are at 4.8, 8.8 and 20.8\,kpc respectively. Right panel: PKV probability distribution of Swift\,J1753.5--0127 using Gaussian systemic radial velocity ($\bar{\gamma}$) distributions with means of $-7$\,$\rm{km~s^{-1}}$, 43\,$\rm{km~s^{-1}}$, 93\,$\rm{km~s^{-1}}$, 143\,$\rm{km~s^{-1}}$ and 193\,$\rm{km~s^{-1}}$ all with a 1$\sigma$ of 50\,$\rm{km~s^{-1}}$. The input distance distribution of 6$\pm$2\,kpc was used instead of the distance posterior distribution derived using the \textit{Gaia} parallax as the former was more tightly constrained. The lowest median amongst all the PKV probability distributions is at 142\,$\rm{km~s^{-1}}$ and the 5$^{th}$ percentiles of all the PKV probability distributions are above 70\,$\rm{km~s^{-1}}$.}
  \label{j1753natal}
  \vspace{0.3cm}
\end{figure*} 
\textit{Gaia}--DR2 measured a parallax of -0.01$\pm$0.13\,mas for this system. As there is a zero-point offset in all \textit{Gaia} parallax measurements of -0.029\,mas \citep{Luri2018}, we use the corrected parallax of 0.02$\pm$0.13\,mas for our simulations. We determined a posterior distribution for the distance as shown in Figure \ref{j1753natal}-Left panel. Comparing this with the distance estimated for this source in the literature (6$\pm$2\,kpc) \citep{Cadolle2007} suggests that the poorly constrained \textit{Gaia} parallax may be overestimating the distance. We thus use a Gaussian distance distribution centred at 6\,kpc with a 1$\sigma$ of 2\,kpc as input for the simulations rather than the distance distribution derived using \textit{Gaia} parallax. Along with the distance distribution, we use Gaussian distributions of the proper motion and five different Gaussian distributions (see Section \ref{Section 5.2} and Table \ref{tab:table5}) as systemic radial velocity ($\gamma$) inputs to calculate the PKV probability distribution (Figure \ref{j1753natal}-Right panel). Even the lowest median amongst the five PKV probability distributions is at 142\,$\rm{km~s^{-1}}$, with the 5$^{th}$ and 95$^{th}$ percentiles as 76\,$\rm{km~s^{-1}}$ and 243\,$\rm{km~s^{-1}}$, respectively (see Table \ref{tab:table4}). Over the past 10\,Gyrs, the system has reached as high as 1.5\,kpc above the Galactic plane. 
\subsection{PKV distributions of \textit{Gaia} DR--2 and archival sources}\label{Section 6.2} 
We used our methodology (see Section \ref{Section 5.1}) to invert the \textit{Gaia}--DR2 parallaxes for nine systems that \textit{Gaia}--DR2 measured parallaxes, and for two sources (V404\,Cyg and VLA\,J2130+12) that had archival parallax measurements . We have summarised the results for these simulations in Table \ref{tab:table4}. Out of these 11 systems, seven also had distance constraints in the literature. We used the distance estimations from the literature ($d_{\rm{lit}}$) as inputs for the MC simulations to obtain another set of PKV probability distributions for these seven systems (1A\,0620-00, GS\,1124--684, GRO\,J1655--40, Swift\,J1753.5--0127, SAX\,J1819--2525, XTE\,J1118+480 and MAXI\,J1820+070). We thus obtain two PKV probability distributions each for these seven systems, but use the PKV probability distributions determined using the literature distance estimates ($d_{\rm{lit}}$) for further analysis of these systems as they were more tightly constrained.\par
We used the distance posterior derived from parallax measurements for our PKV probability distributions for four systems, namely GS\,1354--64, Cyg\,X--1, V404\,Cyg and VLA\,J2130+12, as they did not have better distance estimates in the literature (See Table \ref{tab:table4} and Table \ref{tab:table5}). We tested the sensitivity of the PKV distribution of these four systems to the prior we are using to calculate the distance posterior, using an exponentially decreasing volume density prior \citep{Astraatmadj2016,Gandhi2019}
to determine a revised PKV distribution. The only system showing a marked difference in the PKV distributions from the two priors was GS\,1354--64, with a revised median PKV of $183^{+102}_{-64}$\,$\rm{km~s^{-1}}$ as compared to $213^{+88}_{-78}$\,km\,s$^{-1}$ from the Milky Way prior. However, even this difference was within the uncertainties of the PKV distribution using the Milky Way prior. \par
For V404\,Cyg, we use the archival VLBI proper motion and parallax measurements \citep{Miller-Jones2009a} as they were more precise than the \textit{Gaia}--DR2 measurements. The parallax and proper motion for VLA\,J2130+12 were measured by \citet{Kirsten2014}, though there is no available estimate of its systemic radial velocity. We thus used the expected radial velocity of a source at the Galactocentric distance of VLA\,J2130+12 but in the Galactic plane as the input for estimating its PKV probability distribution. For systems that did not have measured parallaxes, we use the best estimates on distance present in the literature (Table \ref{tab:table5}). \textit{Gaia}--DR2 did not measure proper motions for MAXI\,J1836--194 and GRS\,1915+105, so we also use archival VLBI proper motion measurements for these systems. We have reported the 5$^{th}$, 50$^{th}$ and 95$^{th}$ percentiles for both the distance distribution that we determined using the LMXB Milky Way prior described in Section \ref{Section 5.1}, and for the potential kick velocities for these systems.\par
MAXI\,J1820+070 was first detected in March 2018 \citep{Kawamuro2018} and is still being monitored in the radio and X-ray bands. We obtained an upper limit on the distance of the system as 3.9$\pm$0.6\,kpc by constraining the proper motions of the receding and approaching jets observed in existing VLA data (Bright et al. in prep). For a lower limit on the distance, we used the X-ray flux reported by Swift at the soft-to-hard X-ray spectral state transition. We assume that the state transition happens between 1$\%$ and 4$\%$ of the Eddington luminosity \citep{Maccarone2003,Kalemci2013} to estimate a distance of 1.7\,kpc. We have run our simulations using the distance inferred from the {\it{Gaia}} parallax and also using the distance limits as mentioned above. As this distance is better constrained than the {\it{Gaia}} DR--2 distance, we have used the former for estimating the PKV distribution of MAXI\,J1820+070.
\begin{table*}
\renewcommand{\arraystretch}{1.5}
       \begin{center}
                 \caption{Potential kick velocity (PKV) distributions for systems that have a systemic radial velocity measurement ($\gamma$) in the literature. The measured quantities are the various parameters for the BHXB systems that were available in the literature, and the estimated quantities are the values that have been determined in this work using the MC simulations code described in Section. d${_{\rm{lit}}}$ is the best distance estimate available in the literature for some of these systems. The \textit{Gaia} offset corrected parallax measurement (as described in Section \ref{Section 5.1}) and the distance distribution (d${_{\rm{post}}}$) estimated using the LMXB Milky way density prior are reported. The 5$^{th}$ and 95$^{th}$ percentiles of these distributions are the upper and lower limits on the distance estimated using the Milky Way prior. For distances in the literature that were just a range, a uniform distribution was used as the input to the MC code. For the distances in literature that have been reported with error bars, the input to our MC code was a Gaussian distribution. The PKV is reported as the median of the PKV probability distribution with the lower and upper limits representing the 5$^{th}$ and 95$^{th}$ percentiles. PKV$_{\rm{post}}$ and PKV$_{\rm{lit}}$ are the potential kick velocity distributions using the d${_{\rm{post}}}$ and the d${_{\rm{lit}}}$ as input distance distributions, respectively. The last column is the suggested birth pathway for the BH in each system and could be SN (Supernova), DC (Direct collapse) or U (Unsure). Refer to Section \ref{aggnatal} and Section \ref{snmassloss} for the explanation of the suggested birth pathway. References: [1] \citet{Brown2018}; [2] \citet{Gonzalez2010}; [3] \citet{Cantrell2010}; [4] \citet{Hernandez2008}; [5] \citet{Gelino2006}; [6] \citet{Orosz1996}; [7] \citet{Hynes2005}; [8] \citet{Casares2004}; [9] \citet{Shahbaz1999}; [10] \citet{Hjellming1995}; [11] \citet{Orosz2001}; [12] \citet{MacDonald2014}; [13] \citet{Gies2008}; [14] \citet{Miller-Jones2009b}; [15] \citet{Casares1994}; [16] \citet{Orosz1998}; [17] \citet{Heida2017}; [18] \citet{Russell2014}; [19] \citet{Reid2014}; [20] \citet{Steeghs2013}; [21] This work. \label{tab:table4}}
                \begin{tabular}{l c c c c c c c c l}
                \hline \hline
                 \multicolumn{6}{c}{Measured quantities}  & \multicolumn{4}{c}{Estimated quantities in this work} \\ 
                \cmidrule(lr){1-6}\cmidrule(lr){7-10}
        Source & $\mu_{\alpha}\cos\delta$ & $\mu_{\delta}$  & $\gamma$  & $d_{lit}$ & Parallax & d${_{post}}$ & PKV$_{\rm{post}}$ &  PKV$_{\rm{lit}}$ & BH \\
        &   &   &   &   &   &   &   &  & birth\\
                      & ($\rm{mas\,yr^{-1}}$) & ($\rm{mas\,yr^{-1}}$) & ($\rm{km~s^{-1}}$) & (kpc) & (mas) & (kpc) & ($\rm{km~s^{-1}}$) & ($\rm{km~s^{-1}}$) &\\
         \cmidrule(lr){1-6}\cmidrule(lr){7-10}
         1A\,0620--00       & -0.09$\pm$0.25$^{[1]}$              & -5.20$\pm$0.30$^{[1]}$     & \phantom{1}8.5$\pm$1.8$^{[2]}$  & 1.06$\pm$0.1$^{[3]}$          & 0.67$\pm$0.16$^{[1]}$     & $1.9^{+2.0}_{-0.7}$ & $60^{+55}_{-21}$    & $34^{+08}_{-08}$  & DC\\
         XTE\,J1118+480     & -17.57$\pm$0.34$^{[1]}$ \phantom{1} & -6.98$\pm$0.43$^{[1]}$     & \phantom{11}2.7$\pm$1.1$^{[4]}$& 1.72$\pm$0.10$^{[5]}$         & 0.30$\pm$0.40$^{[1]}$     & $13^{+21}_{-11}$     & $212^{+347}_{-109}$ & $186^{+39}_{-55}$ & SN\\
        GS\,1124--684      & -2.44$\pm$0.61$^{[1]}$              & -0.71$\pm$0.46$^{[1]}$     & 16$\pm$5$^{[6]}$                & 5.9$\pm$0.3$^{[7]}$           & 0.64$\pm$0.34$^{[1]}$     & $5.3^{+7.7}_{-3.5}$ & $101^{+84}_{-54}$   & $133^{+35}_{-34}$ & SN\\
        GS\,1354--64       & -9.38$\pm$2.22$^{[1]}$              & -5.70$\pm$2.26$^{[1]}$     & 103$\pm$4$^{[8]}$\phantom{1}    & --                             & 1.86$\pm$0.58$^{[1]}$     & $7.4^{+9.6}_{-6.3}$ & $213^{+88}_{-78}$   & -- 				  & SN\\
        GRO\,J1655-40      & -4.20$\pm$0.13$^{[1]}$              & -7.44$\pm$0.09$^{[1]}$     & -142$\pm$1.5$^{[9]}$            & 3.2$\pm$0.2$^{[10]}$           & 0.27$\pm$0.08$^{[1]}$     & $6.4^{+4.6}_{-3.2}$ & $137^{+47}_{-36}$   & $140^{+28}_{-36}$ & SN\\
        SAX\,J1819.3--2525 & -0.73$\pm$0.07$^{[1]}$              & 0.42$\pm$0.06$^{[1]}$      & 107$\pm$2.9$^{[11]}$             & 6.2$\pm$0.7$^{[12]}$           & 0.18$\pm$0.04$^{[1]}$     & $7.5^{+1.9}_{-2.2}$ & $189^{+182}_{-87}$  & $122^{+64}_{-28}$ & SN\\
        Cyg\,X--1 		   & -3.88$\pm$0.05$^{[1]}$              & -6.17$\pm$0.05$^{[1]}$     &-5.1$\pm$0.5$^{[13]}$             & --                             & 0.42$\pm$0.03$^{[1]}$     & $2.3^{+0.2}_{-0.3}$ & $30^{+10}_{-10}$    & --    			  & DC\\
        V404 Cyg 		   & -4.99$\pm$0.19$^{[14]}$             & -7.76$\pm$0.21$^{[14]}$    & -0.4$\pm$2.2$^{[15]}$            & --                             & 0.418$\pm$0.024$^{[14]}$  & $2.4^{+0.3}_{-0.2}$ & $43^{+08}_{-09}$    & -- 				  & DC\\
        4U1543--475		   & -7.41$\pm$0.14$^{[1]}$              &  -5.33$\pm$0.10$^{[1]}$    & \phantom{1}-87$\pm$3$^{[16]}$    & 7.5$\pm$1.0$^{[16]}$            & --                        & -- 					& --                  & $124^{+27}_{-27}$ & SN\\
        GX 339--4          & -3.95$\pm$0.07$^{[21]}$\phantom{11}    & -4.71$\pm$0.06$^{[21]}$\phantom{11} & \phantom{1} 26$\pm$2$^{[17]}$    & 5--13$^{[17]}$                  & --                        & -- 					& --                  & $200^{+58}_{-72}$ & SN \\
        MAXI J1836--194    & -2.3$\pm$0.6$^{[18]}$                & -6.1$\pm$1.0$^{[18]}$       & \phantom{11}61$\pm$15$^{[18]}$   & 4--10$^{[18]}$                  & --                        & -- 					& --                  & $162^{+42}_{-47}$ & SN\\
        GRS 1915+105       &  \phantom{1}-2.86$\pm$ 0.07$^{[19]}$ & -6.20$\pm$0.09$^{[19]}$     & \phantom{11}11$\pm$4.5$^{[20]}$  & 8.6$\pm$2.0$^{[19]}$\phantom{1} & --                        & -- 					& --                  & $49^{+37}_{-28}$  & DC\\

                \hline
                \hline
                \end{tabular}
               
\end{center}
\end{table*}

\begin{table*}
\renewcommand{\arraystretch}{1.5}
        \begin{center}
                 \caption{Potential kick velocity (PKV) distributions for systems that do not have a systemic radial velocity measurement ($\gamma$) in the literature. For these systems, we calculated the predicted systemic radial velocity ($\bar{\gamma}$) for pure Galactic rotation at at a projected distance on the Galactic plane (see Section \ref{Section 5.2} for more details). We run the MC simulations code for five possible systemic radial velocity distributions. The measured quantities are the various parameters for the BHXB systems that were available in the literature, and the estimated quantities are the values that have been determined in this work using the MC simulations code described in Section \ref{Section 6}. d${_{\rm{lit}}}$ is the best distance estimate available in the literature for some of these systems. The \textit{Gaia} offset corrected parallax measurement (as described in Section \ref{Section 5.1}) and the distance distribution (d${_{\rm{post}}}$) estimated using the LMXB Milky Way density prior are reported. The 5$^{th}$ and 95$^{th}$ percentiles of these distributions are the upper and lower limits on the distance estimated using the Milky Way prior. For distances in the literature that were just a range, a uniform distribution was used as the input to the MC code. For the distances in literature that have been reported with error bars, the input to our MC code was a Gaussian distribution. The PKV is reported as the median of the PKV probability distribution with the lower and upper limits representing the 5$^{th}$ and 95$^{th}$ percentiles. PKV$_{\rm{post}}$ and PKV$_{\rm{lit}}$ are the potential kick velocity distributions using the d${_{\rm{post}}}$ and the d${_{\rm{lit}}}$ as input distance distributions, respectively. The last column is the suggested birth pathway for the BH in each system and could be SN (Supernova), DC (Direct collapse) or U (Unsure). We have stated that just for the central predicted systemic radial velocity. Refer to Section \ref{aggnatal} and Section \ref{snmassloss} for the explanation of the suggested birth pathway. References: [1] This work; [2] \citet{Della1994}; [3] \citet{Cadolle2007} [4] \citet{Brown2018}; [5] \citet{Kirsten2014}} \label{tab:table5}  
                \begin{tabular}{l c c c c c c c c r}
                \hline \hline
                 \multicolumn{5}{c}{Measured quantities}  & \multicolumn{5}{c}{Estimated quantities in this work} \\ 
                \cmidrule(lr){1-5}\cmidrule(lr){6-10}
        Source & $\mu_{\alpha}\cos\delta$ & $\mu_{\delta}$  & $d_{lit}$ & Parallax & d${_{post}}$ & $\bar{\gamma}$  & PKV$_{\rm{post}}$ &  PKV$_{\rm{lit}}$ & BH birth\\
                      & ($\rm{mas\,yr^{-1}}$) & ($\rm{mas\,yr^{-1}}$) & (kpc) & (mas) & (kpc) & ($\rm{km~s^{-1}}$) & ($\rm{km~s^{-1}}$) & ($\rm{km~s^{-1}}$) &\\ 
        \cmidrule(lr){1-5}\cmidrule(lr){6-10}
         GRS 1716--249 & \phantom{1}-1.7$\pm$1.25$^{[1]}$\phantom{11} & -2.48$\pm$0.75$^{[1]}$\phantom{11} & 2.4$\pm$0.4$^{[2]}$\phantom{1} & -- & -- & -110$\pm$50\phantom{1} & -- & $99^{+68}_{-45}$ & \\
                       &                                      &                            &                                &  &  & -60$\pm$50 & -- & $75^{+54}_{-31}$ &\\
                       &                                      &                            &                                &  &  & -10$\pm$50 & -- & $67^{+41}_{-27}$ & U\\
                       &                                      &                            &                                &  &  & 40$\pm$50 & -- & $75^{+54}_{-32}$ &\\
                       &                                      &                            &                                &  &  & 90$\pm$50 & -- & $100^{+68}_{-47}$ &\\
          Swift\,J1753.5--0127 & 0.84$\pm$0.06$^{[1]}$\phantom{1} & -3.64$\pm$0.06$^{[1]}$\phantom{11} & 6$\pm$2$^{[3]}$ & 0.02$\pm$0.13$^{[4]}$ & $8.8^{+12}_{-4.0}$& -7$\pm$50\phantom{1} & $212^{+114}_{-104}$ & $154^{+116}_{-84}$ &  \\
                               &                          &                            &                  &                       & & 43$\pm$50\phantom{1} &$196^{+118}_{-89}$ &  $142^{+101}_{-66}$ & \\
                               &                          &                            &                  &                       &  & 93$\pm$50\phantom{1} & $240^{+128}_{-115}$ &  $177^{+135}_{-103}$ & SN\\
                               &                          &                            &                  &                       & & 143$\pm$50\phantom{11} & $192^{+118}_{-82}$ & $153^{+88}_{-63}$ &\\
                               &                          &                            &                  &                       & & 193$\pm$50\phantom{11} & $204^{+115}_{-77}$ & $173^{+86}_{-65}$ &\\
          MAXI\,J1820+070 & -3.41$\pm$0.19$^{[4]}$ & -5.90$\pm$0.22$^{[4]}$  & 1.7--3.9$^{[1]}$\phantom{1} & 0.31$\pm$0.11$^{[4]}$ & $4.4^{+5.1}_{-2.0}$ & -67$\pm$50\phantom{11} & $185^{+147}_{-83}$ &  $153^{+128}_{-77}$ & \\
         & & & & & & -17$\pm$50\phantom{11}  & $143^{+104}_{-63}$ &$110^{+97}_{-49}$\\
         & & & & & & 33$\pm$50\phantom{1}  & $111^{+79}_{-41}$ & $86^{+67}_{-30}$ & SN\\
         & & & & &  & 83$\pm$50\phantom{1} & $103^{+62}_{-33}$ & $84^{+50}_{-27}$\\
         & & & & &  & 133$\pm$50\phantom{11} & $109^{+62}_{-33}$ &$98^{+75}_{-34}$\\
         VLA J2130+12 & -0.07$\pm$0.13$^{[5]}$ & -1.26$\pm$0.29$^{[5]}$\phantom{123} & -- & \phantom{123}0.45$\pm$0.08$^{[5]}$ & $2.4^{+3.4}_{-1.8}$ & -90$\pm$50\phantom{11} & $146^{+116}_{-74}$ & --\\
          &  &  & & & & -40$\pm$50\phantom{11} & $101^{+95}_{-47}$ & --\\
          &  &  & & & & 10$\pm$50\phantom{1}   & $82^{+63}_{-32}$& -- & U\\
          &  &  & & & & 60$\pm$50\phantom{1}   & $86^{+60}_{-34}$ & --\\
          &  &  & & & & 110$\pm$50\phantom{11} & $111^{+79}_{-49}$ & --\\
                \hline
                \end{tabular}
                
       \end{center}
\end{table*}

\section{DISCUSSION}\label{Section 7}
\subsection{Potential kick velocity distribution - BHXB population}\label{aggnatal}
\begin{table*}
        \begin{center}
                \caption{Model estimates for the potential kick velocity distribution using unimodal and bimodal gaussian models when all systems are considered and when only systems with well constrained radial systemic velocity ($\gamma$) are considered. Reported values are medians ($\mu$) of the posterior distributions, along with uncertainties ($\sigma$) and all values are in $\rm{km~s^{-1}}$. $w_1$ is the weight of the Gaussian distribution with the mean $\mu_1$. AICc is the corrected Akaike information criterion. A smaller AICc suggests that the model is more favourable. \label{tab:table6}}
                \begin{tabular}{l c c c c c c l}
                \hline \hline
                         &  & $\mu_1$ & $\sigma_1$ & $\mu_2$ & $\sigma_2$ & $w_1$ & AICc \\ \hline 
                        All systems & Unimodal & 107$\pm$16 & 56$\pm$14 & - & - & - & -16.3 \\
                        & Bimodal & 41$\pm$14 & 19$\pm$10 & 136$\pm$17 & 32$\pm$18 & 0.3$\pm$0.1 & -2.2 \\ \hline
                        Systems with $\gamma$ & Unimodal & 112$\pm$22 & 65$\pm$19 & - & - & - & -2.4 \\
                        & Bimodal & 35$\pm$10 & 11$\pm$9 & 145$\pm$15 & 30$\pm$17 & 0.3$\pm$0.1 & 10.7\\ 
                   \hline
                \end{tabular}
        \end{center}
\end{table*}
\begin{figure}
\centering
\includegraphics[width=0.48\textwidth]{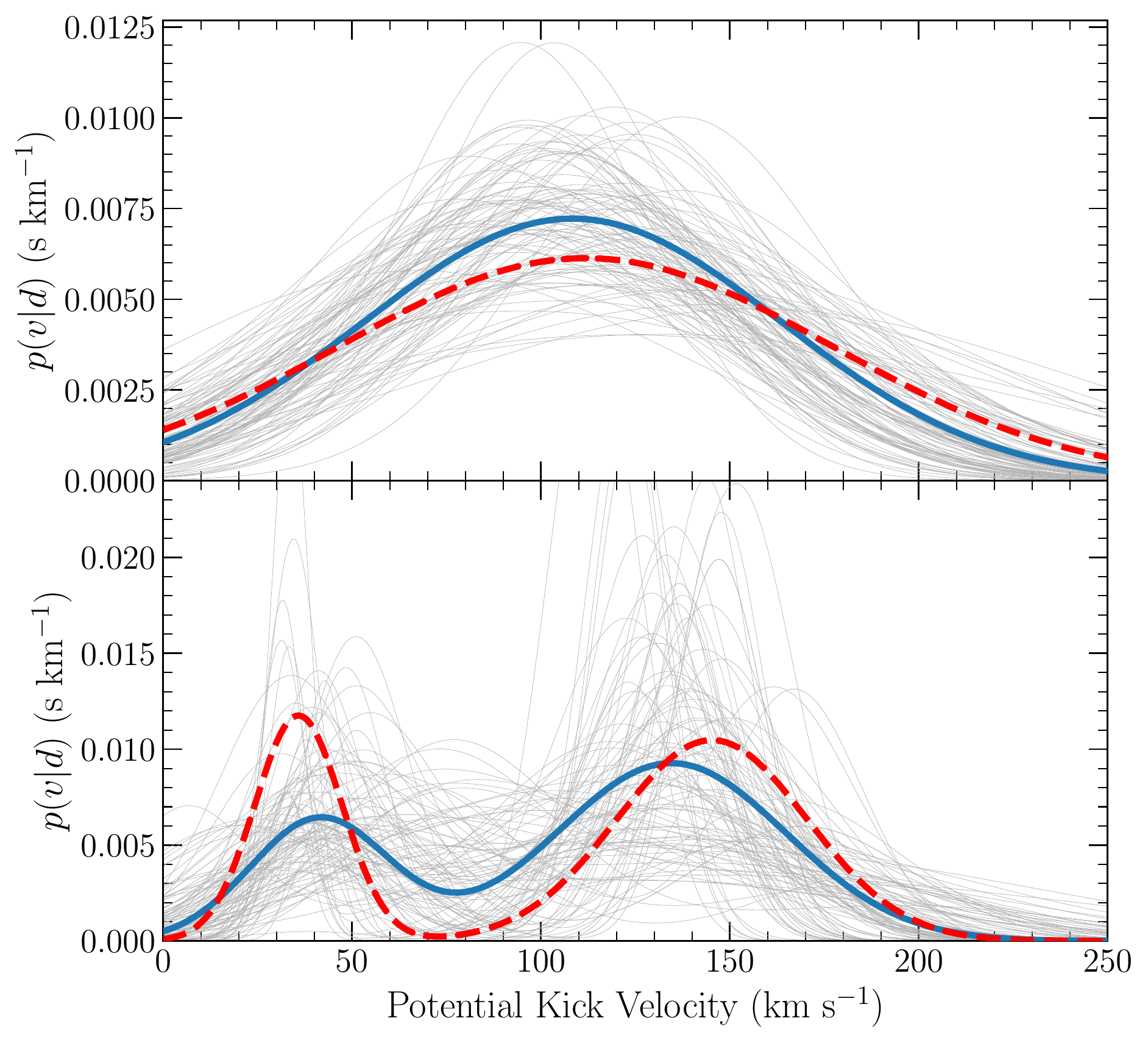}
\caption{Realisations of the inferred unimodal (top) and bimodal (bottom) distributions for potential kick velocities (v), inferred from the data (d). The blue lines represent the model corresponding to the median values from the posterior sample for data from all the systems in the sample, while the red dashed lines represent the model based on the median from the posterior sample for data from the 12 systems with systemic radial velocity ($\gamma$) constraints. The faint gray lines are a small random subset from the posterior MCMC sample for all the systems, as a demonstration of uncertainty.}
\label{Bimodal1}
  \vspace{0.3cm}
\end{figure}
\begin{figure}
\centering
\includegraphics[width=0.48\textwidth]{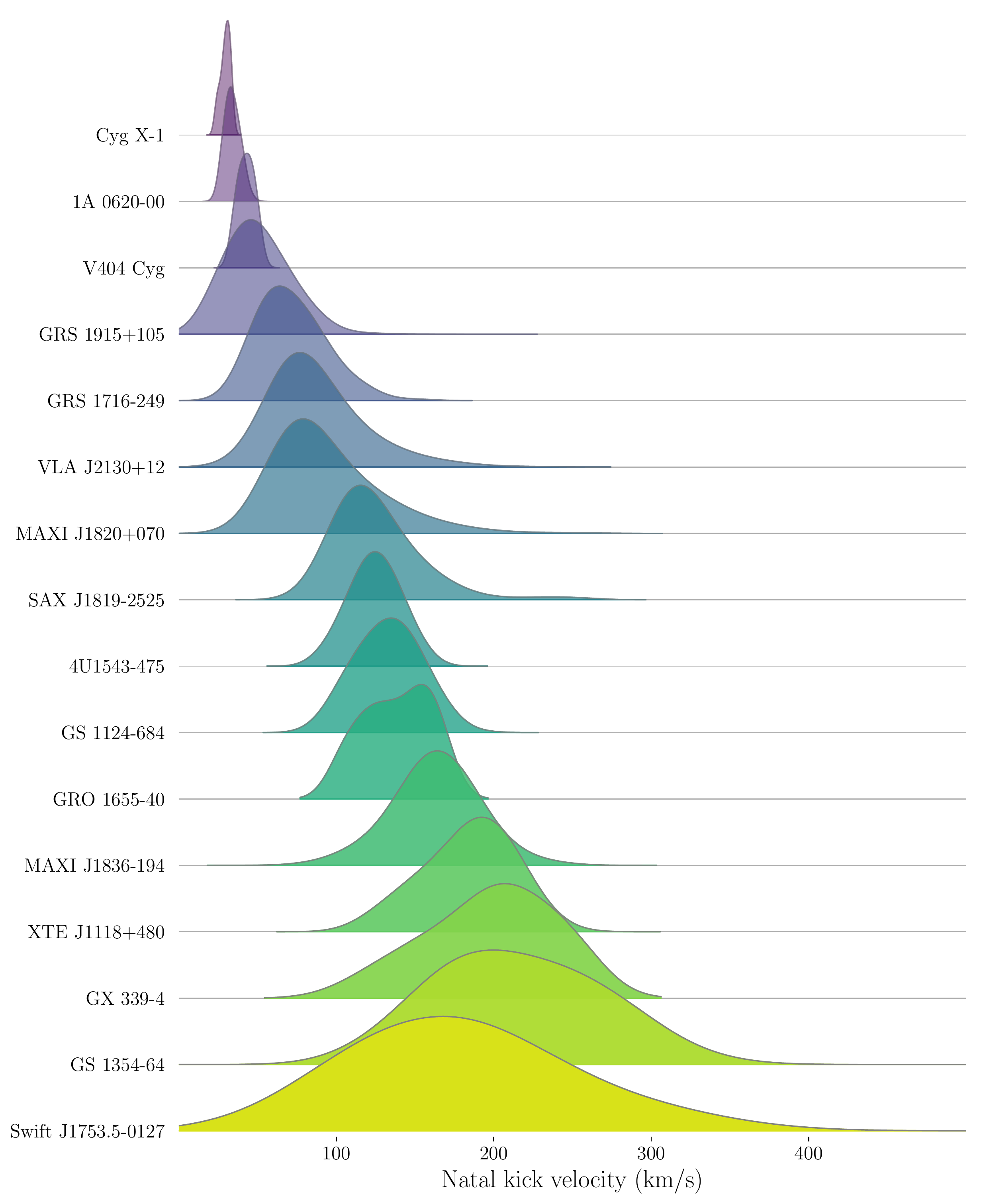}
\caption{PKV probability distributions of the 16 BHXBs in our sample. It can be seen that no single system is responsible for either of the two most prominent peaks in the aggregated natal kick distribution.}
\label{Bimodal2}
 \vspace{0.3cm}
\end{figure}
We investigated the distribution of the potential kick velocities in BHXB systems based on the results in Section \ref{Section 6}. Given the probabilistic nature of these results (probability distributions for the predicted velocity of each system), we follow the Bayesian hierarchical methodology outlined by \citet{Mandel2010} and \citet{Hogg2010} for fitting a distribution model. We evaluate the distribution of posteriors with unimodal (with mean of $\mu$ and standard deviation of $\sigma$) and bimodal (with means of $\mu_1$, $\mu_2$, standard deviations of $\sigma_1$, $\sigma_2$, and weights of $w_1$ and $w_2 = 1-w_1$) gaussian models. For this purpose, we used a Hamiltonian Markov chain Monte Carlo (MCMC) algorithm \citep{Neal2012,Betancourt2017} with No U-Turn Sampling \citep[NUTS;][]{Hoffman2011} as implemented in PyMC3 \citep{Salvatier2016}. We assumed uninformative priors for all parameters in the models and verified convergence in the final samples for all parameters using the Gelman-Rubin diagnostic test \citep{Gelman1992}, with $\hat{R} < 1.001$ for all parameters ($\hat{R}$ closer to one indicates certainty of convergence of the chains). \par
We evaluated the distribution with both models over two sets of velocities, with one set containing all the BHXB systems in our sample (16 systems), and a second set containing only the systems with systemic radial velocities that have been measured in the literature (12 systems). The results for both models in both cases are tabulated in Table~\ref{tab:table6} and plotted in Fig.~\ref{Bimodal1}.\par
Due to the low number of ``events'' (number of systems for which we have kick velocity PDFs), we used the corrected Akaike information criterion \citep[AICc;][]{Akaike1974, Cavanaugh1997, Burnham2002} to compare the models. As tabulated in Table~\ref{tab:table6}, the unimodal distribution is clearly favoured in both cases (a smaller AICc suggests a better model). Removal of systems with loose systemic radial velocity constraints provides a marginal relative improvement to the bimodal model, as the difference between the AICc parameter reduced when evaluated for a sample devoid of loose systemic radial velocity constraints as compared to a sample that consisted of all 16 systems. It is worth noting however, that the clear separation of posterior constraints for $\mu_1$ and $\mu_2$ in the bimodal model (given the uninformative priors) hints at the possibility of a bimodal nature for the distribution of the potential BHXB kicks, but testing this model is currently hampered by the low number of BHXB systems with potential kick velocity constraints.\par
To demonstrate that neither of the two components in the bimodal model are caused by a single source, we plot all individual distributions in Figure \ref{Bimodal2}. We find that there are at least four systems (first four systems in Figure \ref{Bimodal2}) contributing to the lower velocity peak (41\,$\rm{km~s^{-1}}$), which could be consistent with birth of the BH by direct collapse. We find at least nine systems (SAX\,J1819--2525 to Swift\,J1753.5--0127 in Figure \ref{Bimodal2}) clearly contributing to the higher velocity peak (136\,$\rm{km~s^{-1}}$). Better constraints on some of the measured parameters for systems like GS\,1124-684, VLA\,J2130+12, GS\,1352--64, Swift\,J1753.5--0127 and SAX\,J1819--2525 might help in tightening their PKV probability distributions, which could change the form (and interpretation) of the PKV distribution of BHXB population. With the current available data on BHXBs, our results provide the best constraint we can obtain for the PKV distribution of BHXBs.
\subsection{Potential Kick velocity interpretation}
The potential kick velocity (PKV) that we estimate here is not the actual natal kick the BH receives when it is born, but is the potential peculiar velocity of the system, with which it may have been kicked out of the Galactic plane. Calculating the actual BH natal kick from the BHXB peculiar velocity at Galactic plane crossing is complex and depends on modelling the system \citep{Repetto2012,Repetto2017}, taking into account the orbital period, the component masses of the binary system, the direction of the kick and the mass ejected in the SN explosion \citep{Nelemans1999}. It also involves simulating the subsequent dynamical interaction of the sources and evolution of the binary system. These parameters are currently not well constrained for most systems, and thus simulating the evolution of these systems to determine the actual BH natal kick is out of the scope of this paper. Hence the peculiar velocity of the system at Galactic plane crossing is currently the best proxy that we can use for BH natal kicks in a population-wide analysis such as this one. Since most star forming regions are in the Galactic plane, it is probable that most BHXBs form in the Galactic plane. Since we do not know the time of BH birth, we instead consider the velocity of the system every time it crosses the Galactic plane as a potential velocity kick the system might have received. \par
According to \citet{Mignard2000}, stellar velocity dispersions due to Galactic interactions are the order of $\sim50\rm{km~s^{-1}}$ for old systems (and lower for younger systems). Since BHXBs are more massive than these systems, they will suffer even lower velocity dispersions. Thus we classify systems with potential kick velocities higher than the $\sim50\rm{km~s^{-1}}$ limit as systems that might have received strong kicks, and hence the BH in the BHXB might have formed in a SN explosion. We suggest that the systems with a median PKV of less than $\sim50\rm{km~s^{-1}}$ are systems with weak potential kicks, and thus could well have formed by direct collapse rather than a SN explosion. Systems that have a PKV distribution such that their median is $>$50\,$\rm{km~s^{-1}}$ but the 5$^{th}$ is lower than 50\,$\rm{km~s^{-1}}$ do not clearly fall into either category, and hence we could not deduce their likely birth mechanism. We find two such sources in our sample, namely VLA\,J2130+12 and GRS\,1716--249. Ten systems in our sample are likely to have been formed in a SN explosion, whereas four systems could plausibly have formed by direct collapse. We summarise the results of our analysis in Tables \ref{tab:table4} and \ref{tab:table5}.\par
The PKV probability distribution of GRS\,1915+105 (median $\sim$49\,$\rm{kms^{-1}}$) suggests that it could have been born via direct collapse, as the PKV is within the velocity dispersion limits for Galactic scattering ($\sim 50$ km\,s$^{-1}$ for late type stellar systems; \citealt{Mignard2000}). According to our results the v$_{95}$ values for V404\,Cyg and 1A\,0620-00 are lower than the v$_{95}$ for GRS\,1915+105, suggesting that V404\,Cyg and 1A\,0620--00 might also have been born without a strong natal kick. The probability of a compact object receiving a kick between angle $\theta$ and $\theta+d\theta$ during a SN explosion varies as $\frac{1}{2}\sin\theta d\theta$, where $\theta$ is the angle the kick makes with the direction of the peculiar velocity of the system before the explosion. Thus it is also possible that these low PKV systems were born with a SN explosion but their velocity was reduced due to an asymmetry in the explosion counteracting the recoil kick. For Cyg\,X-1, the median of the PKV we obtain is in agreement with the estimation of \citet{Wong2012} and suggests a direct collapse birth (albeit with the caveat that this is a high mass X-ray binary and has never crossed the Galactic plane in its lifetime, so our methodology is not strictly applicable). \par
For our interpretation of PKV as the BH natal kick, we are assuming that none of the systems in our sample were born in GCs. There have been no confirmed BHXBs in GCs, though there are some candidates (\citealt{Strader2012,Camilo2000,Chomiuk2013,Miller-Jones2015}). GCs only contain 5--10\% of the total number of known Galactic X-Ray binaries \citep{Clark1975,Bahramian2014}, so this population is likely to be a small fraction of the total BHXBs in the Galaxy. \par 
The Galactocentric orbits in the past 10\,Gyrs also show that only four (GRO\,1655--40, SAX\,J1819--2525, GX\,339--4 and MAXI\,J1836--194) out of the 16 systems ever get within 2\,kpc of the Galactic centre, which is the truncation radius for a spheroidal bulge model \citep{Dehnen1998}. We note that we have classified these systems as having a SN birth, but due to their possible association with being bulge objects we cannot rule out the high PKV as being due to bulge formation and/or interactions.
\subsection{Comparison with NS natal kicks}\label{nsnatal}
\begin{figure}
\centering
\includegraphics[width=0.48\textwidth]{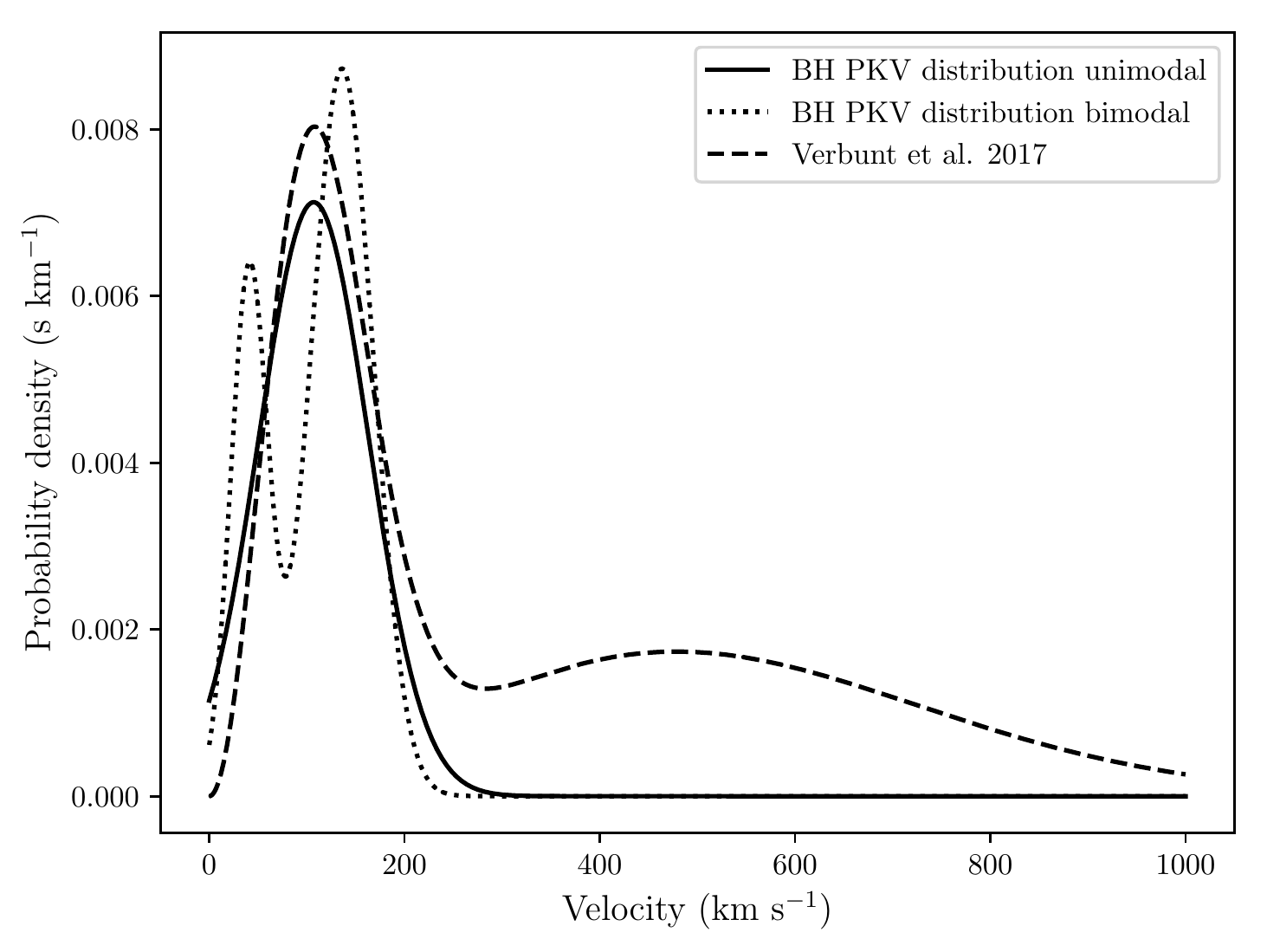}
\caption{Comparison of best fit unimodal and bimodal BHXB PKV distributions to the best pulsar kick velocity distribution. The BHXB PKV unimodal distribution is represented by the solid line and has a median of 107\,$\rm{km~s^{-1}}$. The BHXB PKV bimodal distribution is represented by the dotted line. The best fit, two Maxwellian model for the peculiar velocities of pulsars \citep{Verbunt2017} is denoted by the dashed line. The unimodal BHXB PKV distribution median is close to the low kick pulsar kick velocity component. There is a dearth of extremely high PKV BHs corresponding to the high velocity Maxwellian component for the pulsar velocity distribution.}
\label{NSkicks}
  \vspace{0.3cm}
\end{figure}
The kick velocity distribution of pulsars has been extensively studied \citep[e.g.][]{Hobbs2005, Fryer1998, Arzoumanian2002}, whereas there is no well documented kick velocity measurement catalogue for neutron star X-ray binaries (NS XRBs). Thus to understand the relation between NS and BH natal kicks, we compare our BHXB PKV distribution with the pulsar kick velocity distribution \citep{Verbunt2017}. BHXBs and pulsars suffer from different observational biases. The distance to pulsars is usually smaller than the distance to BHXBs, and thus their detection is less affected by extinction. Pulsars just have their two dimensional velocities measured, there are no systemic velocity measurements for pulsars. Also, it is possible to observe pulsars that might have received strong enough kicks to disrupt the binary they were a part of, whereas our data sample only consists of BHs that are in binaries. Keeping these biases in mind, we compare BHXB and pulsar kick velocities. \par
Theory suggests that fallback of matter ejected in a SN onto the proto-neutron star can give rise to BHs, so we expect similarities in the kick distributions of BHs and NSs. The most recent NS peculiar velocity distribution is fit better with two Maxwellian components than one (V17). Our bimodal Gaussian (Figure \ref{NSkicks}) fit to the data shows that the medians of the two Gaussians in the BHXB PKV distribution (41$\pm$14\,$\rm{km~s^{-1}}$ and 136$\pm$17\,$\rm{km~s^{-1}}$ ) are lower than the NS peculiar velocity peaks (120\,$\rm{km~s^{-1}}$ and 540\,$\rm{km~s^{-1}}$) for the best fit model by V17 by a factor of 3--4. This may suggest that BHs receive weaker natal kicks as compared to NSs by a factor that is similar to the mass ratios of standard BHs and NSs. This is in contrast to the conclusion of \citet{Repetto2012}, where it was shown that BHs and NSs get equally strong kicks. This mismatch of the medians of the two components of the BHXB PKV distribution and pulsar kick velocity distribution could also be because of different kick mechanisms in BHs and NSs. For BHs the low velocity kick is probably due to formation by direct collapse, whereas low NS kicks have been attributed to various mechanisms like ultra-stripped SNe \citep{Tauris2015}, electron capture SNe \citep{Gessner2018} and collapse of low iron mass cores \citep{Podsiadlowski2004}. \par
Our unimodal fit to the BHXB PKV data matches the lower maxwellian component of the pulsar kick velocity distribution (Figure \ref{NSkicks}). However, we are biased against stronger kicks, since they can unbind BHXBs and make them unobservable whereas pulsars can be observed even in an unbound state. Due to this selection bias, we expected the BHXB PKV distribution kicks to be lower than the V17 peculiar velocities. \par
\subsection{Supernova mass loss}\label{snmassloss}
We find that there are at least nine systems of the 16 in our sample set that clearly received strong natal kicks ($>$50$\rm{km~s^{-1}}$). This makes SN birth more probable for these systems than direct collapse BH birth. We note that our sample set may be biased against observing direct collapse BHs. This is because they will not receive strong kicks and will be located close to the Galactic plane, where detecting distant BHXBs is challenging due to high extinction and scattering.

Symmetric mass loss ($\Delta\,M$) during a SN explosion is one of the reasons for a natal kick. However, this mass loss has to be $<0.5(M\rm{_{He}}$+$m$) for the binary to remain bound \citep{Blaauw1961}, where $M\rm{_{He}}$ is the mass of the progenitor helium star and m is the mass of the donor star. Thus the maximum possible recoil velocity due to symmetric mass loss can be estimated by constraining the maximum possible mass ejected in the BHXB system without unbinding the binary. This was given by \citet{Nelemans1999} as
\begin{equation}
\begin{aligned}
v_{\rm{sys}} = 213\,\left(\frac{\Delta M}{\rm{M_{\odot}}}\right)\left(\frac{m}{\rm{M_\odot}}\right)\left(\frac{P_{\rm{re-circ}}}{\rm{days}}\right)^{-1/3}\left(\frac{(M_{\rm{BH}}+m)}{\rm{M_\odot}}\right)^{-5/3} \rm{km\,s^{-1}}
\end{aligned}
\end{equation}
where $\Delta$\,M, $\rm{M_{BH}}$ and m are the mass ejected in the SN, mass of the BH and mass of the donor immediately after the SN, respectively. $\rm{P_{re-circ}}$ is the period of the re-circularised orbit of the system after the formation of the BH, and it is assumed that no mass transfer occurs until the re-circularisation of the orbit is complete. $\rm{v_{sys}}$ is the recoil kick velocity the system should have received due to the symmetric mass loss in the SN explosion. \citet{Nelemans1999} estimated the maximum possible recoil kick velocities for GS\,1124--684 and GRO\,1655--40 and found that they were limited to $<$40$\,\rm{km~s^{-1}}$. Our estimated PKV for all these systems is greater than those estimated by \citet{Nelemans1999} in the case of only symmetric mass ejection. We thus suggest that the observed kick velocity of these systems is not only because of symmetric mass loss in a SN explosion, but due to these systems receiving additional acceleration due to asymmetries associated with the SN explosion \citep{Janka2017}.\par
\citet{Gualandris2005} suggested that XTE\,J1118+480 received an asymmetric kick that put the system in its current Galactocentric orbit. We measured a PKV of 186$\pm$28\,$\rm{km~s^{-1}}$ for XTE\,J1118+480, which agrees with the peculiar velocity measurement of \citet{Gualandris2005}. Systems like 4U1543--475, GS\,1354--64, SAX\,J1819.3--2525 and MAXI\,J1820+070 also have high enough PKVs to suggest asymmetries in the SN explosions, keeping in mind that we are assuming that these systems are similar to the ones already studied in the literature \citep[e.g. ][]{Nelemans1999,Gualandris2005}. The observed metal abundances in XTE\,J1118+480 \citep{Gonazalez2006}, GRO\,J1655--40 \citep{Hernandez2008} and SAX\,J1819.3--2525 \citep{Orosz2001} also point towards contamination of these sources by a SN and support the hypothesis that they were probably born in SN explosions.\par
\subsection{Natal kicks and z-distribution}\label{zdistnatal}
\begin{figure*}
\centering
\includegraphics[width=0.48\textwidth]{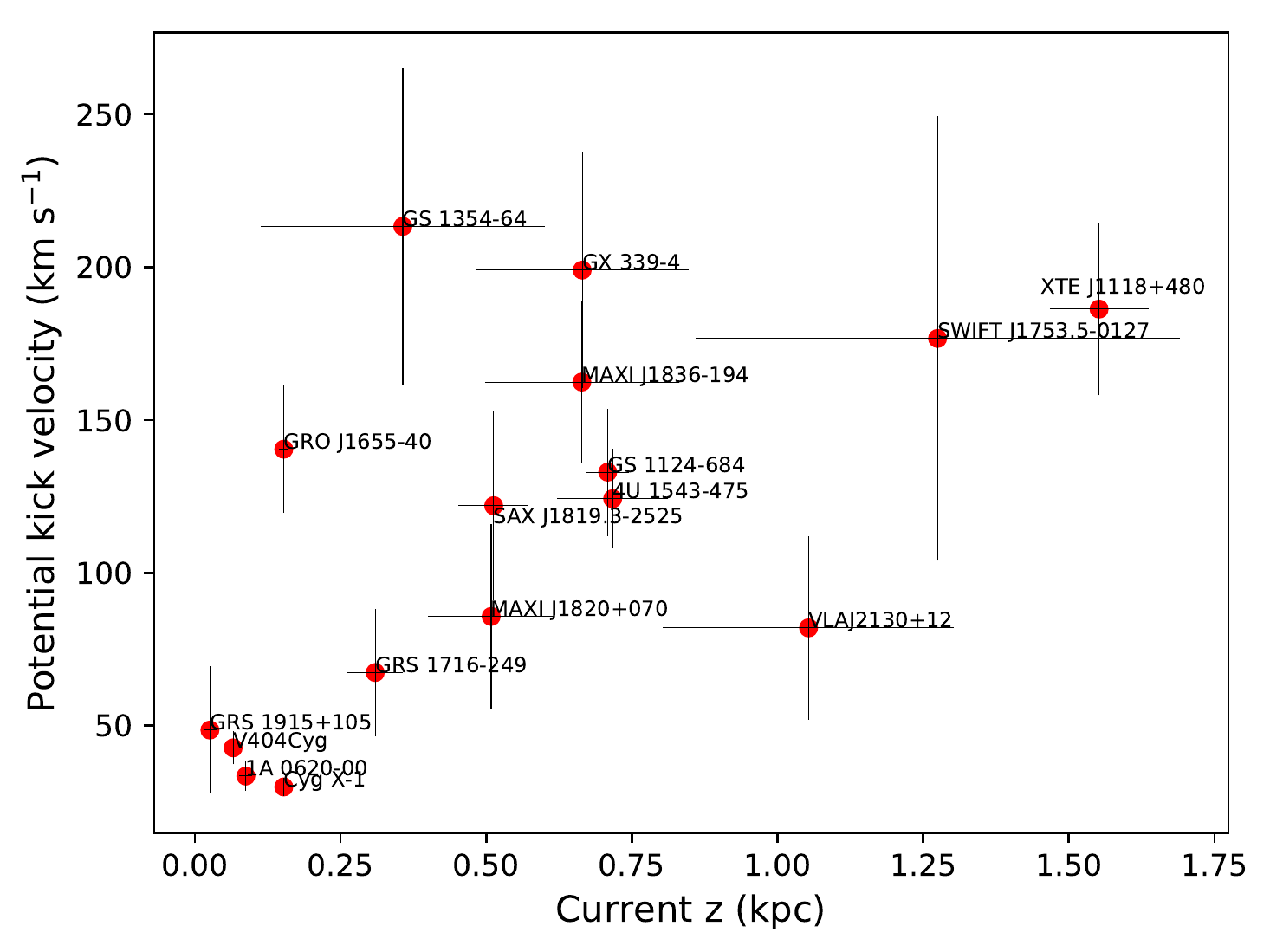}
\includegraphics[width=0.48\textwidth]{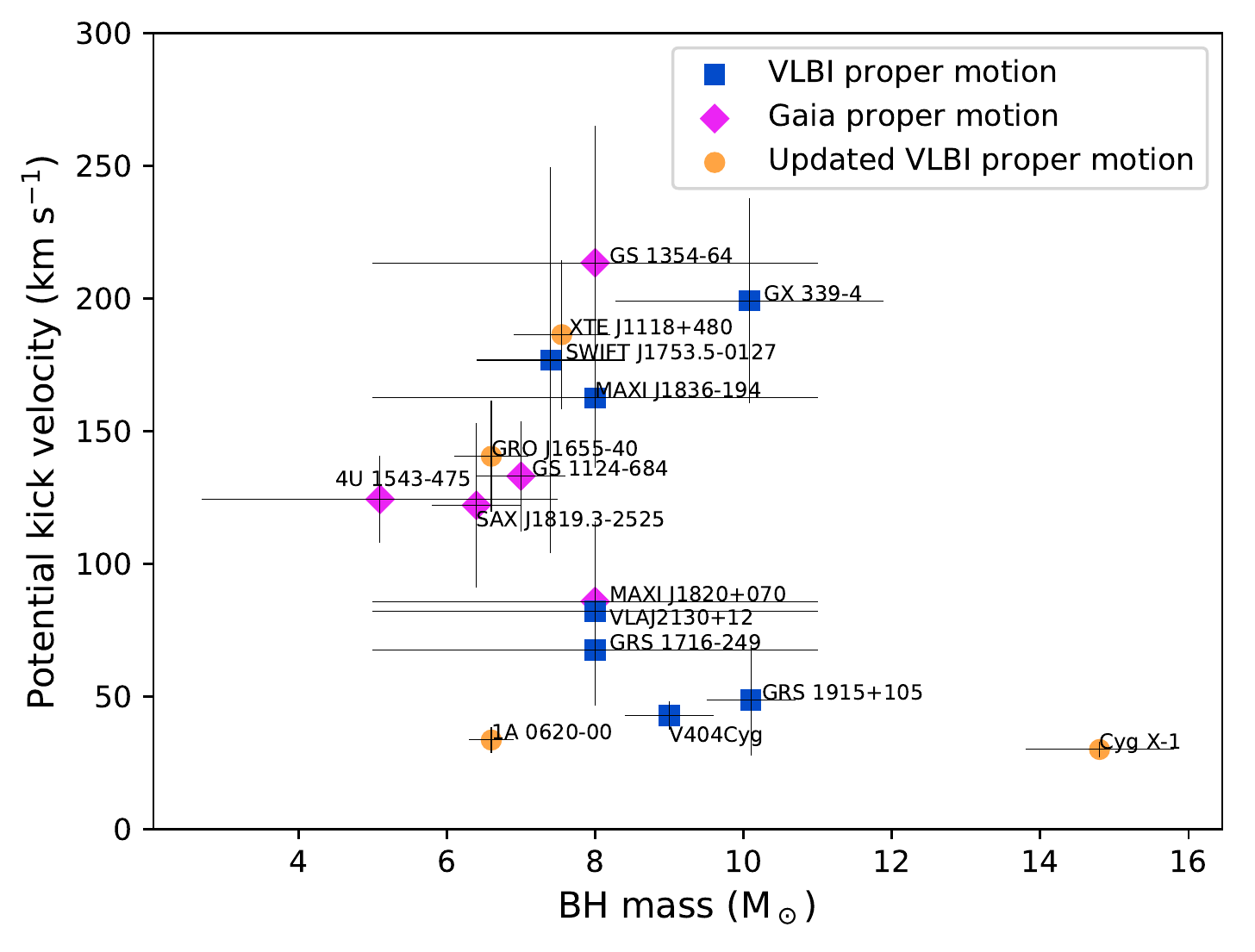}
\caption{\textit{Left}: the variation of PKV with the current height of the system above the Galactic plane. There are five systems with a PKV below 100\,$\rm{km~s^{-1}}$ are within 500\,pc of the Galactic plane. There are two systems that are within 500\,pc but have PKVs $>$100\,$\rm{km~s^{-1}}$. \textit{Right}: the variation of PKVs with black hole mass. The systems marked with blue squares have had their proper motion measured by VLBI and the systems marked with pink diamonds had their proper motion measured by \textit{Gaia}. The systems marked with yellow circles had their proper motions measured by VLBI, but we used the updated proper motions measured by \textit{Gaia} for our simulations. We have assumed a BH mass of 8\,$\rm{M_\odot}$ to represent the systems that do not have dynamical mass measurements \citep{Kreidberg2012}. The error bars are one standard deviation of the PKV probability distributions. No correlation between BH mass and potential kick velocity is seen, even after disregarding the systems without dynamical mass measurements.}
\label{zdist}
  \vspace{0.3cm}
\end{figure*}
\begin{table*}
        \begin{center}
                \caption{Current $\mid$z$\mid$ and PKV estimates for every source, along with the assumed mass used for the correlation test. The PKV is the median of the PKV probability distribution and the error bar is  the 1$\sigma$ deviation of the distribution. Current $\mid$z$\mid$ is the median of the current $\mid$z$\mid$ distribution for the estimated distance distributions for each BHXB and the errors are the 1$\sigma$ of the distributions. GS 1354--64, MAXI J1836--194, GRS 1716--249, MAXI J1820+070 and VLA J2130+12 do not have dynamical mass measurements, so we assumed BH masses of 8M$_{\odot}$ \citep{Kreidberg2012}. \label{tab:table7}}
                \begin{tabular}{l c c c l}
                \hline \hline
        Source & PKV & Current $\mid$z$\mid$ & BH mass & mass Ref  \\
                     & ($\rm{km~s^{-1}}$) & (kpc) & ($\rm{M_{\odot}}$) & \\ \hline         
         1A\,0620--00 &\phantom{1}$34\pm5$ & $0.09\pm0.01$ &\phantom{1}$6.6\pm0.3$ & \citet{Grunsven2017}  \\
         XTE\,J1118+480 &  $186\pm28$ &$1.55\pm0.09$& \phantom{1}$7.55\pm0.65$ & \citet{Khargharia2013}\\
         GS\,1124--684 & $133\pm21$ & $0.71\pm0.04$& \phantom{21.}$7\pm0.6$ & \citet{Casares2007} \\
         GS\,1354--64 (BW Cir) & $213\pm52$ & $0.36\pm0.24$&\phantom{1}-- & -- \\
         4U\,1543--475 &  $124\pm16$ & $0.72\pm0.1$\phantom{1}& \phantom{1}2.7--7.5 & \citet{Orosz1998}\\
         GRO\,J1655-40 & $140\pm21$ & $0.15\pm0.01$ & \phantom{1}$6.6\pm0.5$ & \citet{Casares2007} \\
         GX\,339--4 & $199\pm39$ & $0.66\pm0.18$ & \phantom{11}8.28--11.89 & \citet{Sreehari2018}\\
         GRS\,1716--249 &\phantom{1}$67\pm21$ & $0.31\pm0.05$ & \phantom{1}-- & \citet{Masetti1996}\\
         Swift\,J1753.5--0127 &  $177\pm73$ &  \phantom{1}$1.3\pm0.42$ & $>7.4\pm1.2$\phantom{1} & \citet{Shaw2016}\\
         SAX\,J1819.3--2525 &  $122\pm31$ &$0.51\pm0.06$& \phantom{1}6.4$\pm$0.6 & \citet{MacDonald2014}\\
         MAXI\,J1820+070 & \phantom{1}$86\pm30$ &  $0.51\pm0.11$ &\phantom{1}-- & -- \\
         MAXI\,J1836--194 &  $162\pm27$ & $0.66\pm0.17$& \phantom{1}-- & \citet{Russell2014}\\
         GRS\,1915+105 & \phantom{1}$49\pm21$ &$0.03\pm0.01$& $10.1\pm0.6$ & \citet{Steeghs2013}\\
         Cyg\,X--1 &$30\pm3$ & $0.15\pm0.01$ & $14.8\pm1.0$ & \citet{Orosz2011}\\
         V404\,Cyg &$43\pm5$ & $0.07\pm0.01$&  \phantom{111}$9\pm0.6$ & \citet{Khargharia2010}\\
         VLA\,J2130+12 & \phantom{1}$82\pm30$ &  $1.05\pm0.25$&\phantom{1}-- & -- \\
                \hline
                \end{tabular}
                
\end{center}
\end{table*}
In the past, the distribution of the distance from the Galactic plane ($\mid$z$\mid$) has been used to determine the strength of BH natal kicks \citep[e.g.][]{van1995,Jonker2004,Repetto2017}. We plot the median of the current height probability distributions of all BHXBs in our sample using our distance posterior distributions (Table \ref{tab:table7}) in Figure \ref{zdist}. It can be seen that all systems with a potential kick velocity $<$ $100\,\rm{kms^{-1}}$ are within 500\,pc of the Galactic plane, other than VLA\,J2130+12. We conducted Pearson and Spearman rank correlation tests to check the correlation of the distance to the Galactic plane ($\mid$z$\mid$) with PKVs of BHXBs. We found a potential correlation between the two, with correlation coefficients of $0.51\pm 0.15$ (p=0.02) and $0.56\pm 0.13$ (p=0.01) for Pearson and Spearman Rank tests respectively. This, however does not suggest that systems at small heights from the Galactic plane received low kick velocities because of the two outliers GRO\,J1655--40 and GS\,1354--64. This could be because the systems are passing through the Plane at the present time, which is the case for GRO\,J1655--40 as it is on a path towards the Galactic plane in its orbit \citep{Gandhi2019}. The distance estimate of GS\,1354--64 is poorly constrained and could be the reason for a misleading current height above the Galactic plane. Thus, the strength of PKV a BHXB receives cannot be determined solely by the current $\mid$z$\mid$. Hence, the approach that we take is currently the best that can be done to estimate natal kick distributions.\par
Comparisons between the root mean square distance from the Galactic plane ($z_{rms}$) for BHXBs and NSs have led to contrasting claims about BH natal kicks. NSs and BHs have different observational biases and these contrasting claims could be an artefact of that. \citet{White1996} suggested that BHs get reduced kicks as compared to NSs. \citet{Jonker2004} updated the distances to these systems and showed that BH natal kicks are as strong as those of NSs. We now update the $z_{rms}$ as we have new distances (in some cases better constrained distances) for BHXBs. Our analysis gives the $z_{rms}$ of BHXBs as $\sim$700\,pc, which is higher than those estimated by \citet{White1996} ($\sim$410\,pc) and \citet{Jonker2004} ($\sim$625\,pc), but lower than the most recent estimate by \citet{Repetto2017} ($\sim$980\,pc). If we exclude the halo object XTE\,J1118+480, the $z_{rms}$ for BHXBs drops down to 595\,pc. If we are to believe the relation between $z_{rms}$ and BHXB kick distribution, this would then suggest that NSXBs ($z_{rms}$=1\,kpc \citep{van1995}) receive stronger kicks as compared to BHXBs. We arrived at a similar conclusion when we compared pulsar kick velocities with BHXB PKVs in Section \ref{nsnatal}, but with the caveat of the selection bias of our sample of BHXBs. Hence, even though the current $\mid$z$\mid$ of the BHXBs in our data sample is not a strong proxy for the PKV of individual systems, $z_{rms}$ is a good way to compare the PKV distributions of our sample of BHXBs to the NSs. This could be because systems spend most of their time close to the orbit extrema, and when we average the current heights of a sample, we remove the bias of the epoch of observation. 
\subsection{Natal kicks and black hole mass}
 Attempts have been made to understand if there is any relation between BH mass and the natal kick it receives \citep{Mirabel2017}. This could assist in understanding the correlation between the final pathway a progenitor takes to become a BH and the mass of the BH. The current theoretical understanding does not show a direct correlation and suggests that there is no clean mass cut off for direct collapse formation \citep{Sukhbold2016}. We have obtained observational constraints on the potential kicks of 16 BHXBs (Figure \ref{zdist}), which increases the sample size and updates the estimations of BHXB kicks by \citet{Mirabel2017} and \citet{Gandhi2019} (current peculiar velocity used as proxy in both works). We conducted correlation tests, which indicated negligible correlation between BH mass and kick, with a Pearson Rank coefficient of $-0.23\pm0.17$ (p=0.21) and a Spearman Rank coefficient of $-0.2\pm0.17$ (p=0.40). We used a standard mass of $\rm{8\pm3\,M_{\odot}}$ \citep{Kreidberg2012} for the systems that do not have a dynamically measured mass (see Table \ref{tab:table7} for a summary of masses used). Removing these systems from the test sample gave Pearson and Spearman rank coefficients of $-0.36\pm0.16$ (p=0.20) and $-0.31\pm0.17$ (p=0.28), respectively. Using BHXB potential kick velocities as an indicator of BH natal kick, we suggest that there is no significant dependency of BH mass on natal kick.This is the first time there has been observational evidence to support the simulations of \citet{Sukhbold2016}, although we stress that the large error bars in the masses of the BHs \citep[e.g.][]{Ziolkowski2008} are a caveat.  
\subsection{BHs in Globular clusters}
XTE\,J1118+480 is argued to be a GC escapee because of its high peculiar velocity and high Galactic latitude. We find at least five systems with PKVs comparable to XTE\,J1118+480, making this system non-unique. However, most LMXBs are relatively old systems and thus may have come from low metallicity environments like GCs. So a GC origin of these high PKV systems cannot be ruled out. Since the majority of the systems in our sample get strong kicks, then if this distribution is characteristic of BHs it would imply that the retention fraction may be fairly low in GCs. As globular clusters (GC) are dense stellar environments, they are ideal sites for forming binaries. The escape velocities of GCs are on the order of a few tens of $\rm{km~s^{-1}}$ (\citealt{Gnedin2002}; \citealt{Belczynski2006}). Thus, most of the BHs should be eventually ejected from the GC due to kicks associated with their birth or dynamical interactions (\citealt{Kulkarni1993}; \citealt{Drukier1996}). The recent increase in the number of BHXB candidates detected in GCs challenges this theory (\citealt{Strader2012,Camilo2000,Chomiuk2013,Miller-Jones2015,Giesers2018}). Recent simulations have shown that if GCs retain $\sim$1000 BHs, then these BHs will produce a considerable population of BHXBs in the GCs \citep{Giesler2018}. Simulations that estimate the BHs retained in GCs use various observationally unconstrained distributions ranging from random flat kick distributions up to some value \citep{Sippel2013} through to unconstrained retention fractions \citep{Giesler2018}. Our PKV distribution shows that these assumptions should be used with caution. The range of assumptions for the kick distributions also emphasise the need for a consistent, observationally constrained distribution.\par
According to our results the 20$^{th}$ percentile of the PKV distribution for BHXBs is $\sim$45\,$\rm{km~s^{-1}}$, which is the typical $\rm{v_{esc}}$ for GCs. This suggests that BHXBs, which could have weaker recoil kicks than isolated BHs due to remaining bound, have a very high probability of getting kicked out from GCs. On the contrary, if BHXBs are Blaauw kick dominated, then isolated BHs have a higher chance of being retained in GCs. Thus, at birth, the retention fraction of BHs in GCs could be lower than the current estimates and may need to be updated. However, we note that GCs are old systems with low metallicity, and we have no observational constraints on how metallicity might affect the kick distribution. \par
\subsection{Implications for BH-BH systems}
When a star in the binary undergoes a SN explosion, any asymmetry in the kick could misalign the spin of the remnant BH to the orbital plane. If tidal interactions are then unable to realign the spin, this could potentially give rise to a high degree of misalignment of the spins of the BH to the orbital plane of the BH-BH binary \citep{Gerosa2013}. Recent observations of the GW merger event GW151226 suggested that natal kicks $>$50\,$\rm{km~s^{-1}}$ were needed to explain the inferred spin-orbit misalignment of the binary \citep{OShaughnessy2017}. The spin measurements for three other GW events GW150914, LVT151012 and GW170104 indicated that there is a possibility for these systems to either have low intrinsic spins or large spins that are misaligned with the binary orbit \citep{Farr2017}. 90$\%$ of our sources have PKVs higher than 50\,$\rm{km~s^{-1}}$, thus suggesting that spin-orbit misalignment might be a common phenomenon. \par
Based on LIGO's BH binary merger observations it was suggested that BHs receive high natal kicks $\sim\,\rm{200\,kms^{-1}}$ if the tidal processes do not realign stellar spins \citep{Wysocki2018}. If however tidal processes do realign the spins, kicks on the order of $\>\rm{50\,kms^{-1}}$ are still needed to explain the observations. The bimodality in our natal kick distributions, albeit based on a few sources, could potentially arise from similar tidal effects, though more work needs to done in order to draw definitive conclusions. Theoretical estimates suggest that the majority of BH binaries that will be observed with the third LIGO run will be part of a population in which the first-born BH is slowly spinning, and support low natal kicks for BHXBs  \citep{Bavera2019,Fuller2019}. Measuring PKVs for more BHXBs and future LIGO observations will be essential to reconcile these theoretical expectations of low natal kicks with our observational constraints of high PKVs.
\subsection{BHXB spin-orbit misalignment}
Study of X-ray power density spectra of BHXBs show low frequency quasi periodic oscillations (QPOs) in the power density spectra of almost all BHXB systems \citep{Ingram2016}. Theoretical models suggest that such QPOs could be a result of Lense-Thirring precession \citep{Stella1998}, which is a phenomenon where material out of the BH equatorial plane (e.g. due to misalignment of the BH spin and binary orbit) precesses due to relativistic frame dragging. Any natal kick imparted to a BH at its birth could cause a spin-orbit misalignment in the remnant BHXB. Since re-alignment timescales of BHXBs are usually longer than the ages of these systems \citep{Martin2008}, realignment of the BH spin to the orbital plane is not likely to be common \citep[unless the donor star in the BHXB was intially an intermediate mass donor;][]{Fragos2015}. We found that a majority of the BHXBs in our sample set were strongly kicked, which is consistent with the prevalence of low frequency QPOs and provides strong evidence for spin-orbit misalignment in BHXBs. \par
Extensive efforts to measure the spins of BHs in BHXBs have been made in the past decade. The best available methods \citep{Garcia2018,McClintock2014} often assume that the spin of the BH is aligned with the orbital plane of the binary. Our finding that strong natal kicks are imparted to $\sim$85$\%$ of BHXBs suggests that caution may be warranted in assuming the binary orbital inclination when fitting for BH spin.
\section{Conclusions}
In this paper, we report on measurements of proper motions of three BHXB systems using VLBI. We collated the proper motions, systemic radial velocities and distances of 16 BHXB systems. We developed a methodology to determine potential kick velocity (PKV) distributions for BHXB systems. We also developed a MC simulation code to account for uncertainties in the measured proper motions, systemic radial velocities and distances when determining the potential kick velocity that the BHXB received, providing robust observational constraints on the possible kick velocities of these systems. We estimated PKV probability distributions for sixteen BHXBs and found that $\sim\,75\,\%$ of our sample has a median potential kick velocity of $>\rm{70\,kms^{-1}}$, which we interpreted as a majority of BHXBs acquiring strong kicks when they are born. We combined the PKV probability distributions for these 16 BHXBs to obtain an aggregate PKV distribution. We found that a unimodal Gaussian distribution with a mean of 107$\pm$16\,$\rm{km~s^{-1}}$ fit the data better than a bimodal distribution, which is potentially consistent with the LIGO's BH-BH merger observations and natal kick estimations. Alternatively, the fit suffers from low number of systems in the data sample and hence we could not rule out a bimodal distribution. We found no significant correlation between PKVs and BH mass. We did not find any strict mass cut off for BHs to form with a SN or by direct collapse \par
We conducted Spearman and Pearson rank correlation tests to determine the correlation between the current height above the Galactic plane of a BHXB and the potential kick velocity it received. We found that even though there is a potential correlation between the two (coefficient of $\sim0.5$ for both tests), we should avoid using $\mid$z$\mid$ as a natal kick proxy. We compared our aggregated PKV distribution with the pulsar peculiar velocity distribution \citep{Verbunt2017} and found that BHs may get weaker kicks than NSs by a factor of 3--4. Our finding that BHXB kick velocities are greater than typical escape velocities of GCs favours a large fraction of BHs being kicked out of GCs. The prevalence of strong kicks in our BHXB sample is in agreement with the ubiquity of low frequency QPOs and hence spin-orbit misalignment in almost all BHXBs.

\section{Acknowledgements}
We would like to thank Poshak Gandhi for his useful inputs on the low mass X-ray binary prior to be used for our Bayesian Inference Analysis. PyRAF is a product of the Space Telescope Science Institute, which is operated by AURA for NASA.
The Long Baseline Array is part of the Australia Telescope National Facility which is funded by the Australian Government for operation as a National Facility managed by CSIRO. We would like to thank the Institute of Radio Astronomy and Space Research, AUT University, New Zealand for the use and operational support of their radio telescopes in the collection of data for this work. This work made use of the Swinburne University of Technology software correlator, developed as part of the Australian Major National Research Facilities Programme. The Australian SKA Pathfinder is part of the Australia Telescope National Facility which is managed by CSIRO. Operation of ASKAP is funded by the Australian Government with support from the National Collaborative Research Infrastructure Strategy. ASKAP uses the resources of the Pawsey Supercomputing Centre. Establishment of ASKAP, the Murchison Radio-astronomy Observatory and the Pawsey Supercomputing Centre are initiatives of the Australian Government, with support from the Government of Western Australia and the Science and Industry Endowment Fund. We acknowledge the Wajarri Yamatji people as the traditional owners of the Observatory site.
This work was supported by resources provided by the Pawsey Supercomputing Centre with funding from the Australian Government and the Government of Western Australia. The European VLBI Network is a joint facility of independent European, African, Asian, and North American radio astronomy institutes. Scientific results from data presented in this publication are derived from the following EVN project code(s): EM101. The Long Baseline Observatory is a facility of the National Science Foundation operated under cooperative agreement by Associated Universities, Inc. The GBT is part of the Green Bank Observatory, a facility of the National
Science Foundation operated under a cooperative agreement by Associated Universities, Inc.  This work has made use of data from the European Space Agency (ESA) mission {\it Gaia} (\url{https://www.cosmos.esa.int/gaia}), processed by the {\it Gaia} Data Processing and Analysis Consortium (DPAC,\url{https://www.cosmos.esa.int/web/gaia/dpac/consortium}). Funding for the DPAC has been provided by national institutions, in particular the institutions participating in the {\it Gaia} Multilateral Agreement. JCAM-J is the recipient of an Australian Research Council Future Fellowship (FT140101082) funded by the Australian government . PGJ acknowledges funding from the European Research Council under ERC Consolidator Grant agreement number 647208. MAPT acknowledge support by the Spanish Ministry of Economy, Industry and Competitiveness (MINECO) under grant AYA2017-83216-P and support via a Ram\'on y Cajal Fellowship (RYC-2015-17854). SC acknowledges funding from CNES (french space agency), through MINE, the Multi-wavelength INTEGRAL Network.




\bibliographystyle{mnras}
\bibliography{cat_0} 

\begin{thebibliography}{}
\makeatletter
\relax
\def\mn@urlcharsother{\let\do\@makeother \do\$\do\&\do\#\do\^\do\_\do\%\do\~}
\def\mn@doi{\begingroup\mn@urlcharsother \@ifnextchar [ {\mn@doi@}
  {\mn@doi@[]}}
\def\mn@doi@[#1]#2{\def\@tempa{#1}\ifx\@tempa\@empty \href
  {http://dx.doi.org/#2} {doi:#2}\else \href {http://dx.doi.org/#2} {#1}\fi
  \endgroup}
\def\mn@eprint#1#2{\mn@eprint@#1:#2::\@nil}
\def\mn@eprint@arXiv#1{\href {http://arxiv.org/abs/#1} {{\tt arXiv:#1}}}
\def\mn@eprint@dblp#1{\href {http://dblp.uni-trier.de/rec/bibtex/#1.xml}
  {dblp:#1}}
\def\mn@eprint@#1:#2:#3:#4\@nil{\def\@tempa {#1}\def\@tempb {#2}\def\@tempc
  {#3}\ifx \@tempc \@empty \let \@tempc \@tempb \let \@tempb \@tempa \fi \ifx
  \@tempb \@empty \def\@tempb {arXiv}\fi \@ifundefined
  {mn@eprint@\@tempb}{\@tempb:\@tempc}{\expandafter \expandafter \csname
  mn@eprint@\@tempb\endcsname \expandafter{\@tempc}}}

\bibitem[\protect\citeauthoryear{{Abbott} et~al.,}{{Abbott}
  et~al.}{2016a}]{Abbott2016a}
{Abbott} B.~P.,  et~al., 2016a, \mn@doi [Physical Review X]
  {10.1103/PhysRevX.6.041015}, \href
  {http://adsabs.harvard.edu/abs/2016PhRvX...6d1015A} {6, 041015}

\bibitem[\protect\citeauthoryear{{Abbott} et~al.,}{{Abbott}
  et~al.}{2016b}]{Abbott2016b}
{Abbott} B.~P.,  et~al., 2016b, \mn@doi [\apjl] {10.3847/2041-8205/818/2/L22},
  \href {http://adsabs.harvard.edu/abs/2016ApJ...818L..22A} {818, L22}

\bibitem[\protect\citeauthoryear{{Adams}, {Kochanek}, {Gerke}  \&
  {Stanek}}{{Adams} et~al.}{2017}]{Adams2017a}
{Adams} S.~M.,  {Kochanek} C.~S.,  {Gerke} J.~R.,   {Stanek} K.~Z.,  2017,
  \mn@doi [\mnras] {10.1093/mnras/stx898}, \href
  {http://adsabs.harvard.edu/abs/2017MNRAS.469.1445A} {469, 1445}

\bibitem[\protect\citeauthoryear{{Akaike}}{{Akaike}}{1974}]{Akaike1974}
{Akaike} H.,  1974, IEEE Transactions on Automatic Control, \href
  {http://adsabs.harvard.edu/abs/1974ITAC...19..716A} {19, 716}

\bibitem[\protect\citeauthoryear{{Antonini}, {Gieles}  \&
  {Gualandris}}{{Antonini} et~al.}{2018}]{Antonini2018}
{Antonini} F.,  {Gieles} M.,   {Gualandris} A.,  2018, arXiv e-prints, \href
  {http://adsabs.harvard.edu/abs/2018arXiv181103640A} {}

\bibitem[\protect\citeauthoryear{{Arzoumanian}, {Chernoff}  \&
  {Cordes}}{{Arzoumanian} et~al.}{2002}]{Arzoumanian2002}
{Arzoumanian} Z.,  {Chernoff} D.~F.,   {Cordes} J.~M.,  2002, \mn@doi [\apj]
  {10.1086/338805}, \href {http://adsabs.harvard.edu/abs/2002ApJ...568..289A}
  {568, 289}

\bibitem[\protect\citeauthoryear{{Astraatmadja} \&
  {Bailer-Jones}}{{Astraatmadja} \& {Bailer-Jones}}{2016}]{Astraatmadj2016}
{Astraatmadja} T.~L.,  {Bailer-Jones} C.~A.~L.,  2016, \mn@doi [\apj]
  {10.3847/0004-637X/832/2/137}, \href
  {http://adsabs.harvard.edu/abs/2016ApJ...832..137A} {832, 137}

\bibitem[\protect\citeauthoryear{{Bahramian} et~al.,}{{Bahramian}
  et~al.}{2014}]{Bahramian2014}
{Bahramian} A.,  et~al., 2014, \mn@doi [\apj] {10.1088/0004-637X/780/2/127},
  \href {http://adsabs.harvard.edu/abs/2014ApJ...780..127B} {780, 127}

\bibitem[\protect\citeauthoryear{{Bailer-Jones}}{{Bailer-Jones}}{2015}]{Bailer-Jones2015}
{Bailer-Jones} C.~A.~L.,  2015, \mn@doi [\pasp] {10.1086/683116}, \href
  {http://adsabs.harvard.edu/abs/2015PASP..127..994B} {127, 994}

\bibitem[\protect\citeauthoryear{{Ballet}, {Denis}, {Gilfanov}, {Sunyaev},
  {Harmon}, {Zhang}, {Paciesas}  \& {Fishman}}{{Ballet}
  et~al.}{1993}]{Ballet1993}
{Ballet} J.,  {Denis} M.,  {Gilfanov} M.,  {Sunyaev} R.,  {Harmon} B.~A.,
  {Zhang} S.~N.,  {Paciesas} W.~S.,   {Fishman} G.~J.,  1993, \iaucirc, \href
  {http://adsabs.harvard.edu/abs/1993IAUC.5874....1B} {5874}

\bibitem[\protect\citeauthoryear{{Bassi} et~al.,}{{Bassi}
  et~al.}{2019}]{Bassi2019}
{Bassi} T.,  et~al., 2019, \mn@doi [\mnras] {10.1093/mnras/sty2739}, \href
  {http://adsabs.harvard.edu/abs/2019MNRAS.482.1587B} {482, 1587}

\bibitem[\protect\citeauthoryear{{Bavera} et~al.,}{{Bavera}
  et~al.}{2019}]{Bavera2019}
{Bavera} S.~S.,  et~al., 2019, arXiv e-prints, \href
  {https://ui.adsabs.harvard.edu/abs/2019arXiv190612257B} {p. arXiv:1906.12257}

\bibitem[\protect\citeauthoryear{{Belczynski}, {Bulik}  \&
  {Kalogera}}{{Belczynski} et~al.}{2002}]{Belczynski2002}
{Belczynski} K.,  {Bulik} T.,   {Kalogera} V.,  2002, \mn@doi [\apjl]
  {10.1086/341365}, \href {http://adsabs.harvard.edu/abs/2002ApJ...571L.147B}
  {571, L147}

\bibitem[\protect\citeauthoryear{{Belczynski}, {Sadowski}, {Rasio}  \&
  {Bulik}}{{Belczynski} et~al.}{2006}]{Belczynski2006}
{Belczynski} K.,  {Sadowski} A.,  {Rasio} F.~A.,   {Bulik} T.,  2006, \mn@doi
  [\apj] {10.1086/506186}, \href
  {http://adsabs.harvard.edu/abs/2006ApJ...650..303B} {650, 303}

\bibitem[\protect\citeauthoryear{{Belloni} \& {Motta}}{{Belloni} \&
  {Motta}}{2016}]{Belloni2016}
{Belloni} T.~M.,  {Motta} S.~E.,  2016, in {Bambi} C.,  ed.,  Astrophysics and
  Space Science Library Vol. 440, Astrophysics of Black Holes: From Fundamental
  Aspects to Latest Developments. p.~61 (\mn@eprint {arXiv} {1603.07872}),
  \mn@doi{10.1007/978-3-662-52859-4_2}

\bibitem[\protect\citeauthoryear{{Benacquista} \& {Downing}}{{Benacquista} \&
  {Downing}}{2013}]{Benacquista2013}
{Benacquista} M.~J.,  {Downing} J.~M.~B.,  2013, \mn@doi [Living Reviews in
  Relativity] {10.12942/lrr-2013-4}, \href
  {http://adsabs.harvard.edu/abs/2013LRR....16....4B} {16, 4}

\bibitem[\protect\citeauthoryear{{Betancourt}}{{Betancourt}}{2017}]{Betancourt2017}
{Betancourt} M.,  2017, arXiv e-prints, \href
  {http://adsabs.harvard.edu/abs/2017arXiv170601520B} {}

\bibitem[\protect\citeauthoryear{{Blaauw}}{{Blaauw}}{1961}]{Blaauw1961}
{Blaauw} A.,  1961, \bain, \href
  {http://adsabs.harvard.edu/abs/1961BAN....15..265B} {15, 265}

\bibitem[\protect\citeauthoryear{{Bovy}}{{Bovy}}{2014}]{Bovy2014}
{Bovy} J.,  2014, {galpy: Galactic dynamics package}, Astrophysics Source Code
  Library (\mn@eprint {ascl} {1411.008})

\bibitem[\protect\citeauthoryear{{Brandt}, {Podsiadlowski}  \&
  {Sigurdsson}}{{Brandt} et~al.}{1995}]{Brandt1995}
{Brandt} W.~N.,  {Podsiadlowski} P.,   {Sigurdsson} S.,  1995, \mn@doi [\mnras]
  {10.1093/mnras/277.1.L35}, \href
  {http://adsabs.harvard.edu/abs/1995MNRAS.277L..35B} {277, L35}

\bibitem[\protect\citeauthoryear{{Burnham} \& {Anderson}}{{Burnham} \&
  {Anderson}}{2002}]{Burnham2002}
{Burnham} K.~P.,  {Anderson} D.~R.,  2002, in Model Selection and Multimodel
  Inference. , \mn@doi{10.1007/b97636}

\bibitem[\protect\citeauthoryear{{Buxton}, {Hasan}, {MacPherson}  \&
  {Bailyn}}{{Buxton} et~al.}{2013}]{Buxton2013}
{Buxton} M.,  {Hasan} I.,  {MacPherson} E.,   {Bailyn} C.,  2013, The
  Astronomer's Telegram, \href
  {http://adsabs.harvard.edu/abs/2013ATel.5244....1B} {5244}

\bibitem[\protect\citeauthoryear{{Cadolle Bel} et~al.,}{{Cadolle Bel}
  et~al.}{2007}]{Cadolle2007}
{Cadolle Bel} M.,  et~al., 2007, \mn@doi [\apj] {10.1086/512004}, \href
  {http://adsabs.harvard.edu/abs/2007ApJ...659..549C} {659, 549}

\bibitem[\protect\citeauthoryear{{Camilo}, {Lorimer}, {Freire}, {Lyne}  \&
  {Manchester}}{{Camilo} et~al.}{2000}]{Camilo2000}
{Camilo} F.,  {Lorimer} D.~R.,  {Freire} P.,  {Lyne} A.~G.,   {Manchester}
  R.~N.,  2000, \mn@doi [\apj] {10.1086/308859}, \href
  {http://adsabs.harvard.edu/abs/2000ApJ...535..975C} {535, 975}

\bibitem[\protect\citeauthoryear{{Cantrell} et~al.,}{{Cantrell}
  et~al.}{2010}]{Cantrell2010}
{Cantrell} A.~G.,  et~al., 2010, \mn@doi [\apj] {10.1088/0004-637X/710/2/1127},
  \href {http://adsabs.harvard.edu/a@ARTICLE{2005ApJ...623.1026H, author =
  {{Hynes}, R.~I.}, title = "{The Optical and Ultraviolet Spectral Energy
  Distributions of Short-Period Black Hole X-Ray Transients in Outburst}",
  journal = {\apj}, eprint = {astro-ph/0412531}, keywords = {Accretion,
  Accretion Disks, Stars: Binaries: Close, Stars: Individual: Constellation
  Name: V616 Monocerotis, Stars: Individual: Constellation Name: GU Muscae,
  Stars: Individual: Constellation Name: V2293 Ophiuchi, Stars: Individual:
  Constellation Name: V518 Persei, Stars: Individual: Constellation Name: MM
  Velorum, Stars: Individual: Constellation Name: V406 Vulpeculae}, year =
  2005, month = apr, volume = 623, pages = {1026-1043}, doi = {10.1086/428445},
  adsurl = {http://adsabs.harvard.edu/abs/2005ApJ...623.1026H}, adsnote =
  {Provided by the SAO/NASA Astrophysics Data System} } bs/2010ApJ...710.1127C}
  {710, 1127}

\bibitem[\protect\citeauthoryear{{Casares}}{{Casares}}{2007}]{Casares2007}
{Casares} J.,  2007, in {Karas} V.,  {Matt} G.,  eds,  IAU Symposium Vol. 238,
  Black Holes from Stars to Galaxies -- Across the Range of Masses. pp 3--12
  (\mn@eprint {} {astro-ph/0612312}), \mn@doi{10.1017/S1743921307004590}

\bibitem[\protect\citeauthoryear{{Casares} \& {Charles}}{{Casares} \&
  {Charles}}{1994}]{Casares1994}
{Casares} J.,  {Charles} P.~A.,  1994, \mn@doi [\mnras]
  {10.1093/mnras/271.1.L5}, \href
  {http://adsabs.harvard.edu/abs/1994MNRAS.271L...5C} {271, L5}

\bibitem[\protect\citeauthoryear{{Casares}, {Zurita}, {Shahbaz}, {Charles}  \&
  {Fender}}{{Casares} et~al.}{2004}]{Casares2004}
{Casares} J.,  {Zurita} C.,  {Shahbaz} T.,  {Charles} P.~A.,   {Fender} R.~P.,
  2004, \mn@doi [\apjl] {10.1086/425145}, \href
  {http://adsabs.harvard.edu/abs/2004ApJ...613L.133C} {613, L133}

\bibitem[\protect\citeauthoryear{{Casares} et~al.,}{{Casares}
  et~al.}{2009}]{Casares2009}
{Casares} J.,  et~al., 2009, \mn@doi [\apjs] {10.1088/0067-0049/181/1/238},
  \href {http://adsabs.harvard.edu/abs/2009ApJS..181..238C} {181, 238}

\bibitem[\protect\citeauthoryear{{Cavanaugh}}{{Cavanaugh}}{1997}]{Cavanaugh1997}
{Cavanaugh} J.~E.,  1997, Statistics \& Probability Letters, 33, 201

\bibitem[\protect\citeauthoryear{{Chaty}, {Mirabel}, {Goldoni}, {Mereghetti},
  {Duc}, {Mart{\'{\i}}}  \& {Mignani}}{{Chaty} et~al.}{2002}]{Chaty2002}
{Chaty} S.,  {Mirabel} I.~F.,  {Goldoni} P.,  {Mereghetti} S.,  {Duc} P.-A.,
  {Mart{\'{\i}}} J.,   {Mignani} R.~P.,  2002, \mn@doi [\mnras]
  {10.1046/j.1365-8711.2002.05267.x}, \href
  {http://adsabs.harvard.edu/abs/2002MNRAS.331.1065C} {331, 1065}

\bibitem[\protect\citeauthoryear{{Chomiuk}, {Strader}, {Maccarone},
  {Miller-Jones}, {Heinke}, {Noyola}, {Seth}  \& {Ransom}}{{Chomiuk}
  et~al.}{2013}]{Chomiuk2013}
{Chomiuk} L.,  {Strader} J.,  {Maccarone} T.~J.,  {Miller-Jones} J.~C.~A.,
  {Heinke} C.,  {Noyola} E.,  {Seth} A.~C.,   {Ransom} S.,  2013, \mn@doi
  [\apj] {10.1088/0004-637X/777/1/69}, \href
  {http://adsabs.harvard.edu/abs/2013ApJ...777...69C} {777, 69}

\bibitem[\protect\citeauthoryear{{Chugai}}{{Chugai}}{1984}]{Chugai1984}
{Chugai} N.~N.,  1984, Soviet Astronomy Letters, \href
  {http://adsabs.harvard.edu/abs/1984SvAL...10...87C} {10, 87}

\bibitem[\protect\citeauthoryear{{Clark}}{{Clark}}{1975}]{Clark1975}
{Clark} G.~W.,  1975, \mn@doi [\apjl] {10.1086/181869}, \href
  {http://adsabs.harvard.edu/abs/1975ApJ...199L.143C} {199, L143}

\bibitem[\protect\citeauthoryear{{Corral-Santana}, {Casares},
  {Mu{\~n}oz-Darias}, {Bauer}, {Mart{\'{\i}}nez-Pais}  \&
  {Russell}}{{Corral-Santana} et~al.}{2016}]{Corral-Santana2016}
{Corral-Santana} J.~M.,  {Casares} J.,  {Mu{\~n}oz-Darias} T.,  {Bauer} F.~E.,
  {Mart{\'{\i}}nez-Pais} I.~G.,   {Russell} D.~M.,  2016, \mn@doi [\aap]
  {10.1051/0004-6361/201527130}, \href
  {http://adsabs.harvard.edu/abs/2016A%26A...587A..61C} {587, A61}

\bibitem[\protect\citeauthoryear{{Dehnen} \& {Binney}}{{Dehnen} \&
  {Binney}}{1998}]{Dehnen1998}
{Dehnen} W.,  {Binney} J.,  1998, \mn@doi [\mnras]
  {10.1046/j.1365-8711.1998.01282.x}, \href
  {http://adsabs.harvard.edu/abs/1998MNRAS.294..429D} {294, 429}

\bibitem[\protect\citeauthoryear{{Deller} et~al.,}{{Deller}
  et~al.}{2011}]{Deller2011}
{Deller} A.~T.,  et~al., 2011, \mn@doi [\pasp] {10.1086/658907}, \href
  {http://adsabs.harvard.edu/abs/2011PASP..123..275D} {123, 275}

\bibitem[\protect\citeauthoryear{{Dhawan}, {Mirabel}, {Rib{\'o}}  \&
  {Rodrigues}}{{Dhawan} et~al.}{2007}]{Dhawan2007}
{Dhawan} V.,  {Mirabel} I.~F.,  {Rib{\'o}} M.,   {Rodrigues} I.,  2007, \mn@doi
  [\apj] {10.1086/520111}, \href
  {http://adsabs.harvard.edu/abs/2007ApJ...668..430D} {668, 430}

\bibitem[\protect\citeauthoryear{{Drukier}}{{Drukier}}{1996}]{Drukier1996}
{Drukier} G.~A.,  1996, \mn@doi [\mnras] {10.1093/mnras/280.2.498}, \href
  {http://adsabs.harvard.edu/abs/1996MNRAS.280..498D} {280, 498}

\bibitem[\protect\citeauthoryear{{Fabian}, {Pringle}  \& {Rees}}{{Fabian}
  et~al.}{1975}]{Fabian1975}
{Fabian} A.~C.,  {Pringle} J.~E.,   {Rees} M.~J.,  1975, \mn@doi [\mnras]
  {10.1093/mnras/172.1.15P}, \href
  {http://adsabs.harvard.edu/abs/1975MNRAS.172P..15F} {172, 15p}

\bibitem[\protect\citeauthoryear{{Farr}, {Stevenson}, {Miller}, {Mandel},
  {Farr}  \& {Vecchio}}{{Farr} et~al.}{2017}]{Farr2017}
{Farr} W.~M.,  {Stevenson} S.,  {Miller} M.~C.,  {Mandel} I.,  {Farr} B.,
  {Vecchio} A.,  2017, \mn@doi [\nat] {10.1038/nature23453}, \href
  {http://adsabs.harvard.edu/abs/2017Natur.548..426F} {548, 426}

\bibitem[\protect\citeauthoryear{{Fender}, {Homan}  \& {Belloni}}{{Fender}
  et~al.}{2009}]{Fender2009}
{Fender} R.~P.,  {Homan} J.,   {Belloni} T.~M.,  2009, \mn@doi [\mnras]
  {10.1111/j.1365-2966.2009.14841.x}, \href
  {http://adsabs.harvard.edu/abs/2009MNRAS.396.1370F} {396, 1370}

\bibitem[\protect\citeauthoryear{{Fragos} \& {McClintock}}{{Fragos} \&
  {McClintock}}{2015}]{Fragos2015}
{Fragos} T.,  {McClintock} J.~E.,  2015, \mn@doi [\apj]
  {10.1088/0004-637X/800/1/17}, \href
  {https://ui.adsabs.harvard.edu/abs/2015ApJ...800...17F} {800, 17}

\bibitem[\protect\citeauthoryear{{Fragos}, {Willems}, {Kalogera}, {Ivanova},
  {Rockefeller}, {Fryer}  \& {Young}}{{Fragos} et~al.}{2009}]{Fragos2009}
{Fragos} T.,  {Willems} B.,  {Kalogera} V.,  {Ivanova} N.,  {Rockefeller} G.,
  {Fryer} C.~L.,   {Young} P.~A.,  2009, \mn@doi [\apj]
  {10.1088/0004-637X/697/2/1057}, \href
  {http://adsabs.harvard.edu/abs/2009ApJ...697.1057F} {697, 1057}

\bibitem[\protect\citeauthoryear{{Fryer} \& {Kalogera}}{{Fryer} \&
  {Kalogera}}{2001}]{Fryer2001}
{Fryer} C.~L.,  {Kalogera} V.,  2001, \mn@doi [\apj] {10.1086/321359}, \href
  {http://adsabs.harvard.edu/abs/2001ApJ...554..548F} {554, 548}

\bibitem[\protect\citeauthoryear{{Fryer}, {Burrows}  \& {Benz}}{{Fryer}
  et~al.}{1998}]{Fryer1998}
{Fryer} C.,  {Burrows} A.,   {Benz} W.,  1998, \mn@doi [\apj] {10.1086/305348},
  \href {http://adsabs.harvard.edu/abs/1998ApJ...496..333F} {496, 333}

\bibitem[\protect\citeauthoryear{{Fuller} \& {Ma}}{{Fuller} \&
  {Ma}}{2019}]{Fuller2019}
{Fuller} J.,  {Ma} L.,  2019, arXiv e-prints, \href
  {https://ui.adsabs.harvard.edu/abs/2019arXiv190703714F} {p. arXiv:1907.03714}

\bibitem[\protect\citeauthoryear{{Gaia Collaboration} et~al.,}{{Gaia
  Collaboration} et~al.}{2018}]{Brown2018}
{Gaia Collaboration} et~al., 2018, \mn@doi [\aap]
  {10.1051/0004-6361/201833051}, \href
  {http://adsabs.harvard.edu/abs/2018A%26A...616A...1G} {616, A1}

\bibitem[\protect\citeauthoryear{{Gandhi}, {Rao}, {Johnson}, {Paice}  \&
  {Maccarone}}{{Gandhi} et~al.}{2019}]{Gandhi2019}
{Gandhi} P.,  {Rao} A.,  {Johnson} M. A.~C.,  {Paice} J.~A.,   {Maccarone}
  T.~J.,  2019, \mn@doi [\mnras] {10.1093/mnras/stz438}, \href
  {https://ui.adsabs.harvard.edu/abs/2019MNRAS.485.2642G} {485, 2642}

\bibitem[\protect\citeauthoryear{{Garc{\'{\i}}a} et~al.,}{{Garc{\'{\i}}a}
  et~al.}{2018}]{Garcia2018}
{Garc{\'{\i}}a} J.~A.,  et~al., 2018, \mn@doi [\apj]
  {10.3847/1538-4357/aad231}, \href
  {http://adsabs.harvard.edu/abs/2018ApJ...864...25G} {864, 25}

\bibitem[\protect\citeauthoryear{{Gelino}, {Balman}, {K{\i}z{\i}lo{\v g}lu},
  {Y{\i}lmaz}, {Kalemci}  \& {Tomsick}}{{Gelino} et~al.}{2006}]{Gelino2006}
{Gelino} D.~M.,  {Balman} {\c S}.,  {K{\i}z{\i}lo{\v g}lu} {\"U}.,  {Y{\i}lmaz}
  A.,  {Kalemci} E.,   {Tomsick} J.~A.,  2006, \mn@doi [\apj] {10.1086/500924},
  \href {http://adsabs.harvard.edu/abs/2006ApJ...642..438G} {642, 438}

\bibitem[\protect\citeauthoryear{{Gelman} \& {Rubin}}{{Gelman} \&
  {Rubin}}{1992}]{Gelman1992}
{Gelman} A.,  {Rubin} D.~B.,  1992, \mn@doi [Statistical Science]
  {10.1214/ss/1177011136}, \href
  {http://adsabs.harvard.edu/abs/1992StaSc...7..457G} {7, 457}

\bibitem[\protect\citeauthoryear{{Gerosa}, {Kesden}, {Berti}, {O'Shaughnessy}
  \& {Sperhake}}{{Gerosa} et~al.}{2013}]{Gerosa2013}
{Gerosa} D.,  {Kesden} M.,  {Berti} E.,  {O'Shaughnessy} R.,   {Sperhake} U.,
  2013, \mn@doi [\prd] {10.1103/PhysRevD.87.104028}, \href
  {http://adsabs.harvard.edu/abs/2013PhRvD..87j4028G} {87, 104028}

\bibitem[\protect\citeauthoryear{{Gessner} \& {Janka}}{{Gessner} \&
  {Janka}}{2018}]{Gessner2018}
{Gessner} A.,  {Janka} H.-T.,  2018, \mn@doi [\apj] {10.3847/1538-4357/aadbae},
  \href {http://adsabs.harvard.edu/abs/2018ApJ...865...61G} {865, 61}

\bibitem[\protect\citeauthoryear{{Gies} et~al.,}{{Gies}
  et~al.}{2008}]{Gies2008}
{Gies} D.~R.,  et~al., 2008, \mn@doi [\apj] {10.1086/586690}, \href
  {http://adsabs.harvard.edu/abs/2008ApJ...678.1237G} {678, 1237}

\bibitem[\protect\citeauthoryear{{Giesers} et~al.,}{{Giesers}
  et~al.}{2018}]{Giesers2018}
{Giesers} B.,  et~al., 2018, \mn@doi [\mnras] {10.1093/mnrasl/slx203}, \href
  {http://adsabs.harvard.edu/abs/2018MNRAS.475L..15G} {475, L15}

\bibitem[\protect\citeauthoryear{{Giesler}, {Clausen}  \& {Ott}}{{Giesler}
  et~al.}{2018}]{Giesler2018}
{Giesler} M.,  {Clausen} D.,   {Ott} C.~D.,  2018, \mn@doi [\mnras]
  {10.1093/mnras/sty659}, \href
  {http://adsabs.harvard.edu/abs/2018MNRAS.477.1853G} {477, 1853}

\bibitem[\protect\citeauthoryear{{Gilmore} \& {Reid}}{{Gilmore} \&
  {Reid}}{1983}]{Gilmore1983}
{Gilmore} G.,  {Reid} N.,  1983, \mn@doi [\mnras] {10.1093/mnras/202.4.1025},
  \href {http://adsabs.harvard.edu/abs/1983MNRAS.202.1025G} {202, 1025}

\bibitem[\protect\citeauthoryear{{Gnedin}, {Zhao}, {Pringle}, {Fall}, {Livio}
  \& {Meylan}}{{Gnedin} et~al.}{2002}]{Gnedin2002}
{Gnedin} O.~Y.,  {Zhao} H.,  {Pringle} J.~E.,  {Fall} S.~M.,  {Livio} M.,
  {Meylan} G.,  2002, \mn@doi [\apjl] {10.1086/340319}, \href
  {http://adsabs.harvard.edu/abs/2002ApJ...568L..23G} {568, L23}

\bibitem[\protect\citeauthoryear{{Gonz{\'a}lez Hern{\'a}ndez} \&
  {Casares}}{{Gonz{\'a}lez Hern{\'a}ndez} \& {Casares}}{2010}]{Gonzalez2010}
{Gonz{\'a}lez Hern{\'a}ndez} J.~I.,  {Casares} J.,  2010, \mn@doi [\aap]
  {10.1051/0004-6361/201014088}, \href
  {http://adsabs.harvard.edu/abs/2010A%26A...516A..58G} {516, A58}

\bibitem[\protect\citeauthoryear{{Gonz{\'a}lez Hern{\'a}ndez}, {Rebolo},
  {Israelian}, {Harlaftis}, {Filippenko}  \& {Chornock}}{{Gonz{\'a}lez
  Hern{\'a}ndez} et~al.}{2006}]{Gonazalez2006}
{Gonz{\'a}lez Hern{\'a}ndez} J.~I.,  {Rebolo} R.,  {Israelian} G.,  {Harlaftis}
  E.~T.,  {Filippenko} A.~V.,   {Chornock} R.,  2006, \mn@doi [\apjl]
  {10.1086/505391}, \href {http://adsabs.harvard.edu/abs/2006ApJ...644L..49G}
  {644, L49}

\bibitem[\protect\citeauthoryear{{Gonz{\'a}lez Hern{\'a}ndez}, {Rebolo},
  {Israelian}, {Filippenko}, {Chornock}, {Tominaga}, {Umeda}  \&
  {Nomoto}}{{Gonz{\'a}lez Hern{\'a}ndez} et~al.}{2008}]{Hernandez2008}
{Gonz{\'a}lez Hern{\'a}ndez} J.~I.,  {Rebolo} R.,  {Israelian} G.,
  {Filippenko} A.~V.,  {Chornock} R.,  {Tominaga} N.,  {Umeda} H.,   {Nomoto}
  K.,  2008, \mn@doi [\apj] {10.1086/586888}, \href
  {http://adsabs.harvard.edu/abs/2008ApJ...679..732G} {679, 732}

\bibitem[\protect\citeauthoryear{{Goodman} \& {Hut}}{{Goodman} \&
  {Hut}}{1993}]{Goodman1993}
{Goodman} J.,  {Hut} P.,  1993, \mn@doi [\apj] {10.1086/172200}, \href
  {http://adsabs.harvard.edu/abs/1993ApJ...403..271G} {403, 271}

\bibitem[\protect\citeauthoryear{{Gourgoulhon}}{{Gourgoulhon}}{1991}]{Gourgoulhon1991}
{Gourgoulhon} E.,  1991, \aap, \href
  {http://adsabs.harvard.edu/abs/1991A%26A...252..651G} {252, 651}

\bibitem[\protect\citeauthoryear{{Gourgoulhon} \& {Haensel}}{{Gourgoulhon} \&
  {Haensel}}{1993}]{Gourgoulhon1993}
{Gourgoulhon} E.,  {Haensel} P.,  1993, \aap, \href
  {http://adsabs.harvard.edu/abs/1993A%26A...271..187G} {271, 187}

\bibitem[\protect\citeauthoryear{{Greisen}}{{Greisen}}{2003}]{Greisen2003}
{Greisen} E.~W.,  2003, in {Heck} A.,  ed.,  Astrophysics and Space Science
  Library Vol. 285, Information Handling in Astronomy - Historical Vistas.
  p.~109, \mn@doi{10.1007/0-306-48080-8_7}

\bibitem[\protect\citeauthoryear{{Grimm}, {Gilfanov}  \& {Sunyaev}}{{Grimm}
  et~al.}{2002}]{Grimm2002}
{Grimm} H.-J.,  {Gilfanov} M.,   {Sunyaev} R.,  2002, \mn@doi [\aap]
  {10.1051/0004-6361:20020826}, \href
  {http://adsabs.harvard.edu/abs/2002A%26A...391..923G} {391, 923}

\bibitem[\protect\citeauthoryear{{Gualandris}, {Colpi}, {Portegies Zwart}  \&
  {Possenti}}{{Gualandris} et~al.}{2005}]{Gualandris2005}
{Gualandris} A.,  {Colpi} M.,  {Portegies Zwart} S.,   {Possenti} A.,  2005,
  \mn@doi [\apj] {10.1086/426126}, \href
  {http://adsabs.harvard.edu/abs/2005ApJ...618..845G} {618, 845}

\bibitem[\protect\citeauthoryear{{Heggie} \& {Giersz}}{{Heggie} \&
  {Giersz}}{2014}]{Heggie2014}
{Heggie} D.~C.,  {Giersz} M.,  2014, \mn@doi [\mnras] {10.1093/mnras/stu102},
  \href {http://adsabs.harvard.edu/abs/2014MNRAS.439.2459H} {439, 2459}

\bibitem[\protect\citeauthoryear{{Heida}, {Jonker}, {Torres}  \&
  {Chiavassa}}{{Heida} et~al.}{2017}]{Heida2017}
{Heida} M.,  {Jonker} P.~G.,  {Torres} M.~A.~P.,   {Chiavassa} A.,  2017,
  \mn@doi [\apj] {10.3847/1538-4357/aa85df}, \href
  {http://adsabs.harvard.edu/abs/2017ApJ...846..132H} {846, 132}

\bibitem[\protect\citeauthoryear{{Heise}}{{Heise}}{1999}]{Heise1999}
{Heise} J.,  1999, \mn@doi [Nuclear Physics B Proceedings Supplements]
  {10.1016/S0920-5632(98)00207-2}, \href
  {https://ui.adsabs.harvard.edu/abs/1999NuPhS..69..186H} {69, 186}

\bibitem[\protect\citeauthoryear{{Hjellming} \& {Rupen}}{{Hjellming} \&
  {Rupen}}{1995}]{Hjellming1995}
{Hjellming} R.~M.,  {Rupen} M.~P.,  1995, \mn@doi [\nat] {10.1038/375464a0},
  \href {http://adsabs.harvard.edu/abs/1995Natur.375..464H} {375, 464}

\bibitem[\protect\citeauthoryear{{Hobbs}, {Lorimer}, {Lyne}  \&
  {Kramer}}{{Hobbs} et~al.}{2005}]{Hobbs2005}
{Hobbs} G.,  {Lorimer} D.~R.,  {Lyne} A.~G.,   {Kramer} M.,  2005, \mn@doi
  [\mnras] {10.1111/j.1365-2966.2005.09087.x}, \href
  {http://adsabs.harvard.edu/abs/2005MNRAS.360..974H} {360, 974}

\bibitem[\protect\citeauthoryear{{Hoffman} \& {Gelman}}{{Hoffman} \&
  {Gelman}}{2011}]{Hoffman2011}
{Hoffman} M.~D.,  {Gelman} A.,  2011, arXiv e-prints, \href
  {http://adsabs.harvard.edu/abs/2011arXiv1111.4246H} {}

\bibitem[\protect\citeauthoryear{{Hogg}, {Bovy}  \& {Lang}}{{Hogg}
  et~al.}{2010}]{Hogg2010}
{Hogg} D.~W.,  {Bovy} J.,   {Lang} D.,  2010, arXiv e-prints, \href
  {http://adsabs.harvard.edu/abs/2010arXiv1008.4686H} {}

\bibitem[\protect\citeauthoryear{{Homan} \& {Belloni}}{{Homan} \&
  {Belloni}}{2005}]{Homan2005}
{Homan} J.,  {Belloni} T.,  2005, \mn@doi [\apss] {10.1007/s10509-005-1197-4},
  \href {http://adsabs.harvard.edu/abs/2005Ap%26SS.300..107H} {300, 107}

\bibitem[\protect\citeauthoryear{{Hynes}}{{Hynes}}{2005}]{Hynes2005}
{Hynes} R.~I.,  2005, \mn@doi [\apj] {10.1086/428445}, \href
  {http://adsabs.harvard.edu/abs/2005ApJ...623.1026H} {623, 1026}

\bibitem[\protect\citeauthoryear{{Ingram}, {van der Klis}, {Middleton}, {Done},
  {Altamirano}, {Heil}, {Uttley}  \& {Axelsson}}{{Ingram}
  et~al.}{2016}]{Ingram2016}
{Ingram} A.,  {van der Klis} M.,  {Middleton} M.,  {Done} C.,  {Altamirano} D.,
   {Heil} L.,  {Uttley} P.,   {Axelsson} M.,  2016, \mn@doi [\mnras]
  {10.1093/mnras/stw1245}, \href
  {http://adsabs.harvard.edu/abs/2016MNRAS.461.1967I} {461, 1967}

\bibitem[\protect\citeauthoryear{{Janka}}{{Janka}}{2013}]{Janka2013}
{Janka} H.-T.,  2013, \mn@doi [\mnras] {10.1093/mnras/stt1106}, \href
  {http://adsabs.harvard.edu/abs/2013MNRAS.434.1355J} {434, 1355}

\bibitem[\protect\citeauthoryear{{Janka}}{{Janka}}{2017}]{Janka2017}
{Janka} H.-T.,  2017, \mn@doi [\apj] {10.3847/1538-4357/aa618e}, \href
  {http://adsabs.harvard.edu/abs/2017ApJ...837...84J} {837, 84}

\bibitem[\protect\citeauthoryear{{Johnson} \& {Soderblom}}{{Johnson} \&
  {Soderblom}}{1987}]{Johnson1987}
{Johnson} D.~R.~H.,  {Soderblom} D.~R.,  1987, \mn@doi [\aj] {10.1086/114370},
  \href {http://adsabs.harvard.edu/abs/1987AJ.....93..864J} {93, 864}

\bibitem[\protect\citeauthoryear{{Jonker} \& {Nelemans}}{{Jonker} \&
  {Nelemans}}{2004}]{Jonker2004}
{Jonker} P.~G.,  {Nelemans} G.,  2004, \mn@doi [\mnras]
  {10.1111/j.1365-2966.2004.08193.x}, \href
  {http://adsabs.harvard.edu/abs/2004MNRAS.354..355J} {354, 355}

\bibitem[\protect\citeauthoryear{{Kalemci}, {Din{\c c}er}, {Tomsick}, {Buxton},
  {Bailyn}  \& {Chun}}{{Kalemci} et~al.}{2013}]{Kalemci2013}
{Kalemci} E.,  {Din{\c c}er} T.,  {Tomsick} J.~A.,  {Buxton} M.~M.,  {Bailyn}
  C.~D.,   {Chun} Y.~Y.,  2013, \mn@doi [\apj] {10.1088/0004-637X/779/2/95},
  \href {http://adsabs.harvard.edu/abs/2013ApJ...779...95K} {779, 95}

\bibitem[\protect\citeauthoryear{{Karitskaya} \& {Goranskij}}{{Karitskaya} \&
  {Goranskij}}{1995}]{Karitskaya1995}
{Karitskaya} E.~A.,  {Goranskij} V.~P.,  1995, \mn@doi [\ssr]
  {10.1007/BF00751439}, \href
  {http://adsabs.harvard.edu/abs/1995SSRv...74..489K} {74, 489}

\bibitem[\protect\citeauthoryear{{Kawamuro} et~al.,}{{Kawamuro}
  et~al.}{2018}]{Kawamuro2018}
{Kawamuro} T.,  et~al., 2018, The Astronomer's Telegram, \href
  {http://adsabs.harvard.edu/abs/2018ATel11399....1K} {11399}

\bibitem[\protect\citeauthoryear{{Kelley}, {Ramirez-Ruiz}, {Zemp}, {Diemand}
  \& {Mandel}}{{Kelley} et~al.}{2010}]{Kelley2010}
{Kelley} L.~Z.,  {Ramirez-Ruiz} E.,  {Zemp} M.,  {Diemand} J.,   {Mandel} I.,
  2010, \mn@doi [\apjl] {10.1088/2041-8205/725/1/L91}, \href
  {http://adsabs.harvard.edu/abs/2010ApJ...725L..91K} {725, L91}

\bibitem[\protect\citeauthoryear{{Khargharia}, {Froning}  \&
  {Robinson}}{{Khargharia} et~al.}{2010}]{Khargharia2010}
{Khargharia} J.,  {Froning} C.~S.,   {Robinson} E.~L.,  2010, \mn@doi [\apj]
  {10.1088/0004-637X/716/2/1105}, \href
  {http://adsabs.harvard.edu/abs/2010ApJ...716.1105K} {716, 1105}

\bibitem[\protect\citeauthoryear{{Khargharia}, {Froning}, {Robinson}  \&
  {Gelino}}{{Khargharia} et~al.}{2013}]{Khargharia2013}
{Khargharia} J.,  {Froning} C.~S.,  {Robinson} E.~L.,   {Gelino} D.~M.,  2013,
  \mn@doi [\aj] {10.1088/0004-6256/145/1/21}, \href
  {http://adsabs.harvard.edu/abs/2013AJ....145...21K} {145, 21}

\bibitem[\protect\citeauthoryear{{Khokhlov}, {H{\"o}flich}, {Oran}, {Wheeler},
  {Wang}  \& {Chtchelkanova}}{{Khokhlov} et~al.}{1999}]{Khokhlov1999}
{Khokhlov} A.~M.,  {H{\"o}flich} P.~A.,  {Oran} E.~S.,  {Wheeler} J.~C.,
  {Wang} L.,   {Chtchelkanova} A.~Y.,  1999, \mn@doi [\apjl] {10.1086/312305},
  \href {http://adsabs.harvard.edu/abs/1999ApJ...524L.107K} {524, L107}

\bibitem[\protect\citeauthoryear{{Kirsten}, {Vlemmings}, {Freire}, {Kramer},
  {Rottmann}  \& {Campbell}}{{Kirsten} et~al.}{2014}]{Kirsten2014}
{Kirsten} F.,  {Vlemmings} W.,  {Freire} P.,  {Kramer} M.,  {Rottmann} H.,
  {Campbell} R.~M.,  2014, \mn@doi [\aap] {10.1051/0004-6361/201323239}, \href
  {http://adsabs.harvard.edu/abs/2014A%26A...565A..43K} {565, A43}

\bibitem[\protect\citeauthoryear{{Kreidberg}, {Bailyn}, {Farr}  \&
  {Kalogera}}{{Kreidberg} et~al.}{2012}]{Kreidberg2012}
{Kreidberg} L.,  {Bailyn} C.~D.,  {Farr} W.~M.,   {Kalogera} V.,  2012, \mn@doi
  [\apj] {10.1088/0004-637X/757/1/36}, \href
  {http://adsabs.harvard.edu/abs/2012ApJ...757...36K} {757, 36}

\bibitem[\protect\citeauthoryear{{Kulkarni}, {Hut}  \& {McMillan}}{{Kulkarni}
  et~al.}{1993}]{Kulkarni1993}
{Kulkarni} S.~R.,  {Hut} P.,   {McMillan} S.,  1993, \mn@doi [\nat]
  {10.1038/364421a0}, \href {http://adsabs.harvard.edu/abs/1993Natur.364..421K}
  {364, 421}

\bibitem[\protect\citeauthoryear{{Lindegren} et~al.,}{{Lindegren}
  et~al.}{2016}]{Lindegren2016}
{Lindegren} L.,  et~al., 2016, \mn@doi [\aap] {10.1051/0004-6361/201628714},
  \href {http://adsabs.harvard.edu/abs/2016A%26A...595A...4L} {595, A4}

\bibitem[\protect\citeauthoryear{{Luri} et~al.,}{{Luri}
  et~al.}{2018}]{Luri2018}
{Luri} X.,  et~al., 2018, \mn@doi [\aap] {10.1051/0004-6361/201832964}, \href
  {http://adsabs.harvard.edu/abs/2018A%26A...616A...9L} {616, A9}

\bibitem[\protect\citeauthoryear{{Lyne} \& {Lorimer}}{{Lyne} \&
  {Lorimer}}{1994}]{Lyne1994}
{Lyne} A.~G.,  {Lorimer} D.~R.,  1994, \mn@doi [\nat] {10.1038/369127a0}, \href
  {http://adsabs.harvard.edu/abs/1994Natur.369..127L} {369, 127}

\bibitem[\protect\citeauthoryear{{MacDonald} et~al.,}{{MacDonald}
  et~al.}{2014}]{MacDonald2014}
{MacDonald} R.~K.~D.,  et~al., 2014, \mn@doi [\apj]
  {10.1088/0004-637X/784/1/2}, \href
  {http://adsabs.harvard.edu/abs/2014ApJ...784....2M} {784, 2}

\bibitem[\protect\citeauthoryear{{Maccarone}}{{Maccarone}}{2003}]{Maccarone2003}
{Maccarone} T.~J.,  2003, \mn@doi [\aap] {10.1051/0004-6361:20031146}, \href
  {http://adsabs.harvard.edu/abs/2003A%26A...409..697M} {409, 697}

\bibitem[\protect\citeauthoryear{{Mandel}}{{Mandel}}{2010}]{Mandel2010}
{Mandel} I.,  2010, \mn@doi [\prd] {10.1103/PhysRevD.81.084029}, \href
  {http://adsabs.harvard.edu/abs/2010PhRvD..81h4029M} {81, 084029}

\bibitem[\protect\citeauthoryear{{Mandel}}{{Mandel}}{2016}]{Mandel2016}
{Mandel} I.,  2016, \mn@doi [\mnras] {10.1093/mnras/stv2733}, \href
  {http://adsabs.harvard.edu/abs/2016MNRAS.456..578M} {456, 578}

\bibitem[\protect\citeauthoryear{{Markert}, {Canizares}, {Clark}, {Lewin},
  {Schnopper}  \& {Sprott}}{{Markert} et~al.}{1973}]{Markert1973}
{Markert} T.~H.,  {Canizares} C.~R.,  {Clark} G.~W.,  {Lewin} W.~H.~G.,
  {Schnopper} H.~W.,   {Sprott} G.~F.,  1973, \mn@doi [\apjl] {10.1086/181290},
  \href {http://adsabs.harvard.edu/abs/1973ApJ...184L..67M} {184, L67}

\bibitem[\protect\citeauthoryear{{Martin}, {Reis}  \& {Pringle}}{{Martin}
  et~al.}{2008}]{Martin2008}
{Martin} R.~G.,  {Reis} R.~C.,   {Pringle} J.~E.,  2008, \mn@doi [\mnras]
  {10.1111/j.1745-3933.2008.00545.x}, \href
  {http://adsabs.harvard.edu/abs/2008MNRAS.391L..15M} {391, L15}

\bibitem[\protect\citeauthoryear{{Masetti}, {Bianchini}, {Bonibaker}, {della
  Valle}  \& {Vio}}{{Masetti} et~al.}{1996}]{Masetti1996}
{Masetti} N.,  {Bianchini} A.,  {Bonibaker} J.,  {della Valle} M.,   {Vio} R.,
  1996, \aap, \href {http://adsabs.harvard.edu/abs/1996A%26A...314..123M} {314,
  123}

\bibitem[\protect\citeauthoryear{{McClintock}, {Narayan}  \&
  {Steiner}}{{McClintock} et~al.}{2014}]{McClintock2014}
{McClintock} J.~E.,  {Narayan} R.,   {Steiner} J.~F.,  2014, \mn@doi [\ssr]
  {10.1007/s11214-013-0003-9}, \href
  {http://adsabs.harvard.edu/abs/2014SSRv..183..295M} {183, 295}

\bibitem[\protect\citeauthoryear{{Mignard}}{{Mignard}}{2000}]{Mignard2000}
{Mignard} F.,  2000, \aap, \href
  {http://adsabs.harvard.edu/abs/2000A%26A...354..522M} {354, 522}

\bibitem[\protect\citeauthoryear{{Miller-Jones}, {Jonker}, {Nelemans},
  {Portegies Zwart}, {Dhawan}, {Brisken}, {Gallo}  \& {Rupen}}{{Miller-Jones}
  et~al.}{2009a}]{Miller-Jones2009a}
{Miller-Jones} J.~C.~A.,  {Jonker} P.~G.,  {Nelemans} G.,  {Portegies Zwart}
  S.,  {Dhawan} V.,  {Brisken} W.,  {Gallo} E.,   {Rupen} M.~P.,  2009a,
  \mn@doi [\mnras] {10.1111/j.1365-2966.2009.14404.x}, \href
  {http://adsabs.harvard.edu/abs/2009MNRAS.394.1440M} {394, 1440}

\bibitem[\protect\citeauthoryear{{Miller-Jones}, {Jonker}, {Dhawan}, {Brisken},
  {Rupen}, {Nelemans}  \& {Gallo}}{{Miller-Jones}
  et~al.}{2009b}]{Miller-Jones2009b}
{Miller-Jones} J.~C.~A.,  {Jonker} P.~G.,  {Dhawan} V.,  {Brisken} W.,  {Rupen}
  M.~P.,  {Nelemans} G.,   {Gallo} E.,  2009b, \mn@doi [\apjl]
  {10.1088/0004-637X/706/2/L230}, \href
  {http://adsabs.harvard.edu/abs/2009ApJ...706L.230M} {706, L230}

\bibitem[\protect\citeauthoryear{{Miller-Jones} et~al.,}{{Miller-Jones}
  et~al.}{2015}]{Miller-Jones2015}
{Miller-Jones} J.~C.~A.,  et~al., 2015, \mn@doi [\mnras]
  {10.1093/mnras/stv1869}, \href
  {http://adsabs.harvard.edu/abs/2015MNRAS.453.3918M} {453, 3918}

\bibitem[\protect\citeauthoryear{{Miller-Jones} et~al.,}{{Miller-Jones}
  et~al.}{2018}]{Miller-Jones2018}
{Miller-Jones} J.~C.~A.,  et~al., 2018, \mn@doi [\mnras]
  {10.1093/mnras/sty1775}, \href
  {http://adsabs.harvard.edu/abs/2018MNRAS.479.4849M} {479, 4849}

\bibitem[\protect\citeauthoryear{{Mirabel}}{{Mirabel}}{2017}]{Mirabel2017}
{Mirabel} F.,  2017, \mn@doi [\nar] {10.1016/j.newar.2017.04.002}, \href
  {http://adsabs.harvard.edu/abs/2017NewAR..78....1M} {78, 1}

\bibitem[\protect\citeauthoryear{{Mirabel} \& {Rodrigues}}{{Mirabel} \&
  {Rodrigues}}{2003}]{Mirabel2003}
{Mirabel} I.~F.,  {Rodrigues} I.,  2003, \mn@doi [Science]
  {10.1126/science.1083451}, \href
  {http://adsabs.harvard.edu/abs/2003Sci...300.1119M} {300, 1119}

\bibitem[\protect\citeauthoryear{{Mirabel}, {Dhawan}, {Mignani}, {Rodrigues}
  \& {Guglielmetti}}{{Mirabel} et~al.}{2001}]{Mirabel2001}
{Mirabel} I.~F.,  {Dhawan} V.,  {Mignani} R.~P.,  {Rodrigues} I.,
  {Guglielmetti} F.,  2001, \mn@doi [\nat] {10.1038/35093060}, \href
  {http://adsabs.harvard.edu/abs/2001Natur.413..139M} {413, 139}

\bibitem[\protect\citeauthoryear{{Mirabel}, {Mignani}, {Rodrigues}, {Combi},
  {Rodr{\'{\i}}guez}  \& {Guglielmetti}}{{Mirabel} et~al.}{2002}]{Mirabel2002}
{Mirabel} I.~F.,  {Mignani} R.,  {Rodrigues} I.,  {Combi} J.~A.,
  {Rodr{\'{\i}}guez} L.~F.,   {Guglielmetti} F.,  2002, \mn@doi [\aap]
  {10.1051/0004-6361:20021440}, \href
  {http://adsabs.harvard.edu/abs/2002A%26A...395..595M} {395, 595}

\bibitem[\protect\citeauthoryear{{Neal}}{{Neal}}{2012}]{Neal2012}
{Neal} R.~M.,  2012, arXiv e-prints, \href
  {http://adsabs.harvard.edu/abs/2012arXiv1206.1901N} {}

\bibitem[\protect\citeauthoryear{{Negoro} et~al.,}{{Negoro}
  et~al.}{2016}]{Negoro2016}
{Negoro} H.,  et~al., 2016, The Astronomer's Telegram, \href
  {http://adsabs.harvard.edu/abs/2016ATel.9876....1N} {9876}

\bibitem[\protect\citeauthoryear{{Nelemans}, {Tauris}  \& {van den
  Heuvel}}{{Nelemans} et~al.}{1999}]{Nelemans1999}
{Nelemans} G.,  {Tauris} T.~M.,   {van den Heuvel} E.~P.~J.,  1999, \aap, \href
  {http://adsabs.harvard.edu/abs/1999A%26A...352L..87N} {352, L87}

\bibitem[\protect\citeauthoryear{{O'Shaughnessy}, {Gerosa}  \&
  {Wysocki}}{{O'Shaughnessy} et~al.}{2017}]{OShaughnessy2017}
{O'Shaughnessy} R.,  {Gerosa} D.,   {Wysocki} D.,  2017, \mn@doi [Physical
  Review Letters] {10.1103/PhysRevLett.119.011101}, \href
  {http://adsabs.harvard.edu/abs/2017PhRvL.119a1101O} {119, 011101}

\bibitem[\protect\citeauthoryear{{Orosz}, {Bailyn}, {McClintock}  \&
  {Remillard}}{{Orosz} et~al.}{1996}]{Orosz1996}
{Orosz} J.~A.,  {Bailyn} C.~D.,  {McClintock} J.~E.,   {Remillard} R.~A.,
  1996, \mn@doi [\apj] {10.1086/177698}, \href
  {http://adsabs.harvard.edu/abs/1996ApJ...468..380O} {468, 380}

\bibitem[\protect\citeauthoryear{{Orosz}, {Jain}, {Bailyn}, {McClintock}  \&
  {Remillard}}{{Orosz} et~al.}{1998}]{Orosz1998}
{Orosz} J.~A.,  {Jain} R.~K.,  {Bailyn} C.~D.,  {McClintock} J.~E.,
  {Remillard} R.~A.,  1998, \mn@doi [\apj] {10.1086/305620}, \href
  {http://adsabs.harvard.edu/abs/1998ApJ...499..375O} {499, 375}

\bibitem[\protect\citeauthoryear{{Orosz} et~al.,}{{Orosz}
  et~al.}{2001}]{Orosz2001}
{Orosz} J.~A.,  et~al., 2001, \mn@doi [\apj] {10.1086/321442}, \href
  {http://adsabs.harvard.edu/abs/2001ApJ...555..489O} {555, 489}

\bibitem[\protect\citeauthoryear{{Orosz}, {McClintock}, {Aufdenberg},
  {Remillard}, {Reid}, {Narayan}  \& {Gou}}{{Orosz} et~al.}{2011}]{Orosz2011}
{Orosz} J.~A.,  {McClintock} J.~E.,  {Aufdenberg} J.~P.,  {Remillard} R.~A.,
  {Reid} M.~J.,  {Narayan} R.,   {Gou} L.,  2011, \mn@doi [\apj]
  {10.1088/0004-637X/742/2/84}, \href
  {http://adsabs.harvard.edu/abs/2011ApJ...742...84O} {742, 84}

\bibitem[\protect\citeauthoryear{{Palmer}, {Barthelmey}, {Cummings}, {Gehrels},
  {Krimm}, {Markwardt}, {Sakamoto}  \& {Tueller}}{{Palmer}
  et~al.}{2005}]{Palmer2005}
{Palmer} D.~M.,  {Barthelmey} S.~D.,  {Cummings} J.~R.,  {Gehrels} N.,  {Krimm}
  H.~A.,  {Markwardt} C.~B.,  {Sakamoto} T.,   {Tueller} J.,  2005, The
  Astronomer's Telegram, \href
  {http://adsabs.harvard.edu/abs/2005ATel..546....1P} {546}

\bibitem[\protect\citeauthoryear{{Podsiadlowski}, {Langer}, {Poelarends},
  {Rappaport}, {Heger}  \& {Pfahl}}{{Podsiadlowski}
  et~al.}{2004}]{Podsiadlowski2004}
{Podsiadlowski} P.,  {Langer} N.,  {Poelarends} A.~J.~T.,  {Rappaport} S.,
  {Heger} A.,   {Pfahl} E.,  2004, \mn@doi [\apj] {10.1086/421713}, \href
  {http://adsabs.harvard.edu/abs/2004ApJ...612.1044P} {612, 1044}

\bibitem[\protect\citeauthoryear{{Postnov} \& {Yungelson}}{{Postnov} \&
  {Yungelson}}{2014}]{Postnov2014}
{Postnov} K.~A.,  {Yungelson} L.~R.,  2014, \mn@doi [Living Reviews in
  Relativity] {10.12942/lrr-2014-3}, \href
  {http://adsabs.harvard.edu/abs/2014LRR....17....3P} {17, 3}

\bibitem[\protect\citeauthoryear{{Pradel}, {Charlot}  \& {Lestrade}}{{Pradel}
  et~al.}{2006}]{Pradel2006}
{Pradel} N.,  {Charlot} P.,   {Lestrade} J.-F.,  2006, \mn@doi [\aap]
  {10.1051/0004-6361:20053021}, \href
  {http://adsabs.harvard.edu/abs/2006A%26A...452.1099P} {452, 1099}

\bibitem[\protect\citeauthoryear{{Reid}, {McClintock}, {Narayan}, {Gou},
  {Remillard}  \& {Orosz}}{{Reid} et~al.}{2011}]{Reid2011}
{Reid} M.~J.,  {McClintock} J.~E.,  {Narayan} R.,  {Gou} L.,  {Remillard}
  R.~A.,   {Orosz} J.~A.,  2011, \mn@doi [\apj] {10.1088/0004-637X/742/2/83},
  \href {http://adsabs.harvard.edu/abs/2011ApJ...742...83R} {742, 83}

\bibitem[\protect\citeauthoryear{{Reid}, {McClintock}, {Steiner}, {Steeghs},
  {Remillard}, {Dhawan}  \& {Narayan}}{{Reid} et~al.}{2014}]{Reid2014}
{Reid} M.~J.,  {McClintock} J.~E.,  {Steiner} J.~F.,  {Steeghs} D.,
  {Remillard} R.~A.,  {Dhawan} V.,   {Narayan} R.,  2014, \mn@doi [\apj]
  {10.1088/0004-637X/796/1/2}, \href
  {http://adsabs.harvard.edu/abs/2014ApJ...796....2R} {796, 2}

\bibitem[\protect\citeauthoryear{{Repetto}, {Davies}  \&
  {Sigurdsson}}{{Repetto} et~al.}{2012}]{Repetto2012}
{Repetto} S.,  {Davies} M.~B.,   {Sigurdsson} S.,  2012, \mn@doi [\mnras]
  {10.1111/j.1365-2966.2012.21549.x}, \href
  {http://adsabs.harvard.edu/abs/2012MNRAS.425.2799R} {425, 2799}

\bibitem[\protect\citeauthoryear{{Repetto}, {Igoshev}  \& {Nelemans}}{{Repetto}
  et~al.}{2017}]{Repetto2017}
{Repetto} S.,  {Igoshev} A.~P.,   {Nelemans} G.,  2017, \mn@doi [\mnras]
  {10.1093/mnras/stx027}, \href
  {http://adsabs.harvard.edu/abs/2017MNRAS.467..298R} {467, 298}

\bibitem[\protect\citeauthoryear{{Reynolds}, {Fraser}  \& {Gilmore}}{{Reynolds}
  et~al.}{2015}]{Reynolds2015}
{Reynolds} T.~M.,  {Fraser} M.,   {Gilmore} G.,  2015, \mn@doi [\mnras]
  {10.1093/mnras/stv1809}, \href
  {http://adsabs.harvard.edu/abs/2015MNRAS.453.2885R} {453, 2885}

\bibitem[\protect\citeauthoryear{{Rodriguez}, {Haster}, {Chatterjee},
  {Kalogera}  \& {Rasio}}{{Rodriguez} et~al.}{2016}]{Rodriguez2016}
{Rodriguez} C.~L.,  {Haster} C.-J.,  {Chatterjee} S.,  {Kalogera} V.,   {Rasio}
  F.~A.,  2016, \mn@doi [\apjl] {10.3847/2041-8205/824/1/L8}, \href
  {http://adsabs.harvard.edu/abs/2016ApJ...824L...8R} {824, L8}

\bibitem[\protect\citeauthoryear{{Russell}, {Soria}, {Motch}, {Pakull},
  {Torres}, {Curran}, {Jonker}  \& {Miller-Jones}}{{Russell}
  et~al.}{2014}]{Russell2014}
{Russell} T.~D.,  {Soria} R.,  {Motch} C.,  {Pakull} M.~W.,  {Torres} M.~A.~P.,
   {Curran} P.~A.,  {Jonker} P.~G.,   {Miller-Jones} J.~C.~A.,  2014, \mn@doi
  [\mnras] {10.1093/mnras/stt2480}, \href
  {http://adsabs.harvard.edu/abs/2014MNRAS.439.1381R} {439, 1381}

\bibitem[\protect\citeauthoryear{{Russell} et~al.,}{{Russell}
  et~al.}{2015}]{Russell2015}
{Russell} T.~D.,  et~al., 2015, \mn@doi [\mnras] {10.1093/mnras/stv723}, \href
  {http://adsabs.harvard.edu/abs/2015MNRAS.450.1745R} {450, 1745}

\bibitem[\protect\citeauthoryear{{Sagert} \& {Schaffner-Bielich}}{{Sagert} \&
  {Schaffner-Bielich}}{2008}]{Sagert2008}
{Sagert} I.,  {Schaffner-Bielich} J.,  2008, \mn@doi [\aap]
  {10.1051/0004-6361:20078530}, \href
  {http://adsabs.harvard.edu/abs/2008A%26A...489..281S} {489, 281}

\bibitem[\protect\citeauthoryear{{Salvatier}, {Wiecki{\^a}}  \&
  {Fonnesbeck}}{{Salvatier} et~al.}{2016}]{Salvatier2016}
{Salvatier} J.,  {Wiecki{\^a}} T.~V.,   {Fonnesbeck} C.,  2016, {PyMC3: Python
  probabilistic programming framework}, Astrophysics Source Code Library
  (\mn@eprint {ascl} {1610.016})

\bibitem[\protect\citeauthoryear{{Shahbaz}}{{Shahbaz}}{2003}]{Shahbaz2003}
{Shahbaz} T.,  2003, \mn@doi [\mnras] {10.1046/j.1365-8711.2003.06258.x}, \href
  {http://adsabs.harvard.edu/abs/2003MNRAS.339.1031S} {339, 1031}

\bibitem[\protect\citeauthoryear{{Shahbaz}, {van der Hooft}, {Casares},
  {Charles}  \& {van Paradijs}}{{Shahbaz} et~al.}{1999}]{Shahbaz1999}
{Shahbaz} T.,  {van der Hooft} F.,  {Casares} J.,  {Charles} P.~A.,   {van
  Paradijs} J.,  1999, \mn@doi [\mnras] {10.1046/j.1365-8711.1999.02481.x},
  \href {http://adsabs.harvard.edu/abs/1999MNRAS.306...89S} {306, 89}

\bibitem[\protect\citeauthoryear{{Shahbaz}, {Fender}  \& {Charles}}{{Shahbaz}
  et~al.}{2001}]{Shahbaz2001}
{Shahbaz} T.,  {Fender} R.,   {Charles} P.~A.,  2001, \mn@doi [\aap]
  {10.1051/0004-6361:20011042}, \href
  {http://adsabs.harvard.edu/abs/2001A%26A...376L..17S} {376, L17}

\bibitem[\protect\citeauthoryear{{Shapiro} et~al.,}{{Shapiro}
  et~al.}{1979}]{Shapiro1979}
{Shapiro} I.~I.,  et~al., 1979, \mn@doi [\aj] {10.1086/112565}, \href
  {http://adsabs.harvard.edu/abs/1979AJ.....84.1459S} {84, 1459}

\bibitem[\protect\citeauthoryear{{Shaw}, {Charles}, {Casares}  \&
  {Hern{\'a}ndez Santisteban}}{{Shaw} et~al.}{2016}]{Shaw2016}
{Shaw} A.~W.,  {Charles} P.~A.,  {Casares} J.,   {Hern{\'a}ndez Santisteban}
  J.~V.,  2016, \mn@doi [\mnras] {10.1093/mnras/stw2092}, \href
  {http://adsabs.harvard.edu/abs/2016MNRAS.463.1314S} {463, 1314}

\bibitem[\protect\citeauthoryear{{Sigurdsson} \& {Hernquist}}{{Sigurdsson} \&
  {Hernquist}}{1993}]{Sigurdsson1993a}
{Sigurdsson} S.,  {Hernquist} L.,  1993, \mn@doi [\nat] {10.1038/364423a0},
  \href {http://adsabs.harvard.edu/abs/1993Natur.364..423S} {364, 423}

\bibitem[\protect\citeauthoryear{{Sigurdsson} \& {Phinney}}{{Sigurdsson} \&
  {Phinney}}{1993}]{Sigurdsson1993b}
{Sigurdsson} S.,  {Phinney} E.~S.,  1993, \mn@doi [\apj] {10.1086/173190},
  \href {http://adsabs.harvard.edu/abs/1993ApJ...415..631S} {415, 631}

\bibitem[\protect\citeauthoryear{{Sippel} \& {Hurley}}{{Sippel} \&
  {Hurley}}{2013}]{Sippel2013}
{Sippel} A.~C.,  {Hurley} J.~R.,  2013, \mn@doi [\mnras]
  {10.1093/mnrasl/sls044}, \href
  {http://adsabs.harvard.edu/abs/2013MNRAS.430L..30S} {430, L30}

\bibitem[\protect\citeauthoryear{{Sreehari}, {Iyer}, {Radhika}, {Nandi}  \&
  {Mandal}}{{Sreehari} et~al.}{2018}]{Sreehari2018}
{Sreehari} H.,  {Iyer} N.,  {Radhika} D.,  {Nandi} A.,   {Mandal} S.,  2018,
  preprint, \href {http://adsabs.harvard.edu/abs/2018arXiv181104341S} {}
  (\mn@eprint {arXiv} {1811.04341})

\bibitem[\protect\citeauthoryear{{Steeghs}, {McClintock}, {Parsons}, {Reid},
  {Littlefair}  \& {Dhillon}}{{Steeghs} et~al.}{2013}]{Steeghs2013}
{Steeghs} D.,  {McClintock} J.~E.,  {Parsons} S.~G.,  {Reid} M.~J.,
  {Littlefair} S.,   {Dhillon} V.~S.,  2013, \mn@doi [\apj]
  {10.1088/0004-637X/768/2/185}, \href
  {http://adsabs.harvard.edu/abs/2013ApJ...768..185S} {768, 185}

\bibitem[\protect\citeauthoryear{{Stella} \& {Vietri}}{{Stella} \&
  {Vietri}}{1998}]{Stella1998}
{Stella} L.,  {Vietri} M.,  1998, \mn@doi [\apjl] {10.1086/311075}, \href
  {http://adsabs.harvard.edu/abs/1998ApJ...492L..59S} {492, L59}

\bibitem[\protect\citeauthoryear{{Strader}, {Chomiuk}, {Maccarone},
  {Miller-Jones}  \& {Seth}}{{Strader} et~al.}{2012}]{Strader2012}
{Strader} J.,  {Chomiuk} L.,  {Maccarone} T.~J.,  {Miller-Jones} J.~C.~A.,
  {Seth} A.~C.,  2012, \mn@doi [\nat] {10.1038/nature11490}, \href
  {http://adsabs.harvard.edu/abs/2012Natur.490...71S} {490, 71}

\bibitem[\protect\citeauthoryear{{Sukhbold}, {Ertl}, {Woosley}, {Brown}  \&
  {Janka}}{{Sukhbold} et~al.}{2016}]{Sukhbold2016}
{Sukhbold} T.,  {Ertl} T.,  {Woosley} S.~E.,  {Brown} J.~M.,   {Janka} H.-T.,
  2016, \mn@doi [\apj] {10.3847/0004-637X/821/1/38}, \href
  {http://adsabs.harvard.edu/abs/2016ApJ...821...38S} {821, 38}

\bibitem[\protect\citeauthoryear{{Tauris}, {Langer}  \&
  {Podsiadlowski}}{{Tauris} et~al.}{2015}]{Tauris2015}
{Tauris} T.~M.,  {Langer} N.,   {Podsiadlowski} P.,  2015, \mn@doi [\mnras]
  {10.1093/mnras/stv990}, \href
  {http://adsabs.harvard.edu/abs/2015MNRAS.451.2123T} {451, 2123}

\bibitem[\protect\citeauthoryear{{Tetarenko}, {Sivakoff}, {Heinke}  \&
  {Gladstone}}{{Tetarenko} et~al.}{2016a}]{Tetarenko2016a}
{Tetarenko} B.~E.,  {Sivakoff} G.~R.,  {Heinke} C.~O.,   {Gladstone} J.~C.,
  2016a, \mn@doi [\apjs] {10.3847/0067-0049/222/2/15}, \href
  {http://adsabs.harvard.edu/abs/2016ApJS..222...15T} {222, 15}

\bibitem[\protect\citeauthoryear{{Tetarenko} et~al.,}{{Tetarenko}
  et~al.}{2016b}]{Tetarenko2016}
{Tetarenko} B.~E.,  et~al., 2016b, \mn@doi [\apj] {10.3847/0004-637X/825/1/10},
  \href {http://adsabs.harvard.edu/abs/2016ApJ...825...10T} {825, 10}

\bibitem[\protect\citeauthoryear{{The LIGO Scientific Collaboration} \& {The
  Virgo Collaboration}}{{The LIGO Scientific Collaboration} \& {The Virgo
  Collaboration}}{2018}]{LIGO2018}
{The LIGO Scientific Collaboration} {The Virgo Collaboration} 2018, arXiv
  e-prints, \href {http://adsabs.harvard.edu/abs/2018arXiv181112940T} {}

\bibitem[\protect\citeauthoryear{{Verbunt}, {Igoshev}  \& {Cator}}{{Verbunt}
  et~al.}{2017}]{Verbunt2017}
{Verbunt} F.,  {Igoshev} A.,   {Cator} E.,  2017, \mn@doi [\aap]
  {10.1051/0004-6361/201731518}, \href
  {http://adsabs.harvard.edu/abs/2017A%26A...608A..57V} {608, A57}

\bibitem[\protect\citeauthoryear{{White} \& {van Paradijs}}{{White} \& {van
  Paradijs}}{1996}]{White1996}
{White} N.~E.,  {van Paradijs} J.,  1996, \mn@doi [\apjl] {10.1086/310380},
  \href {http://adsabs.harvard.edu/abs/1996ApJ...473L..25W} {473, L25}

\bibitem[\protect\citeauthoryear{{Wijnands} et~al.,}{{Wijnands}
  et~al.}{2006}]{Wijnands2006}
{Wijnands} R.,  et~al., 2006, \mn@doi [\aap] {10.1051/0004-6361:20054129},
  \href {http://adsabs.harvard.edu/abs/2006A%26A...449.1117W} {449, 1117}

\bibitem[\protect\citeauthoryear{{Willems}, {Henninger}, {Levin}, {Ivanova},
  {Kalogera}, {McGhee}, {Timmes}  \& {Fryer}}{{Willems}
  et~al.}{2005}]{Willems2005}
{Willems} B.,  {Henninger} M.,  {Levin} T.,  {Ivanova} N.,  {Kalogera} V.,
  {McGhee} K.,  {Timmes} F.~X.,   {Fryer} C.~L.,  2005, \mn@doi [\apj]
  {10.1086/429557}, \href {http://adsabs.harvard.edu/abs/2005ApJ...625..324W}
  {625, 324}

\bibitem[\protect\citeauthoryear{{Wong}, {Valsecchi}, {Fragos}  \&
  {Kalogera}}{{Wong} et~al.}{2012}]{Wong2012}
{Wong} T.-W.,  {Valsecchi} F.,  {Fragos} T.,   {Kalogera} V.,  2012, \mn@doi
  [\apj] {10.1088/0004-637X/747/2/111}, \href
  {http://adsabs.harvard.edu/abs/2012ApJ...747..111W} {747, 111}

\bibitem[\protect\citeauthoryear{Woodburn, Natusch, Weston, Thomasson, Godwin,
  Granet  \& Gulyaev}{Woodburn et~al.}{2015}]{Woodburn2015}
Woodburn L.,  Natusch T.,  Weston S.,  Thomasson P.,  Godwin M.,  Granet C.,
  Gulyaev S.,  2015, \mn@doi [Publications of the Astronomical Society of
  Australia] {10.1017/pasa.2015.13}, 32, e017

\bibitem[\protect\citeauthoryear{{Woosley} \& {Weaver}}{{Woosley} \&
  {Weaver}}{1995}]{Woosley1995}
{Woosley} S.~E.,  {Weaver} T.~A.,  1995, \mn@doi [\apjs] {10.1086/192237},
  \href {http://adsabs.harvard.edu/abs/1995ApJS..101..181W} {101, 181}

\bibitem[\protect\citeauthoryear{{Wysocki}, {Gerosa}, {O'Shaughnessy},
  {Belczynski}, {Gladysz}, {Berti}, {Kesden}  \& {Holz}}{{Wysocki}
  et~al.}{2018}]{Wysocki2018}
{Wysocki} D.,  {Gerosa} D.,  {O'Shaughnessy} R.,  {Belczynski} K.,  {Gladysz}
  W.,  {Berti} E.,  {Kesden} M.,   {Holz} D.~E.,  2018, \mn@doi [\prd]
  {10.1103/PhysRevD.97.043014}, \href
  {http://adsabs.harvard.edu/abs/2018PhRvD..97d3014W} {97, 043014}

\bibitem[\protect\citeauthoryear{{Xu}, {Zhang}, {Reid}, {Zheng}  \&
  {Wang}}{{Xu} et~al.}{2019}]{Xu2019}
{Xu} S.,  {Zhang} B.,  {Reid} M.~J.,  {Zheng} X.,   {Wang} G.,  2019, \mn@doi
  [\apj] {10.3847/1538-4357/ab0e83}, \href
  {https://ui.adsabs.harvard.edu/abs/2019ApJ...875..114X} {875, 114}

\bibitem[\protect\citeauthoryear{{Yan}, {Zhang}, {Zhang}, {Stiele}  \&
  {Yu}}{{Yan} et~al.}{2014}]{Yan2014}
{Yan} Z.,  {Zhang} W.,  {Zhang} H.,  {Stiele} H.,   {Yu} W.,  2014, The
  Astronomer's Telegram, \href
  {http://adsabs.harvard.edu/abs/2014ATel.6649....1Y} {6649}

\bibitem[\protect\citeauthoryear{{Zi{\'o}{\l}kowski}}{{Zi{\'o}{\l}kowski}}{2008}]{Ziolkowski2008}
{Zi{\'o}{\l}kowski} J.,  2008, Chinese Journal of Astronomy and Astrophysics
  Supplement, \href {http://adsabs.harvard.edu/abs/2008ChJAS...8..273Z} {8,
  273}

\bibitem[\protect\citeauthoryear{{Zurita}, {Durant}, {Torres}, {Shahbaz},
  {Casares}  \& {Steeghs}}{{Zurita} et~al.}{2008}]{Zurita2008}
{Zurita} C.,  {Durant} M.,  {Torres} M.~A.~P.,  {Shahbaz} T.,  {Casares} J.,
  {Steeghs} D.,  2008, \mn@doi [\apj] {10.1086/588721}, \href
  {http://adsabs.harvard.edu/abs/2008ApJ...681.1458Z} {681, 1458}

\bibitem[\protect\citeauthoryear{{della Valle}, {Mirabel}  \&
  {Rodriguez}}{{della Valle} et~al.}{1994}]{Della1994}
{della Valle} M.,  {Mirabel} I.~F.,   {Rodriguez} L.~F.,  1994, \aap, \href
  {http://adsabs.harvard.edu/abs/1994A%26A...290..803D} {290, 803}

\bibitem[\protect\citeauthoryear{{van Grunsven}, {Jonker}, {Verbunt}  \&
  {Robinson}}{{van Grunsven} et~al.}{2017}]{Grunsven2017}
{van Grunsven} T.~F.~J.,  {Jonker} P.~G.,  {Verbunt} F.~W.~M.,   {Robinson}
  E.~L.,  2017, \mn@doi [\mnras] {10.1093/mnras/stx2071}, \href
  {http://adsabs.harvard.edu/abs/2017MNRAS.472.1907V} {472, 1907}

\bibitem[\protect\citeauthoryear{{van Paradijs} \& {White}}{{van Paradijs} \&
  {White}}{1995}]{van1995}
{van Paradijs} J.,  {White} N.,  1995, \mn@doi [\apjl] {10.1086/309558}, \href
  {http://adsabs.harvard.edu/abs/1995ApJ...447L..33V} {447, L33}

\makeatother
\end{thebibliography}



\appendix


\bsp	
\label{lastpage}
\end{document}